\documentclass{vgtc}                          % final (conference style)
\ifpdf%                                % if we use pdflatex
  \pdfoutput=1\relax                   % create PDFs from pdfLaTeX
  \usepackage{graphicx}                % allow us to embed graphics files
  \DeclareGraphicsExtensions{.pdf,.png,.jpg,.jpeg} % for pdflatex we expect .pdf, .png, or .jpg files
\else%                                 % else we use pure latex
  \usepackage{graphicx}                % allow us to embed graphics files
  \DeclareGraphicsExtensions{.eps}     % for pure latex we expect eps files
\fi%

\graphicspath{{figures/}{pictures/}{images/}{./}} % where to search for the images

\usepackage{microtype}                 % use micro-typography (slightly more compact, better to read)
\PassOptionsToPackage{warn}{textcomp}  % to address font issues with \textrightarrow
\usepackage{textcomp}                  % use better special symbols
\usepackage{mathptmx}                  % use matching math font
\usepackage{times}                     % we use Times as the main font
\usepackage{cite}                      % needed to automatically sort the references
\usepackage{tabu}                      % only used for the table example
\usepackage{booktabs}                  % only used for the table example
\usepackage{url}
\usepackage{subfig}
\usepackage{multirow}
\usepackage{balance}
\usepackage{amsmath}
\usepackage{amssymb}
\usepackage[dvipsnames]{xcolor}
\newcommand{\tabfig}[2]{\parbox[c]{1.5em}{\includegraphics[width=#1 in]{#2}}}
\usepackage{tabularx}
\usepackage{array}
\newcolumntype{L}[1]{>{\raggedright\let\newline\\\arraybackslash\hspace{0pt}}m{#1}}
\newcolumntype{C}[1]{>{\centering\let\newline\\\arraybackslash\hspace{0pt}}m{#1}}
\newcolumntype{R}[1]{>{\raggedleft\let\newline\\\arraybackslash\hspace{0pt}}m{#1}}
\onlineid{1075}

\vgtccategory{Research}

\title{Listen2Scene: Interactive material-aware binaural sound propagation for reconstructed 3D scenes}

\author{Anton Ratnarajah\thanks{e-mail: jeran@umd.edu}\\ %
        \scriptsize University of Maryland, College Park  %
\and Dinesh Manocha\thanks{e-mail: dmanocha@umd.edu}\\ %
     \scriptsize University of Maryland, College Park} %
\vgtcinsertpkg
\teaser{
  \centering
   \vspace{-0.3cm}
  \includegraphics[width=\linewidth]{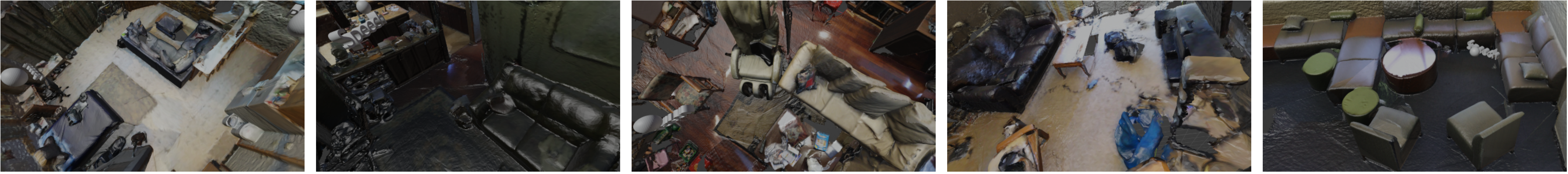} %mesh_net1.eps
  \caption{ We show the 3D reconstructed real-world environments with various levels of complexity that are used to evaluate our learning-based real-time sound propagation method and our audio rendering quality. In practice, our Listen2Scene approach is two orders of magnitude faster than the interactive geometric sound propagation algorithm. In our supplementary demo video, we show that the overall sound quality of Listen2Scene is very similar to the interactive geometric sound propagation algorithm.} %WHAT ARE THE NEW AND NOVEL COMPONENTS
  \label{audio_scene}
}

\abstract {We present an end-to-end binaural audio rendering approach (Listen2Scene) for virtual reality (VR) and augmented reality (AR) applications. We propose a novel neural-network-based binaural sound propagation method to generate acoustic effects for indoor 3D models of real environments. Any clean audio or dry audio can be convolved with the generated acoustic effects to render audio corresponding to the real environment. We propose a graph neural network that uses both the material and the topology information of the 3D scenes and generates a scene latent vector. Moreover, we use a conditional generative adversarial network (CGAN) to generate acoustic effects from the scene latent vector. Our network can handle holes or other artifacts in the reconstructed 3D mesh model. We present an efficient cost function for the generator network to incorporate spatial audio effects. Given the source and the listener position, our learning-based binaural sound propagation approach can generate an acoustic effect in 0.1 milliseconds on an NVIDIA GeForce RTX 2080 Ti GPU. We have evaluated the accuracy of our approach with binaural acoustic effects generated using an interactive geometric sound propagation algorithm and captured real acoustic effects / real-world recordings. We also performed a perceptual evaluation and observed that the audio rendered by our approach is more plausible than audio rendered using prior learning-based and geometric-based sound propagation algorithms. We quantitatively evaluated the accuracy of our approach using statistical acoustic parameters, and energy decay curves. The demo videos, code and dataset are available online \footnote{\url{https://anton-jeran.github.io/Listen2Scene/}}.
} % end of abstract

\CCScatlist{
  \CCScatTwelve{Computing methodologies}{Machine learning}{Machine learning approaches}{Learning latent representations};
}

\begin{document}


\firstsection{Introduction}

\maketitle
\label{into}
Recent advances in computer vision and  3D reconstruction algorithms have made it possible to generate 3D models of real scenes in real-time~\cite{bundle,scannet}. These reconstructed 3D models are used for ray-tracing simulation~\cite{ray-trace_sim}, surveying~\cite{3d_survey}, visual analysis or interactive walkthroughs of buildings~\cite{rent3d}. Furthermore, many tools or systems are available to transform real-life spaces into digital models~\cite{ScanNet++}, which offer higher visual fidelity than panoramic scans. The resulting static 3D models are used to generate immersive 3D experiences for VR or AR applications.

Many reconstructed models corresponding to apartments, houses, offices, public places, malls, or tourist attractions consist of multiple sound sources (e.g., human speaker, dishwasher, telephone, music). In order to improve the sense of the presence for a user, it is important to augment the visual realism with acoustic effects generated by these sources. It is well known that a user's sense of presence in VR or AR environments can be improved by generating plausible sounds
 ~\cite{larsson2002better}. The resulting acoustic effects vary based on the location of each source, the listener and the environment characteristics~\cite{liu2022sound}. In practice,
the acoustic effects in VR or AR environments can be modeled using
impulse responses (IRs), which capture how sound propagates from a source location to the position of the receiver in a given scene.
IRs contain the necessary information for acoustic scene analysis such as the early reflections, late reverberation, arrival time, energy of direct and indirect sound, etc. The IR can be convolved with any dry sound (real or virtual) to apply the desired acoustic effects. Binaural IR characterizes the sound propagation from the sound source to the left and right ears of the listener. Unlike monaural IRs, binaural IRs have sufficient spatial information to locate the sound source accurately. Therefore binaural impulse responses (BIRs) give an immersive experience in AR and VR applications. It turns out that recording the BIRs in real scenes can be challenging and needs special capturing hardware. Furthermore, these BIRs need to be recaptured if the source or listener position changes.

% DEFINE WHAT ARE BINURAL IRs AND WHY IT IS IMPORTANT TO COMPUTE OR CAPTURE THEM (WITH APPROPRIATE REFERENCES)?   

In synthetic scenes, the IRs can be computed in real-time using sound propagation algorithms~\cite{liu2022sound,Savioja2010REALTIME3F}. However, current propagation algorithms are limited to synthetic
scenes where an exact geometric representation of the scene and acoustic material properties are known as apriori. On the other hand, generating a large number of high-quality IRs for complex 3D real scenes in real-time remains a challenging problem~\cite{soundspaces2}.

%However, recording the IRs of realworld scenes can be challenging, especially for interactive applications.
%Many times special recording hardware is needed to record the IRs.
%Furthermore, the IR is a function of the source and listener positions
%and it needs to be re-recorded as either position changes.
%where a user can using mobile devices\cite{neuralrecon} and commodity hardware~\cite{bundle,scannet}, applications that can create a virtual tour for apartments are becoming popular. The virtual tour allows the potential tenant to virtually walk through the apartments and get a more accurate feeling of the apartment when compared to images~\cite{rent3d}. Most virtual tours do not consider the way the sound propagates in the 3D reconstructed apartment. For instance, when the laundry room is placed near the bedroom, the potential buyer can understand whether the washing machine sound will affect their sleep if they can hear the washing machine sound when they virtually walk through the apartment. 

%There can be multiple sound sources (e.g., speaker, dishwasher, telephone etc.) working at the same time in an apartment and the way we hear the sound sources varies with the location of the sound sources, listener, and the 3D scene~\cite{schissler2011gsound:}.

%Our goal is to enable the potential tenant to hear the desired combination of sound sources when virtually walking through the apartment in real time. 

Recently, neural-network-based sound propagation methods to generate IRs have been proposed for interactive audio rendering applications~\cite{fastrir,mesh2ir,luo,inras}. After training, the network can be used to generate a large number of IRs for 3D scenes. However, current learning methods have some limitations. They only deal with the mesh geometry, compute monaural IRs, and do not consider the acoustic material properties of the objects in the 3D scene. The material acoustic properties depend on the surface roughness, thickness and acoustic impedance~\cite{material1,material2}.  
The materials in the 3D scene strongly influence the overall accuracy of the IR by controlling the amount of sound absorption and scattering when propagating sound waves interact with each surface in the scene. Moreover, current methods may not be directly applied to reconstructed 3D scenes with significant holes. 
%Our proposed approach is robust and can handle, significant holes by generating and merging the convex hull of the 3D scene with the reconstructed 3D scene.

%In sound propagation algorithms, the material's acoustic properties are represented using the frequency depended absorption and scattering coefficients.

% YOU SHOULD BRIEFLY EXPLAIN HOW THEY ARE REPRESENTED BASED ON ABSORPTION AND SCATTERING COEFFICIENTS, WHICH ARE FREQUENCY DEPENDENT (AS EXPLAINED IN THE GWA PAPER).
% YOU SHOULD ALSO EXPLAIN THAT THESE METHODS MAY NOT DIRECTLY APPLY TO RECONSTRUCTED REALWORLD MODELS WHICH HAVE NOISE OR HOLES OR OTHER ISSUES. %The accuracy of the IRs  The materials in the 3D scene strongly influence the overall accuracy of the IR by controlling the amount of sound absorption and scattering when the sound touches each surface.

% IS THIS SENTENCE USEFUL, IF SO MAKE IT PART OF MAIN CONTRIBUTIONS We propose a neural network that takes the 3D scene represented using a mesh, the material information of each object present in the 3D scene, and the source and listener positions as input and generates the acoustic impulse responses (IRs) on the fly (Figure 1).

\textbf{Main Results:} We present a novel neural-network-based sound propagation method to render audio for real indoor 3D scenes in real-time. Our approach is general and can generate BIRs for arbitrary topologies and material properties in the 3D scenes, based on the source and listener locations. Our sound propagation network comprises a graph neural network to encode the 3D scene materials and the topology, and a conditional generative adversarial network (CGAN) conditioned on the encoded 3D scene to generate the BIRs. The CGAN consists of a generator and a discriminator network. Some of the novel components of our work include: \\

% We present a novel neural-network-based binaural IR generator Listen2Scene, that can be controlled using real 3D scenes (THIS SENTENCE IS VAGUE; YOU DON'T CONTROL A NETWORK; INSTEAD MAKE IT CLEAR WHAT IS THE INPUT TO YOUR METHOD, AS YOU EMPHASIZE THE END-TO-END NATURE)  captured using commodity hardware with arbitrary topology and material properties, and source and listener positions and generates BIR (BIR) THIS SENTENCE IS TOO LONG, WRITE THE INPUT AND OUTPUT OF YOUR NETWORK AS PART OF A SEPARATE SENTENCE.
% Our network makes no assumptions about the quality of the reconstructed 3D scene (i.e., no holes in the 3D scene).  IS THE RECONSTRUCTION STEP PART OF YOUR NETWORK OR A PREPROCESS.. CAN YOU SAY SOME NEW CHARACTERISCS OF YOUR NETWORK (IS IT A COMBINATION OF ENCODER AND GENERATIVE, OR GRAPH BASED OR OTHER KEY CHARACTERISTIC)
% Some of the novel components of our work include: \\
% \textbf{Material-aware learning-based method:} We calculate the material properties such as the sound absorption and scattering coefficients of each vertex in the 3D scene from the semantic labels. IS THE SEMANTIC LABEL AN INPUT OR YOU COMPUTE THEM. DO YOU ACCOUNT FOR FREQUENCY-DEPENDENT EFFECTS We propose an efficient approach to input material properties in our Listen2Scene architecture, and quantitatively and qualitatively show that adding material information improves the quality of generated BIRs for a 3D scene.\\

\textbf{1. Material-aware learning-based method:}  We represent the material's acoustic properties using the frequency-dependent absorption and scattering coefficients. We calculate these material properties using average sound absorption and scattering coefficients for each vertex in a 3D scene from the input semantic labels of the 3D model and acoustic material databases. We propose an efficient approach to incorporate material properties in our Listen2Scene architecture. Our method results in 48\% better accuracy over prior learning methods in terms of acoustic characteristics of the IRs. \\

%and quantitatively and qualitatively show that adding material information improves the quality of generated BIRs for a 3D scene.\\
% \textbf{BIR synthesis:} We presemt a simple and efficient cost function (HOW IS THIS COST FUNCTION USED IN THE OVERALL NETWORK) to incorporate the difference in the time-of-arrival of sound arriving in left and right ears (interaural time difference)~\cite{ITD} HOW IS THIS RELATED TO HRTF OR SPATIAL AUDIOand sound level difference in both ears caused by the barrier created by the head when the sound is arriving (interaural level difference). \\
\textbf{2. Binaural Impulse Response (BIR) Generation:} We present a simple and efficient cost function to the generator network in our CGAN to incorporate spatial acoustic effects such as the difference in the time-of-arrival of sound arriving in left and right ears (interaural time difference)~\cite{ITD} and sound level difference in both ears caused by the barrier created by the head when the sound is arriving (interaural level difference). \\

% \textbf{Auralizing low-quality 3D reconstructions:} 3D reconstructions of real scenes using commodity hardware (e.g., ScanNet) usually have holes. The holes will prevent the sound from reflecting back to the listener and cause unrealistic acoustic effects. We propose a preprocessing approach to close the holes in the 3D reconstruction and interpolate the material properties of the closed holes in our end-to-end pipeline. \\
% IS THIS PART OF THE NETWORK
% \textbf{Large high-quality IR dataset:} We present an approach to create high-quality BIRs for real 3D scenes captured using commodity hardware.  We generate 1 million high-quality BIRs using the geometric-sound propagation~\cite{geo_int1,pygsound} for around 1500 3D scenes in the ScanNet dataset. Among 1 million BIRs, we randomly sampled 200,000 BIRs to train our network. We will release the full BIR dataset upon publication to support learning-based audio-visual research. \\
\textbf{3. Perceptual evaluation:} We performed a user study to evaluate the benefits of our proposed audio rendering approach. We rendered audio for 5 real environments with different levels of complexity with the number of vertices in the selected environments varying from 0.5 million to 2.5 million (Fig.~\ref{audio_scene}) and asked the participants to choose between our proposed approach and the baseline methods. More than 67\% and 45\% of the participants observed that the audio rendered from Listen2Scene is more plausible than the prior learning-based approach MESH2IR and interactive geometric-based sound propagation algorithm respectively. We also compared the audio rendering using our approach with recorded IRs~\cite{bras} where the materials are an independent variable. \\
%DID YOU COMPARE THE QUALITY WITH ANY REAL-WORLD RECORDINGS OR WITH GEOMETRIC SOUND PROPAGATION METHODS? \\
%HOW DOES THIS COMPLEXITY CHANGES?

\textbf{4. Novel Dataset:} We generate 1 million high-quality BIRs using the geometric-based sound propagation method~\cite{geo_int1} for around 1500 3D real scenes in the ScanNet dataset~\cite{scannet}. Among 1 million BIRs, we randomly sampled 200,000 BIRs to train our network. We release the full BIR dataset in the wav format~\footnote{\url{https://drive.usercontent.google.com/download?id=1FnBadVRQvtV9jMrCz_F-U_YwjvxkK8s0&authuser=0}}.

We have evaluated the accuracy of our approach using the captured BIRs from the BRAS dataset~\cite{bras} and synthetic BIRs generated using the geometric propagation approach for real scenes not used during training. Our network is capable of generating $10,000$ BIRs per second for a given 3D scene on an NVIDIA GeForce RTX 2080 Ti GPU. In practice, we observe two orders of magnitude performance improvement over interactive sound propagation algorithms. 
%Our method results in more than 48\% accuracy over prior learning methods in terms of the acoustic characteristics of IRs.  % and can handle dynamic sources. 

% More than 32\% to 60\% (THIS NUMBER IS VAGUE; ARE YOU SAYING ONLY 60\% ARE BETTER)
%of the participants feel that the acoustic effects generated using our approach are more plausible than MESH2IR.

% MENTION ABOUT ACCURACY ANALYSIS ON BRAS DATASETS; COMPARISON OF QUALITY WITH GSOUND

% EVALUATION ON A LARGE NUMBER OF BENCHMARKS. WHAT IS THE RUNNING TIME OF THE NETWORK TO COMPUTE THE BIRs
%\vspace{-0.3cm}
\section{Related Works}
\label{related}

% \subsection{Sound Propagation and IR Computation}
\textbf{Sound Propagation and IR Computation:} The IRs can be computed using wave-based~\cite{wave2,wave3,waven1,waven2} or geometric~\cite{geometric1,geometric2,schissler2011gsound} sound propagation algorithms. The wave-based algorithms are computationally expensive, and their runtime is proportional to the third or fourth power of the highest simulation frequency~\cite{wave_raghu1}. For interactive applications, IRs are precomputed for a 3D scene grid and IRs are calculated at run time for different listener positions using efficient interpolation techniques~\cite{interpolate1,interpolate2}. Geometric sound propagation algorithms are based on ray tracing or its variants and can be used for interactive applications~\cite{geo_int1,soundspaces2}. They can handle dynamic scenes and work well for high frequencies. Many hybrid combinations of geometric and wave-based methods have been proposed~\cite{diffraction_kernel}. These methods are increasingly used for games and VR applications and can take tens of milliseconds to compute each IR on commodity hardware.\\

% Geometric sound simulation algorithms are a less complex alternative to the wave-based approach. 
% Interactive geometric simulation algorithms are proposed for large 3D scenes with multiple sources~\cite{geo_int1,soundspaces2}. These interactive algorithms compromise the quality of the simulated IR by reducing the number of rays used during simulation.
% Low-quality 3D reconstructions of real scenes (e.g., ScanNet) usually have holes. The holes will prevent the sound to reflect back to the listener and cause unrealistic acoustic effects. We propose an approach close the holes in the 3D reconstruction and interpolate the material properties of the closed holes. Our proposed network can generate BIRs for a given 3D scene with binaural effects such as ITD and ILD. We also present an approach to creating BIR dataset using a geometric based apporach for the ScanNet dataset.
% \subsection{Learning-based IR synthesis}
\textbf{Learning-based sound propagation:} Learning-based sound propagation methods for IR computation have been proposed to generate IRs based on a single image of the environment~\cite{image2reverb,imageRIR,fewshot}, reverberant speech signal~\cite{anton_meta,filtered_noise}, or shoe-box shaped room geometry~\cite{fastrir}. Neural networks are also used to translate synthetic IRs to real IRs and to augment IRs~\cite{ts-rir,ir-gan} and estimate room acoustic parameters~\cite{par1,para2,para3}. Learning-based approaches are proposed to learn the implicit representation of IRs for a given 3D scene and predict IRs for new locations on the same training scene~\cite{luo,inras}.  MESH2IR~\cite{mesh2ir} is a sound propagation network that takes the complete 3D mesh of a 3D scene and the source and the listener positions as input and generates monaural IRs in real-time on a high-end GPU. However, the audio rendered using these learning-based sound propagation methods may not be smooth and can have artifacts. Prior learning-based binaural sound propagation methods require a few BIRs captured in a new 3D scene to generate new BIRs for different source and listener locations in the same 3D scene~\cite{fewshot}. Our learning-based sound propagation method is more accurate and general than prior methods.\\

%In addition to geometry, the materials in the 3D scene contribute to the accuracy of IR synthesis. Learning-based approaches are much faster than prior geometric or wave-based propagation methods. \\
%
% \subsection{real Scenes}
\textbf{Real Scenes}. The materials in the real scene influence the acoustic effects corresponding to the scene. The material information can be estimated from images and videos of real scenes and given as input to sound propagation algorithms using material acoustic coefficients~\cite{pygsound,roundrobin,schissler2017acoustic}. 
%and these coefficients are not readily available for real scenes. 
Other methods are based on capturing reference audio samples or IRs in real scenes and the simulated IRs are adjusted to match the materials using reference audios or IRs~\cite{scene-aware1,scene-aware2,scene-aware3}. In recent works, real scenes are annotated using crowd-sourcing~\cite{scannet} and material acoustic coefficients can be estimated by mapping the real scenes' annotated material labels to materials in the existing acoustic coefficient database~\cite{soundspaces2,GWA}. As compared to these methods, our approach is either significantly faster or generates higher-quality acoustic effects in real scenes.

\section{Model Representation and Dataset Generation} % and Material Computation}
\begin{figure}[h!]
  \centering
  \includegraphics[width=0.9\linewidth]{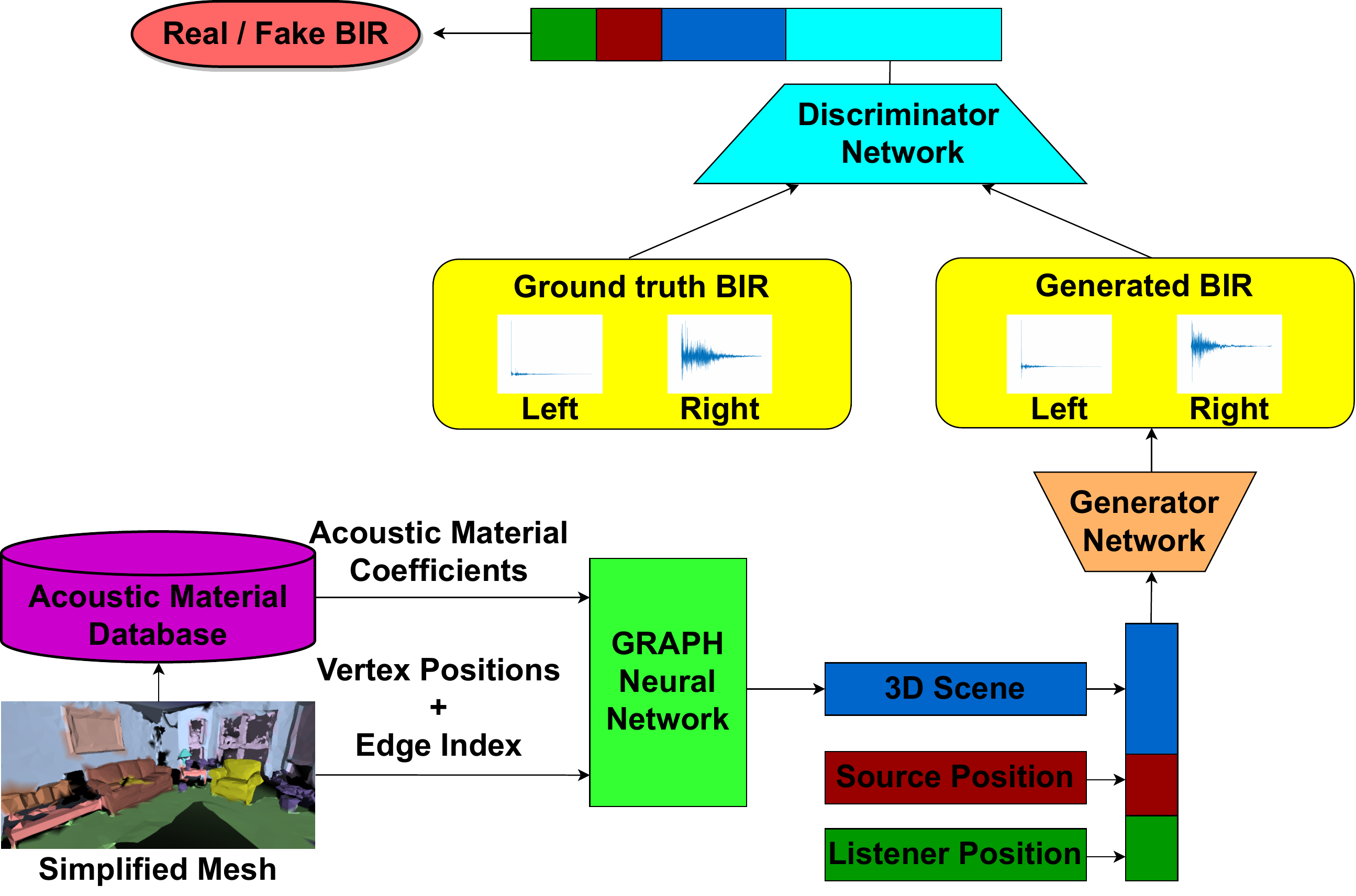} %mesh_net1.eps
  \caption{The overall sound propagation architecture of our Listen2Scene method: The simplified 3D scene mesh with material annotations is passed to the acoustic material database to estimate the acoustic material coefficients (absorption and scattering coefficient). We pass the acoustic material coefficients, vertex positions, and edge index to our graph neural network (Fig.~\ref{graph_network}) to encode the 3D scene into a latent vector. Our generator network takes the 3D Scene and listener and source positions as input and generates a corresponding BIR. The discriminator network discriminates between the generated BIR and the ground truth BIR during training.} %WHAT ARE THE NEW AND NOVEL COMPONENTS
  \vspace{-0.3cm}
  \label{overall_archi}
\end{figure}

Our approach is designed for real scenes. We use 3D reconstructed scenes from the RGB-D data captured using commodity devices (e.g., iPad and Microsoft Kinect). These reconstructed 3D scenes are segmented and the objects in the 3D scene can be annotated by crowdsourcing~\cite{scannet,bundle}. Our goal is to use these mesh representations and semantic information to generate plausible acoustic effects. An overview of our approach is given in Fig.~\ref{overall_archi}.

% YOU ARE DEFINING SYMBOLS IN THIS PARAGRAPH: CAN YOU USE A HEADING LIKE "Notation" BEFORE THAT PARAGRPH

We preprocess the annotated 3D scene to close the holes in the reconstructed 3D scene and simplify the 3D scene by reducing the number of faces. We perform mesh simplification using graph processing to reduce the complexity of the 3D scene input into our network. We represent the simplified 3D scene as a graph $GN$ and input $GN$ to our graph neural network $Net_{GR}$ (Fig.~\ref{graph_network}) to encode the input 3D scene as an 8-dimensional latent vector. Then we pass the encoded 3D scene latent vector along with the listener position $LP$ and the source position $SP$ to our generator network $Net_{GN}$ (Fig.~\ref{overall_archi}) to generate binaural impulse response $BIR$ (Equation~\ref{goal}).
\vspace{-0.2cm}
% and input the simplified 3D scene ($SN$)  YOU SUDDENTLY MENTION THE SIMPLIFIED SCENE? WHY DO YOU NEED SIMPLIFICATION? HOW IS THIS SIMPLIFIED SCENE REPRESENTED to our graph neural network $Net_{GR}$  to encode the 3D scene. YOU USE TWO SUBSCRIPTS (GR and GN) FOR THE NETWORK? CAN YOU DEFINE AND EXPLAIN THESE NOTATIONS FIRST? WHAT EXACTLY DOES A NETWORK $N_{GR}$ REPRESENTS Then we pass the encoded 3D scene latent vector WHAT IS LATENT VECTOR AND WHAT DOES IT REALLY REPRESENTS along with the listener position $LP$ and the source position $SP$ to our CGAN network $Net_{GN}$ to generate binaural IR $BIR$ (Equation~\ref{goal}).
{
%\belowdisplayskip 0.4\belowdisplayshortskip
\begin{equation}\label{goal}
\small
\begin{aligned}[b]
   BIR = Net_{GN}(Net_{GR}(GN),LP,SP).
\end{aligned}
\end{equation}
% CAN YOU CLERL 
} 

We rendered audio $S_{R}$ for the given spatial locations of the receiver and listener in a given 3D scene at time $t$ by convolving the corresponding $BIR$ with any clean or dry audio signal $S_{C}$ (Equation~\ref{eq:render}). 
{
% \belowdisplayskip 0.4\belowdisplayshortskip
\begin{equation}\label{eq:render}
\small
\begin{aligned}[b]
  S_{R}[t] = S_{C}[t] \circledast BIR[t].
\end{aligned}
\end{equation}
% CAN YOU CLERL 
} 

\subsection{Dataset Creation}
\label{dataset_create}
% WHAT IS THE GOAL OF THIS DATASET CREATION PROCESSING

There aren't real-world and synthetic BIR datasets for a wide range of real 3D scenes captured using commodity hardware available to train our Listen2Scene. 
Therefore we create synthetic BIRs using a geometric simulator~\cite{pygsound} for 3D reconstructed real-world scenes in the ScanNet dataset~\cite{scannet} to train our Listen2Scene. We preprocess the 3D meshes and assign meaningful acoustic material properties to each object and surface in the 3D scene (\S~\ref{mesh_preprocess}). Next, we sample source and listener positions and simulate BIRs using the geometric simulator (\S~\ref{geometric_simulate}).

\subsubsection{Mesh Preprocessing and Material Assignment}
\label{mesh_preprocess}
% The ScanNet dataset contains vertex-level segmented mesh with the semantic annotation (i.e., instance-level object category labels such as dish rack, wall, laundry basket etc.). 
The ScanNet dataset contains vertex-level segmented mesh. To make the dataset compatible with a geometric-based sound propagation system, we convert vertex-level segmentation of the 3D scene to face-level segmentation of the 3D scene. Face-level segmentation is used to assign material acoustic coefficients to each surface in the 3D scene. Many of the meshes in the ScanNet dataset have holes in the surface boundary and the ceiling is not present. The holes can prevent some of the sound rays from reflecting back to the listener and result in generating unrealistic acoustic effects using the ray tracing-based geometric sound propagation algorithm. We compute the convex hull of the overall 3D scene mesh and merge it with the original mesh to close the holes in the outer surface boundaries. We fill small holes on internally separated spaces using fill\_holes() function in the trimesh library~\cite{trimesh}.
% IS THAT FOR THE OVERALL MESH OF THE SCENE OR EACH OBJECT
%%%%%%%%%%%%%%%%%SIGGRAPH Para%%%%%%%%%%%%%%%%%%%%%%%%%%%%
% The ScanNet dataset contains the object labels for every 3D scene. We use the absorption coefficient acoustic database with more than 2,000 materials properties~\cite{kling_2018} and assign the acoustic absorption coefficient for each individual surface or object materials in the 3D scenes using the acoustic material assignment approach proposed in GWA~\cite{GWA}. In addition to absorption coefficients, we need scattering coefficients for geometric sound propagation. The scattering coefficients are not available in the acoustic database~\cite{kling_2018}. Therefore, we adapt the sampling approach proposed in GWA~\shortcite{GWA}. We fit a Gaussian distribution by calculating the mean and standard deviation of 37 sets of scattering coefficients collected from the BRAS benchmark~\cite{bras} and we sample randomly from the distribution for every 3D scene.
%%%%%%%%%%%%%%%%%%%%%%%%%%%%%%%%%%%%%%%%%%%%%%%%%%%%%%%%%%%%%%%%%%%%%%%%%%%%%%%%%%%%
% We fit a normal distribution GIVE MORE DETAILS.. HOW ARE THESE 37 SCATTERING COEFFICIENTS GIVEN? WHAT IS THE GOAL OF COMPUTING A NORMAL DISTRIBUTIONS using 37 sets of scattering coefficients collected from the BRAS benchmark~\cite{bras} and we sample  WHAT IS THE SAMPLING CRITERIA from the distribution for every 3D scene.

The ScanNet dataset also contains the semantic annotation (i.e., instance-level object category labels such as dish rack, wall, laundry basket etc.) for every 3D scene. We use the absorption coefficient acoustic database with more than 2000 materials~\cite{kling_2018} to get the absorption coefficient of each material in the ScanNet 3D reconstructions. We do not always find exact ScanNet object labels in the acoustic database. Therefore, we use the natural language processing (NLP) technique to find the closest matching material in the acoustic database for every ScanNet object label and assign its absorption coefficient to the ScanNet object label. To find the closest matching material, we encode the object labels in ScanNet and material names in the acoustic database into fixed-length sentence embeddings~\cite{mishra2019survey}. Transformer-based sentence embedding vectors are close in cosine similarity distance for sentences with similar meaning and outperform in many NLP tasks~\cite{sentence-transformer}. We use the Microsoft pre-trained sentence transformer model to encode materials into 768-dimensional sentence embedding. We use the cosine similarity of the ScanNet object labels and materials in the acoustic database and assign the closest materials absorption coefficients to the objects in the ScanNet. 

In addition to absorption coefficients, we need scattering coefficients for geometric sound propagation. The scattering coefficients are not available in the acoustic database~\cite{kling_2018}. Therefore, we adapt the sampling approach proposed in GWA~\cite{GWA}. We fit a Gaussian distribution by calculating the mean and standard deviation of 37 sets of scattering coefficients collected from the BRAS benchmark~\cite{bras} and we sample randomly from the distribution for every 3D scene.

% The ScanNet dataset contains the object labels (e.g., dish rack, wall, laundry basket etc.) for every 3D scene. We use the absorption coefficient acoustic database with more than 2000 materials~\cite{kling_2018} and map the acoustic absorption coefficient to the 3D scenes in ScanNet using the approach proposed in GWA~\cite{GWA}. 

% When multiple materials have the same cosine similarity, we randomly choose the material in the acoustic database. 
\subsubsection{Geometric Sound Propagation}
\label{geometric_simulate}
For every 3D scene, we perform grid sampling with 1m spacing in all three dimensions. We also ensure that there is a minimum gap of $0.2$ m between the sampled position and objects in the scene to prevent collisions. The number of grid samples varies with the dimension of the 3D scene. We randomly place $10$ sources in the grid sampled locations and the rest of the samples are assigned to listener locations. We perform geometric simulations for every combination of listener and source positions. We use $20,000$ rays for geometric propagation and the simulation stops when the maximum depth of specular or diffuse reflection is 2000 or the ray energy is below the hearing threshold. 
\section{OUR LEARNING APPROACH}
In this section, we present the details of our learning method.
Our approach learns to generate BIRs for 3D reconstructed real scenes, which may have noise or holes. We first present our approach to representing the topology and material details of the 3D scene using our graph neural network (\S~\ref{3d_represent}). Next, we present our overall architecture, which takes the 3D scene and generates plausible BIRs and training details (\S~\ref{bir}).
% BIR helps to locate the direction of the sound source in addition to the distance. For example, when the source is closer to the left ear, there will be a significant time difference between the sound arriving at the left and right ears (interaural time difference). Also, the sound arriving in the left ear will be more intense than the sound arriving in the right ear because of the attenuation of sound when it passes through our head (interaural level difference). The materials present in the room control the reverberation effects of the room. For example, when the room is covered with sound-absorbing materials (e.g., curtain, carpet, cushion, etc.) we hear fewer echoes. Also, when the material surface is rough, the sound energy will scatter in all directions; on smooth materials, there will be a mirror-like reflection of sound waves from the surface.   
\begin{figure}[h!] %t-top, b-bottom, h-exactly where I put %figure (These letters are called 'float')
\centering
\subfloat[3D reconstructed mesh.]{\includegraphics[width=0.4\columnwidth]{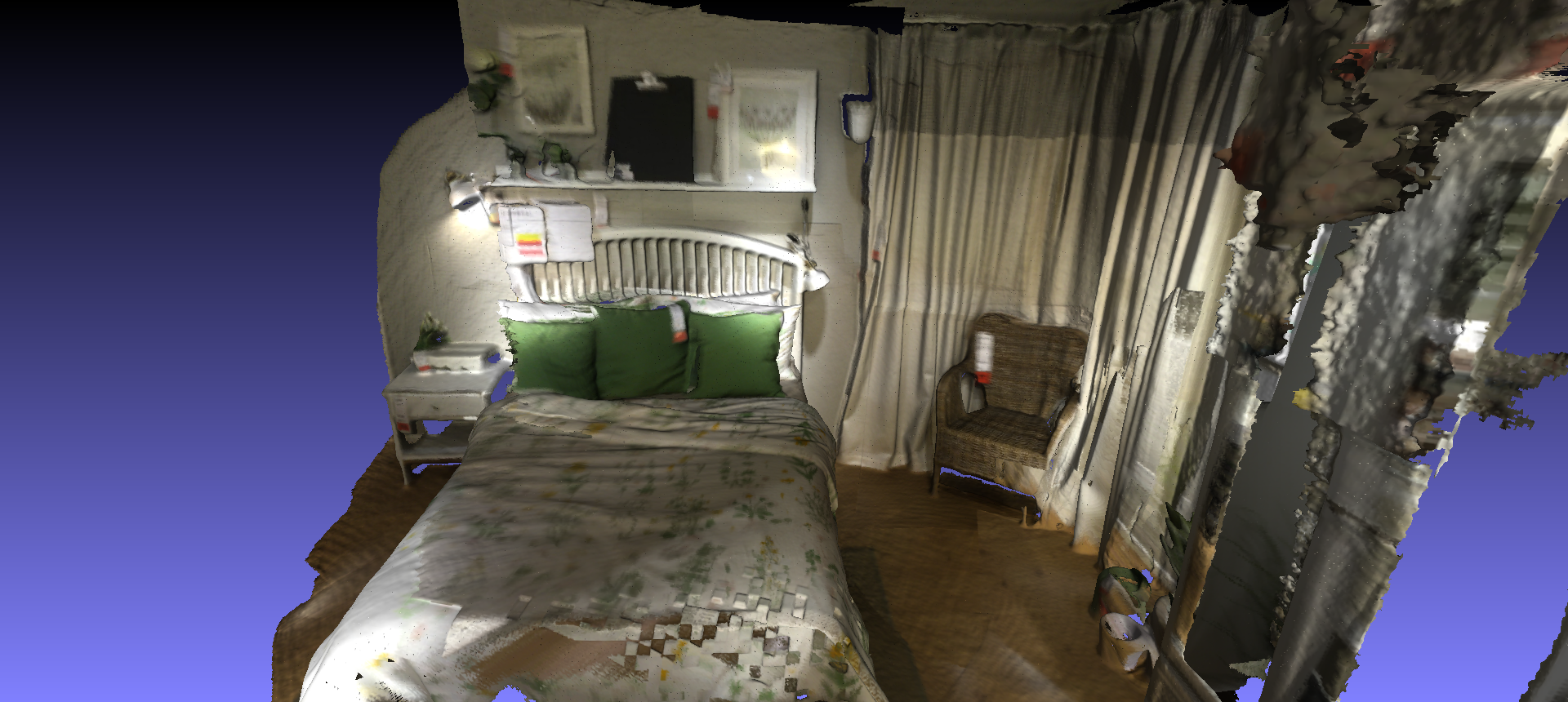}}%original00.png
\quad
\subfloat[Segmented mesh.]{\includegraphics[width=0.4\columnwidth]{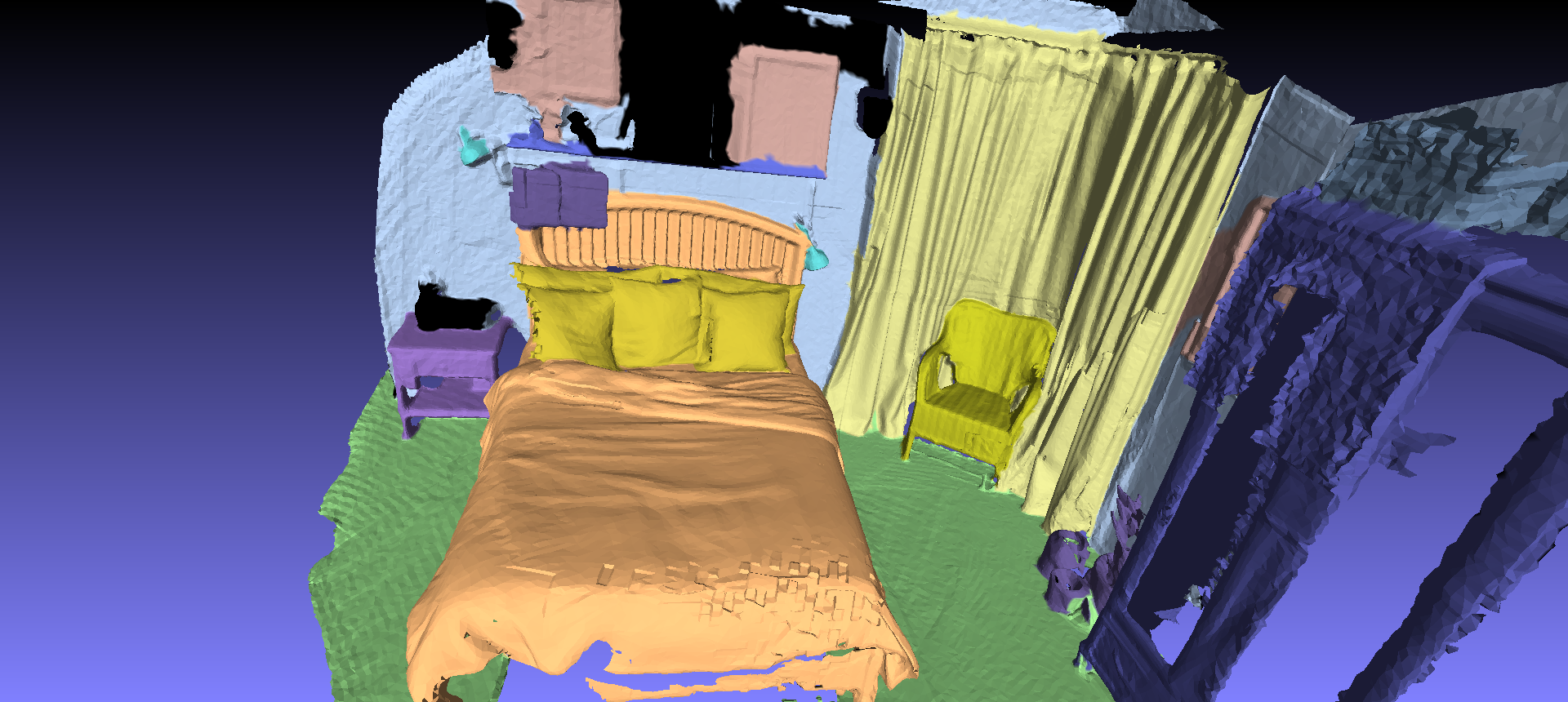}} %segment00.png
\quad
\subfloat[Closed segmented mesh.]{\includegraphics[width=0.4\columnwidth]{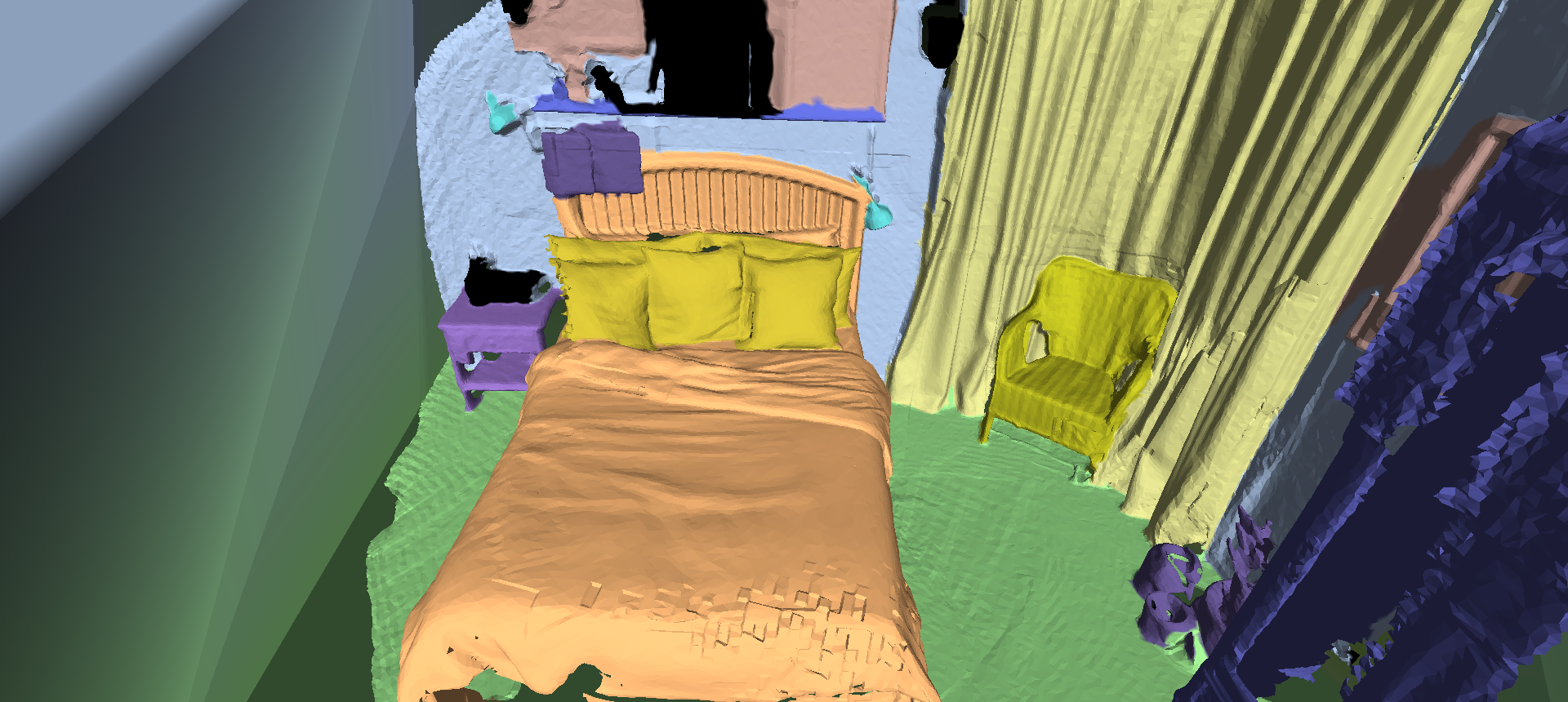}}
\quad
\subfloat[Mesh after simplification.]{\includegraphics[width=0.4\columnwidth]{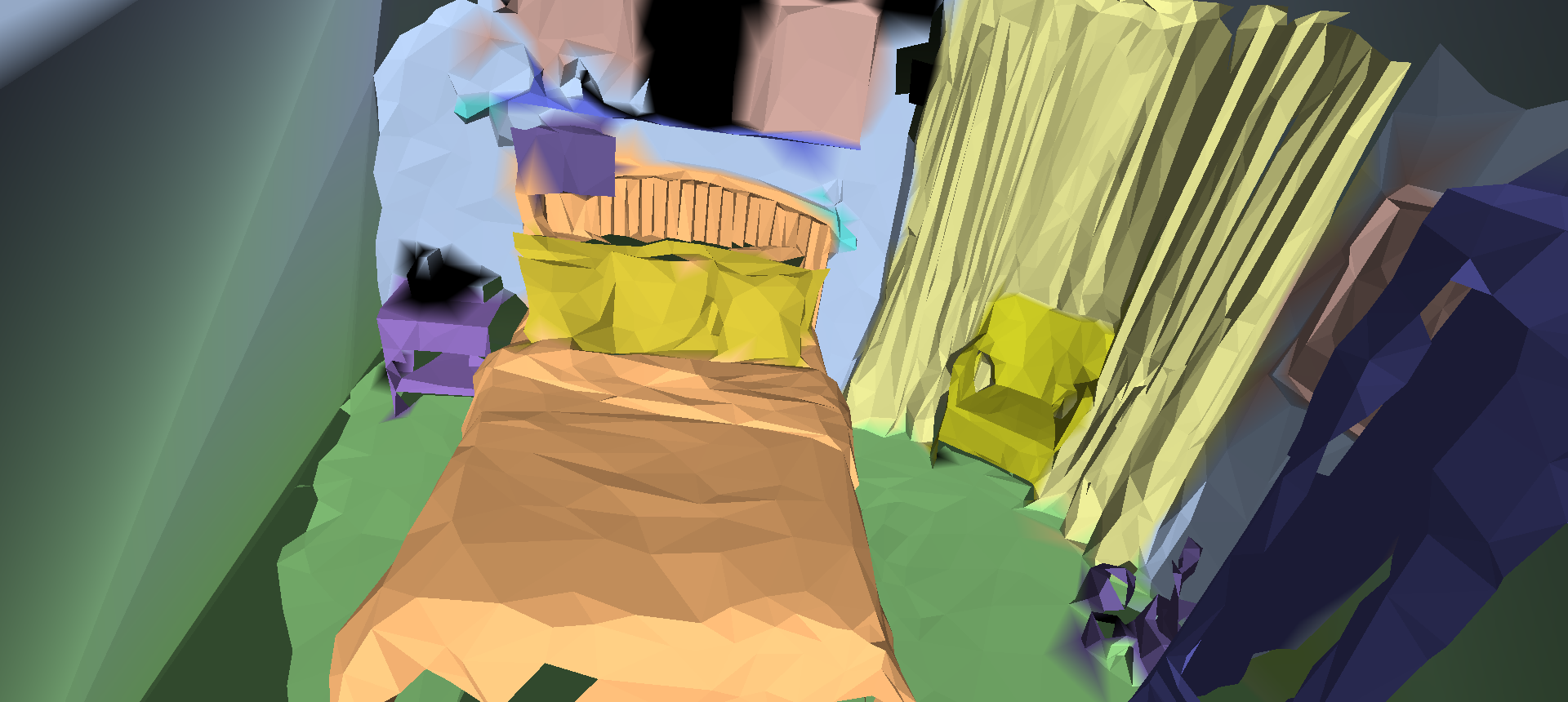}} %simplified00.png
\caption{The 3D reconstruction of the real scene from the ScanNet (a); object category-level segmentation of the 3D scene with each category is represented by a different color (b); the modified mesh after closing the holes using convex hull (c); the simplified mesh with object-level segmentation information preserved (d); we observe that high-level object shapes (e.g., bed, office chair, wooden table, etc.) and materials are preserved even after simplifying the mesh to 2.5\% of the original size.} %(DOES EACH COLOR REPRESENTS A SEPARATE OBJECT, WITH DIFFERENT ACOUSTIC MATERIAL CHARACTERISICS);
% CAN YOU SHOW THE IRs COMPUTED USING ORIGINAL AND SIMPLIFIED MESH. YOU WANT TO SHOW THAT THE IRs DO NOT CHANGE MUCH.
\label{mesh_simplification}
\vspace{-0.3cm}
\end{figure}
\begin{figure}[h!]
  \centering
  \includegraphics[width=0.9\linewidth]{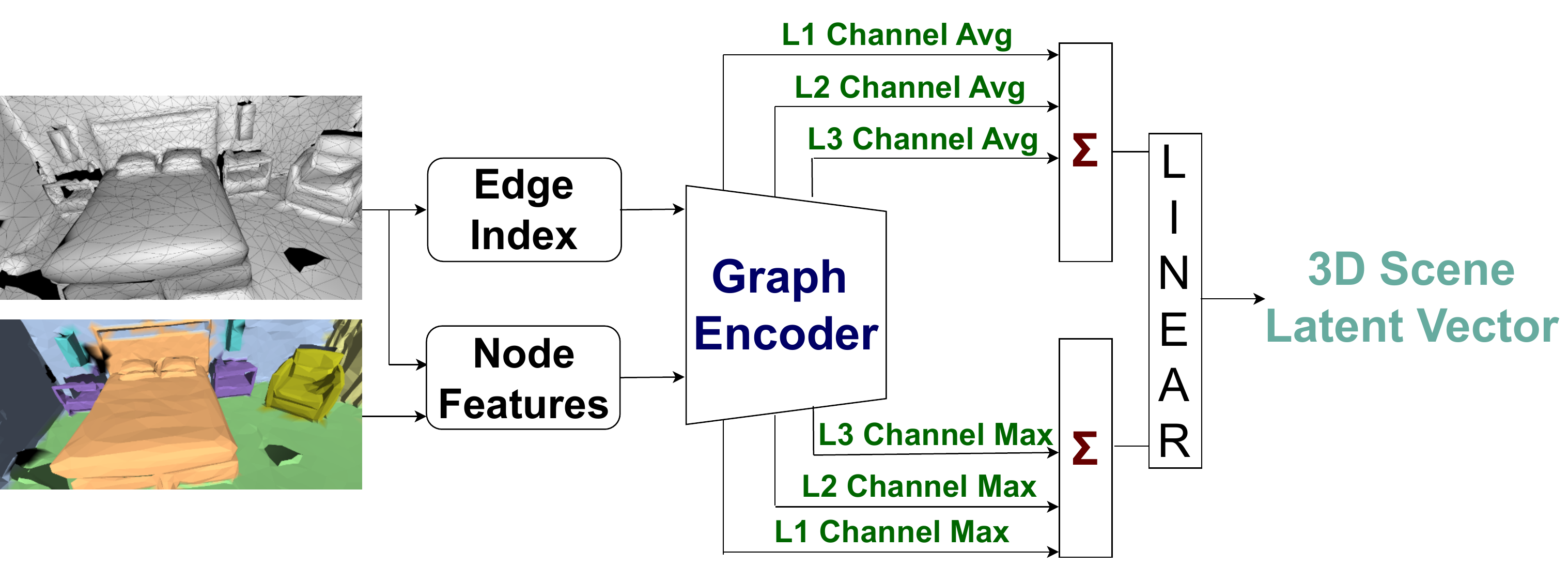} %mesh_net1.eps
  \caption{Our network architecture represents a 3D scene as an 8-dimensional latent vector. The vertex positions and material properties are combined to produce the node features. We pass the edge index and node features from the 3D scene as input to the graph encoder. The graph encoder consists of 3 graph layers (L1, L2, and L3). The channel-wise average and the channel-wise maximum of the node features in each layer are aggregated and passed to linear layers. Linear layers output a 3D scene latent vector.} %WHAT ARE THE NEW AND NOVEL COMPONENTS %In each layer, we have a graph convolutional layer and a graph pool layer (THESE TWO ARE NOT SHOWN). %WHAT ARE THESE LINEAR LAYERS USED FOR
  \label{graph_network}
  \vspace{-0.3cm}
\end{figure}
\subsection{3D Scene Representation}
\label{3d_represent}
The ScanNet dataset represents the RGB-D data collected from the 3D scene in the form of a 3D  mesh. The shapes of the objects in the 3D scene are represented using the vertices and triangular faces in the 3D Cartesian coordinates. The ScanNet dataset also provides object category labels at the vertex level. We perform the mesh pre-processing and material assignment approach as mentioned in \S~\ref{mesh_preprocess}.  % (ARE THESE LABELS RELATED TO VISUAL OR OTHER MATERIALS; OR THEY ARE ACOUSTIC MATERIAS LIKE ABSORPTION COEFFICIENTS) of objects 
% The number of faces in the ScanNet dataset varies from 18471 to 981117. 
To reduce the size and complexity of the data passed to the neural network while preserving high-level object details, we adapt and modify prior work~\cite{mesh2ir} by performing mesh simplifications using PyMeshlab's implementation of the quadratic-based edge collapse mesh simplification algorithm~\cite{pymeshlab}. We simplify the meshes to have only 2.5\% of the initial number of faces. The mesh simplification algorithm can simplify the mesh while preserving the vertex-level segmentation of the mesh (Fig.~\ref{mesh_simplification}). The simplified meshes typically have around 10,000 faces.

In Fig.~\ref{mesh_simplification}, we observe that segmented mesh interpolates the nearby materials to the closed holes (e.g., holes near the floor are assigned to materials of the floor and the material is represented in green). We observe that even after mesh simplification to 2.5\% of the original size, high-level object structures are preserved.

The triangular mesh of the 3D scene can be represented using graph $G =\langle V,E \rangle$, where $V$ represents the 3D Cartesian coordinates of the set of vertices/nodes and $E$ is the connectivity of each node (edge index). The vertex coordinates of three dimensions are features of the node in a graph. To add the material properties of the 3D scene, we increase the node feature dimension to five. The material properties can be represented using the material's absorption coefficient and scattering coefficient. The absorption coefficient represents how much sound can be absorbed by the material. Metal absorbs the least sound and has a very low coefficient. A cushion is a sound-absorbing material and has a high coefficient. The scattering coefficient represents the roughness of the material's surface. When the surface is rough, the sound will be scattered in all directions and has a high coefficient; smooth surfaces have a low coefficient value.
% (WHAT ARE THE TWO ADDITIONAL DIMENSIONS, ACOUSTIC AND SCATTERING COEFFICIENTS; DON'T YOU USE DIFFERENT VALUES FOR EACH BAND. IF THERE ARE k BANDS, THERE WOULD BE 2K MATERIAL BANDS)
The absorbing and scattering coefficients are frequency-dependent coefficients. The coefficients are defined for the 8-octave bands between 62.5 Hz and 8000 Hz. To reduce the dimensionality of the coefficients, we calculate the average coefficients by taking the coefficients at 500 Hz and 1000 Hz. We show the benefit of our approach of calculating average coefficients at 500 Hz and 1000 Hz in \S~\ref{acc_analysis}. In many practical applications, the average value of room acoustics parameters like reverberation time is used for analysis instead of all the values at different octave bands~\cite{scene-aware3,soundspaces2,ts-rir}. We increase the node features $V$ by combining $(x,y,z)$ Cartesian coordinates of the vertex with the average absorption coefficient $ab$ and average scattering coefficient $sc$ ($V = [x,y,z,ab,sc]$).

% {
% \belowdisplayskip 0.4\belowdisplayshortskip
% \begin{equation}\label{node_features}
% \begin{aligned}[b]
%    V = (x,y,z,ab,sc).
% \end{aligned}
% \end{equation}
% } 

We input node features and edge index to the graph encoder network to encode the 3D scene to a low dimensional space. The encoder network has 3 layers. In each layer, the graph convolution layer~\cite{GCN} is used to encode the node features (Equation~\ref{overall_networ3}). We gradually reduce the size of the graph by dropping the number of node features to 0.6 times the original number of node features in each layer using the graph pooling layer.  %(WHY 3 LAYERS)

In Equation~\ref{overall_networ3}, the adjacency matrix representing the edge index of the 3D scene ($A$) and the identity matrix $I$ are aggregated to calculate $\hat{A}$  ($\hat{A}  = A + I $). Each column of $\hat{A}$ is summed to get diagonal matrix $\hat{D}$ ($\hat{D}_{ii}  = \sum_{j} \hat{A}_{ij}$). $W^{(n)}$ is a trainable weight matrix for layer $n$. Node features at layers $n$ and $n+1$ are $N_{F}^{(n)}$ and $N_{F}^{(n+1)}$, respectively. %WHAT IS THE GOAL OF USING SUCH AN ADJANCENCY MATRIX? WHAT DOES IT REPRESENT IN THE OVERALL LEARNING PIPELINE
{
% \belowdisplayskip 0.4\belowdisplayshortskip
\begin{equation}\label{overall_networ3}
\small
\begin{aligned}[b]
   N_{F}^{(n+1)} = \sigma(\hat{D}^{-\frac{1}{2}}\hat{A}\hat{D}^{-\frac{1}{2}}N_{F}^{(n)}W^{(n)}),
\end{aligned}
\end{equation}
} 
%CAN YOU SHOW THESE LAYERS IN Figure 4? 
The output of the graph convolution layer is passed to the graph pooling layers~\cite{Kpool1,Kpool2} to simplify the graph by reducing the node features and edge index. The graph pooling layer initially calculates the square of the adjacency matrix ($A_{new}^{(n)} = A^{(n)} A^{(n)}$) to increase the graph connectivity and is used to choose the top N node features. The adjacency matrix $A_{new}^{(n)}$ prevents isolated edges in the graph encoded 3D scene when choosing top N node features from the input graph and discarding other features. %WHAT DOES N CORRESPONDS TO (GIVE SOME INTUITION)

We calculate the channel-wise average and channel-wise maximum of the output node features in each graph layer in the graph encoder network. We aggregate the channel-wise average and channel-wise maximum separately over the 3 layers. We concatenate the aggregated maximum and aggregated average values and pass them as input to a set of linear layers. We concatenate the learned features in each layer to ensure that the linear layers use all the learned features to construct an accurate 3D scene latent vector of dimension 8 as an output from the linear layer. %HOW ARE CHANNEL WISE AVERAGE AND CHANNEL WISE MAXIMUM CMOPUTED (NOT SHOWN IN Figure 4) 
% %\vspace{-0.3cm}
\subsection{BIR Generation}
\label{bir} %WHY DO YOU USE SUCH CGAN (MOTIVATE THE CHOICE)
We use a one-dimensional modified conditional generative adversarial network (CGAN) to generate BIRs. The standard CGAN architectures~\cite{CGAN1,CGAN2} generate multiple different samples corresponding to input condition $y$ by changing the input random noise vector $z$. In our CGAN architecture, we only input the condition $y$ to generate a single precise output. Our CGAN network takes a 3D scene latent vector as the input condition and generates a single precise BIR. We propose a novel cost function to trigger the network to generate binaural effects such as interaural level difference (ILD) and interaural time difference (ITD) accurately. 
%The previously proposed energy decay relief (ED) cost function~\cite{mesh2ir} does not capture the energy decay in the latter part of the BIR accurately. Therefore, we present a modified ED cost function. 
 % WHY DO YOU CHOOSE 48K SAMPLING RATE? WHAT ARE THE IMPLICATIONS OF HIGHER OR LOWER SAMPLING RATES

We extend the IR preprocessing approach proposed in MESH2IR to make the network learn to generate BIRs with large variations of standard deviation (SD) efficiently. In \S~\ref{dataset_create}, we generate high-fidelity BIRs with a sampling rate of 48,000 Hz. We initially downsample the BIRs to 16,000 Hz to represent a longer duration of BIRs. We train our network to generate around 0.25 seconds (3968 samples) of BIR to reduce the complexity of the network. Our architecture can be easily modified to train the network to generate any duration of BIRs. The complexity of our network changes linearly with the duration of generated BIRs. We calculate the SD of the BIR and divide the BIR with SD to have fewer variations over training samples. We replicate the SD 128 times and concatenate it towards the end of the BIR. Therefore, each channel of the preprocessed BIR will have 4096 samples (3968+128). We train our network to generate preprocessed BIRs. Later, we can recover the original BIR by removing SD represented in the last 128 samples, getting the average of SD values, and multiplying the first 3968 samples by the average SD value. We get the average SD over 128 samples to reduce the error of the recovered SD.

Our CGAN architecture consists of a generator network ($G$) and a discriminator network ($D$) (Fig.~\ref{overall_archi}). We pass the 3D scene information $\Gamma_{S}$ consisting of mesh topology and materials of the 3D scenes represented using a latent vector, and the listener and source position as an input to $G$. We train the $G$ and the $D$ in our CGAN architecture using our created BIRs (\S~\ref{dataset_create}) and $\Gamma_{S}$ in the data distribution $p_{data}$. We train $G$ to minimize the objective function $\mathcal{L}_{G}$ and the $D$ to maximize the objective function $\mathcal{L}_{D}$ alternatingly.

\textbf{Generator Objective Function ($\mathcal{L}_{G}$) : } The $\mathcal{L}_{G}$ is minimized during training to generate accurate BIRs for the given condition $\Gamma_{S}$.
The $\mathcal{L}_{G}$ (Equation~\ref{generator_loss}) consists of modified CGAN error ($\mathcal{L}_{CGAN}$), BIR error ($\mathcal{L}_{BIR}$), ED error ($\mathcal{L}_{ED}$), and mean square error ($\mathcal{L}_{MSE}$). The contribution of each individual error is controlled using the weights $\lambda_{BIR}$, $\lambda_{ED}$ and $\lambda_{MSE}$:
{
% \belowdisplayskip 0.1\belowdisplayshortskip
\begin{equation}\label{generator_loss}
\small
\begin{aligned}[b]
    \mathcal{L}_{G} = \mathcal{L}_{CGAN} + \lambda_{BIR} \; \mathcal{L}_{BIR} + \lambda_{ED} \; \mathcal{L}_{ED} + \lambda_{MSE} \; \mathcal{L}_{MSE}.
\end{aligned}
\end{equation}
}

The modified CGAN error is minimized when the BIRs generated using $G$ are difficult to differentiate from the ground truth BIRs by $D$ for each 3D scene $\Gamma_{S}$:
{
% \belowdisplayskip 0.1\belowdisplayshortskip
\begin{equation}\label{CGAN_loss}
\small
\begin{aligned}[b]
    \mathcal{L}_{CGAN} = \mathbb{E}_{\Gamma_{S} \sim p_{data}}[\log(1 - D(G(\Gamma_{S}),\Gamma_{S}))].
\end{aligned}
\end{equation}
} 
The time of arrival of the direct signal and the magnitude levels of the left and right channels of the BIRs vary significantly with the direction of the sound source. To make sure the network captures the relative variation of the IRs in the left and right channels, we propose the BIR error formulation.   
{
% \belowdisplayskip 0.1\belowdisplayshortskip
\begin{equation}\label{bir_loss}
\small
\begin{aligned}[b]
    \mathcal{L}_{BIR} = \mathbb{E}_{(B_{G},\Gamma_{S}) \sim p_{data}}[\mathbb{E}[((B_{LN}(\Gamma_{S},s) - B_{RN}(\Gamma_{S},s)) %\\
    - (B_{LG}(\Gamma_{S},s) - B_{RG}(\Gamma_{S},s)))^{2}]],
\end{aligned}
\end{equation}

}
where $B_{LN}$ and $B_{RN}$ are the left and right channels of the BIRs generated using our network and $B_{LG}$ and $B_{RG}$ are the left and right channels of the ground truth BIRs.

The energy remaining in the BIR ($b$) with respect to the time $t_{i}$ seconds and at frequency band with center frequency $f_{c}$ Hz (Equation~\ref{EDRelief}) is described using energy decay relief (ED)~\cite{EDR,EDC}. In Equation~\ref{EDRelief}, the bin $c$ of the short-time Fourier transform of $b$ at time $t$ is defined as $H(b,t,c)$. The ED curves decay smoothly over time and they can be converted into an "equivalent IR"~\cite{energycurve1}. In previous works~\cite{mesh2ir,anton_meta}, it is observed that ED helps the model to converge.
{
%\belowdisplayskip 0.4\belowdisplayshortskip
\begin{equation}\label{EDRelief}
\small
\begin{aligned}[b]
  ED(b,t_{i},f_c)  = \sum_{t=i}^T |H(b,t,c)|^2.
\end{aligned}
\end{equation}
} 
The ED curves reduce exponentially over time. In previous works~\cite{mesh2ir}, the mean square error (MSE) between the ED curves of the ground truth BIR ($B_{G}$) and the generated BIR ($B_{N}$) is calculated. This approach does not capture the latter part of ED curves accurately. Therefore we compare the log of the ED curves between ground truth and generated BIRs for each sample ($s$) as follows:
{
% \belowdisplayskip 0.1\belowdisplayshortskip
\begin{equation}\label{ed_loss}
\small
\begin{aligned}[b]
    \mathcal{L}_{ED} = \mathbb{E}_{(B_{G},\Gamma_{S}) \sim p_{data}} [\mathbb{E}_{c \sim C}[\mathbb{E}[(\log(ED(B_{G}(\Gamma_{S}),c,s)) %\\
    - \log(ED(B_{N}(\Gamma_{S}),c,s)))^{2}]]].
\end{aligned}
\end{equation}

}
To capture the structures of the BIR, we also calculate MSE error in the time domain. For each 3D scene $\Gamma_{S}$ we compare $B_{G}$ and $B_{N}$ over the samples ($s$) of BIR as follows:
{
% \belowdisplayskip 0.4\belowdisplayshortskip
\begin{equation}\label{mse_loss}
\small
\begin{aligned}[b]
    \mathcal{L}_{MSE} = \mathbb{E}_{(B_{G},\Gamma_{S}) \sim p_{data}}[\mathbb{E}[(B_{G}(\Gamma_{S},s) - B_{N}(\Gamma_{S},s))^{2}]].
\end{aligned}
\end{equation}
}

\textbf{Discriminator Objective Function ($\mathcal{L}_{D}$) : } 
The discriminator ($D$) is trained to maximize the objective function $\mathcal{L}_{D}$ (Equation~\ref{discriminator_loss}) to differentiate the ground truth BIR ($B_{G}$) and the BIR generated using the generator ($G$) during training for each 3D scene $\Gamma_{S}$.
{
% \belowdisplayskip 0.4\belowdisplayshortskip
\begin{equation}\label{discriminator_loss}
\small
\begin{aligned}[b]
    \mathcal{L}_{D} = \mathbb{E}_{(B_{G},\Gamma_{S}) \sim p_{data}}[\log(D(B_{G}(\Gamma_{S}),\Gamma_{S}))] 
    + \mathbb{E}_{\Gamma_{S} \sim p_{data}}[\log(1 - D(G(\Gamma_{S}),\Gamma_{S}))].
\end{aligned}
\end{equation}
}

\textbf{Network Architecture and Training:}
We extend the standard time domain Generator ($G$) and Discriminator ($D$) architectures proposed for monaural IR generation~\cite{fastrir,mesh2ir}. We modify $G$ to take our 3D scene latent vector of 8 dimensions (Fig.~\ref{graph_network}) and the source and listener positions in 3D Cartesian coordinates. Our $G$ takes 14-dimensional conditional vectors and generates 4096x2 preprocessed BIR as output.
We also modify our $D$ to differentiate between two channel ground truth and generated BIRs. We train all networks with a batch size of 96 using an RMSprop optimizer. The hyperparameter is chosen manually by looking at how the network converges at the initial epochs. We initially started with a learning rate of 8 x $10^{-5}$, and the learning rate decayed to 0.7 of its previous value every 7 epochs. We trained our network for 100 epochs. %We will release the code upon publication.

\section{Ablation Experiments}
\label{ablation}
We perform ablation experiments to analyze the contribution of our proposed BIR error (Equation~\ref{bir_loss}) and Energy Decay (ED) error (Equation~\ref{ed_loss}) in training our network. We also analyze the performance of the network with and without closing holes in the 3D mesh. We generated 900 BIRs for 20 real testing environments for our ablation study.\\
\vspace{-0.3cm}
\subsection{BIR Error}
\begin{figure}[h!]
  \centering
  \includegraphics[width=0.8\linewidth]{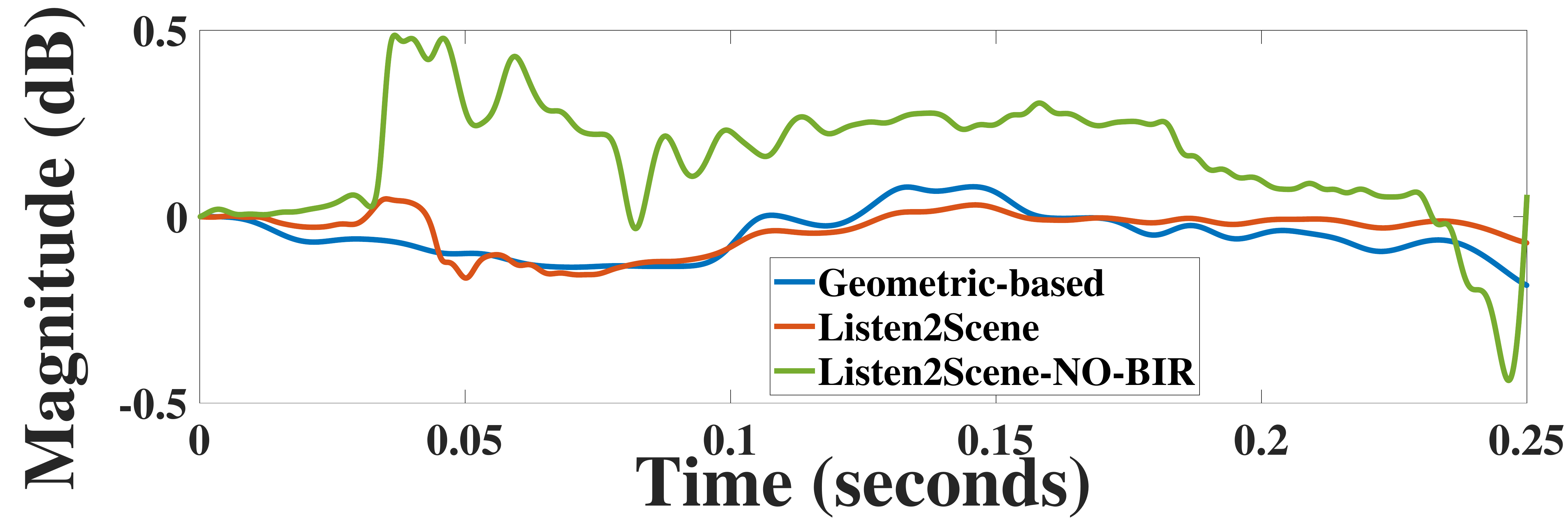} %mesh_net1.eps
  \caption{The normalized difference in energy decay (ED) curves of left and right channels of BIR. The BIRs are generated using the geometric method, Listen2Scene and Listen2Scene-No-BIR (Listen2Scene trained without BIR error). We observe that the ED curve difference of Listen2Scene closely matches the geometric method.} %when compared to Listen2Scene-No-BIR %WHAT ARE THE NEW AND NOVEL COMPONENTS %In each layer, we have a graph convolutional layer and a graph pool layer (THESE TWO ARE NOT SHOWN). %WHAT ARE THESE LINEAR LAYERS USED FOR
  \label{BIR_ablation}
  % \vspace{-0.3cm}
\end{figure}

Our BIR error (Equation~\ref{bir_loss}) helps to generate binaural acoustic effects by incorporating magnitude level differences between the left and right channels of BIRs. In Fig.~\ref{BIR_ablation}, we plot the difference between the left and right channels of the ED curve of BIRs generated using a geometric-based approach, Listen2Scene and Listen2Scene approach trained without BIR error (Listen2Scene-No-BIR). We can observe that incorporating BIR error reduces the gap between the geometric approach (ground truth) and our Listen2Scene.\\

\vspace{-0.5cm}
\subsection{ED Error}
\begin{figure}[h!]
  \centering
  \includegraphics[width=0.8\linewidth]{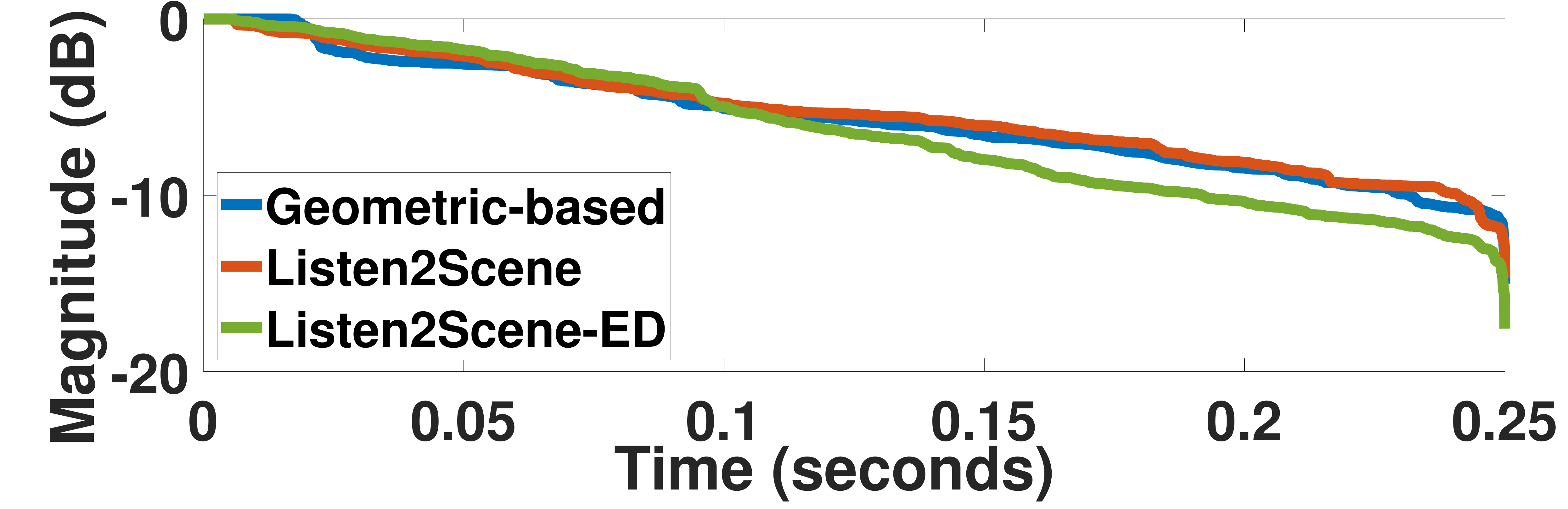} %mesh_net1.eps
  \caption{The normalized energy decay (ED) curve of the BIRs (left channel) generated using the geometric-based method, Listen2Scene and Listen2Scene-ED (Listen2Scene trained with ED error proposed in MESH2IR~\cite{mesh2ir}) at 2000 Hz. We can see that the ED curve of Listen2Scene matches the geometric method for the entire duration while the ED curve of Listen2Scene-ED starts diverging after 0.1 seconds.} %WHAT ARE THE NEW AND NOVEL COMPONENTS %In each layer, we have a graph convolutional layer and a graph pool layer (THESE TWO ARE NOT SHOWN). %WHAT ARE THESE LINEAR LAYERS USED FOR
  \label{ED_ablation}
  \vspace{-0.3cm}
\end{figure}

We trained our Listen2Scene network with the ED error proposed in MESH2IR~\cite{mesh2ir} (Listen2Scene-ED) and our proposed ED error (Equation~\ref{ed_loss}). We calculated the MSE between the normalized ED curves of the ground truth BIRs from the geometric-based approach and the generated BIRs over the center frequencies 125Hz, 500Hz, 1000Hz, 2000Hz and 4000 Hz covering voice frequency and reported in Table~\ref{ED_table_loss}. We can see that MSE of the normalized ED curves in the testing environment is low for our proposed ED error (Listen2Scene). Fig.~\ref{ED_ablation}, shows the normalized ED curves of the left channel BIR from the geometric-based method, Listen2Scene and Listen2Scene-ED at 2000Hz. We can see that the ED curve of Listen2Scene-ED diverges from the geometric-based method after 0.1 seconds.\\

\begin{table}[h!]
  \setlength{\tabcolsep}{1.8pt}
	\caption{The MSE error between the normalized energy decay (ED) curves of the ground truth BIRs from the geometric sound propagation algorithm and the generated BIRs from our Listen2Scene and Listen2Scene trained with ED error proposed in MESH2IR~\cite{mesh2ir} (Listen2Scene-ED). We calculate the MSE over the center frequencies 125Hz, 500Hz, 1000Hz, 2000Hz and 4000 Hz. The best results are shown in~\textbf{bold}}
    %\vspace{-0.3cm}
	\label{ED_table_loss}
	\centering
	\begin{tabular}{@{}lcccccc@{}}	%c-no.of columns. |-verticle line between columns
		\toprule
		 \textbf{Method}
			& \multicolumn{5}{c}{\textbf{Frequency}}\\
\cmidrule(r{1pt}){2-6} 
		&	\textbf{125Hz}	 &\textbf{500Hz}	&\textbf{1000Hz} &\textbf{2000Hz}&\textbf{4000Hz} \\  
		\midrule

	Listen2Scene-ED &2.58 & 3.28&3.99& 4.16 & 4.23 \\
	\textbf{Listen2Scene}&  \textbf{2.50} & \textbf{2.93} & \textbf{3.54}& \textbf{3.56} & \textbf{3.56} \\

		\bottomrule
	\end{tabular}
	% %\vspace{-0.3cm}
\end{table} 

\vspace{-0.5cm}
\subsection{Closed and Open Mesh Models}

\begin{figure}[h!] %t-top, b-bottom, h-exactly where I put %figure (These letters are called 'float')
\centering
\subfloat[Geometric-based.]{\includegraphics[width=0.45\columnwidth]{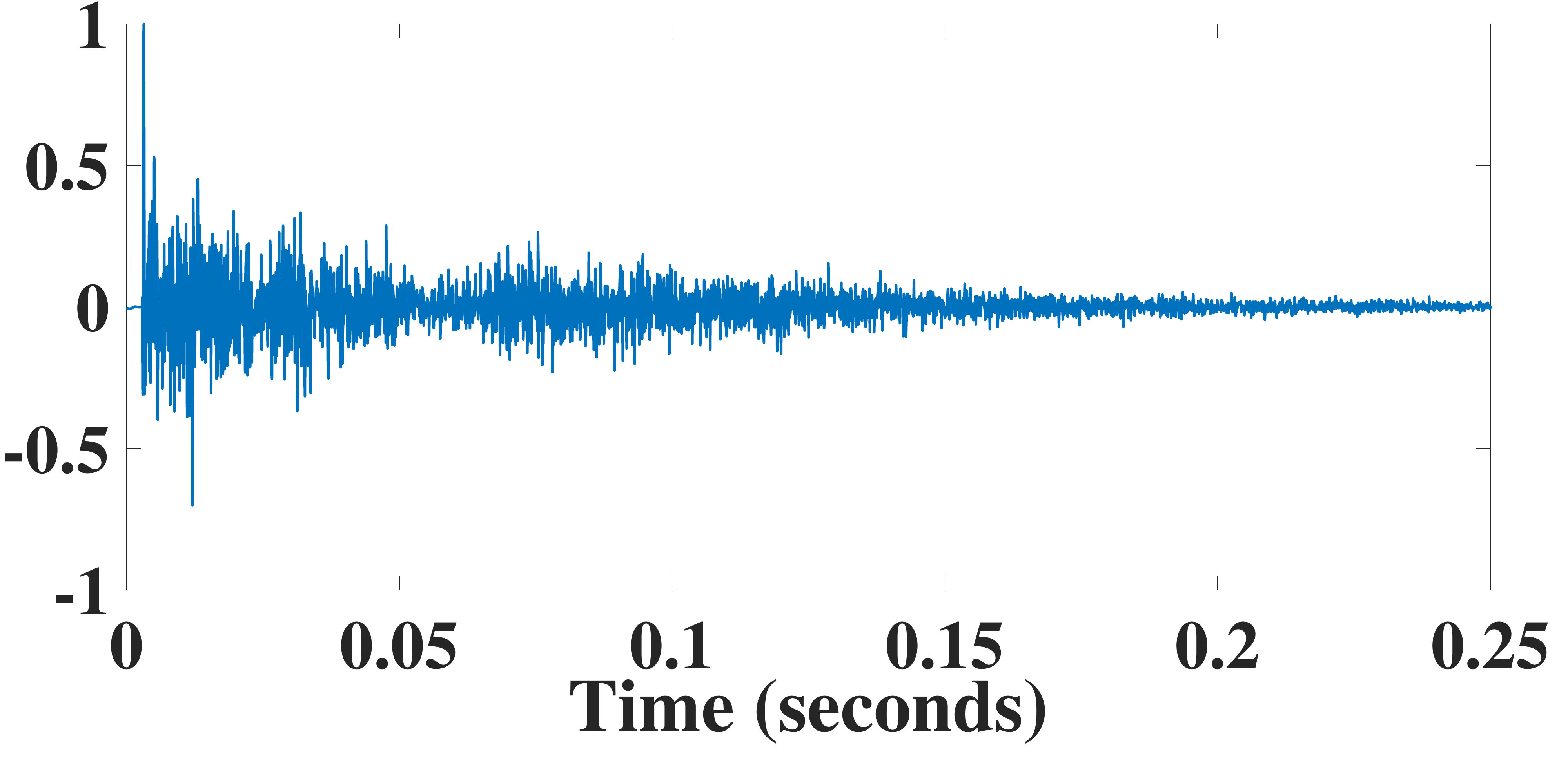}}%original00.png
\quad
\subfloat[Listen2Scene-Hole.]{\includegraphics[width=0.45\columnwidth]{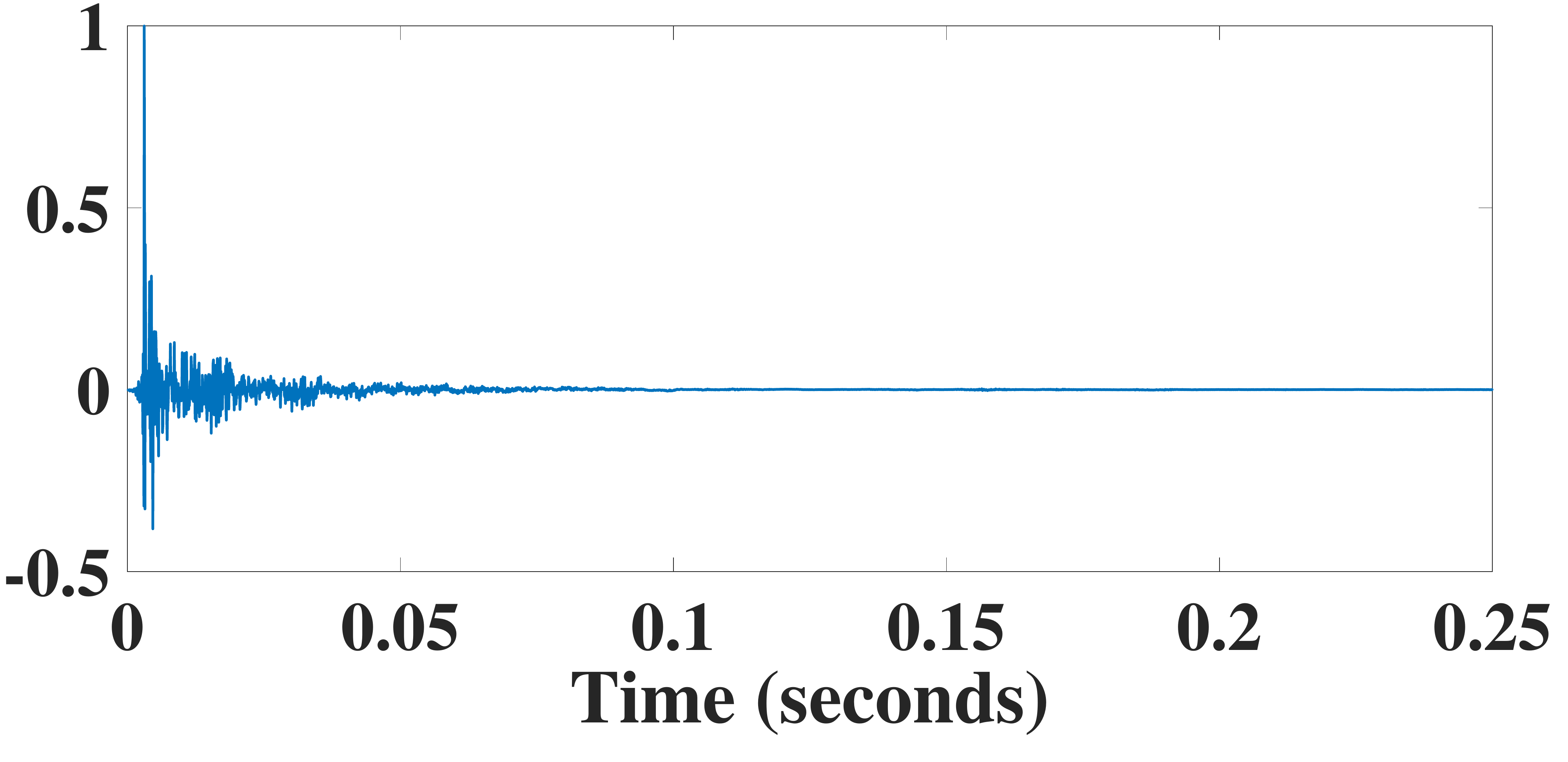}} %segment00.png

\caption{The left channel of the BIR generated using a geometric-based sound propagation algorithm and our Listen2Scene approach without closing the holes (Listen2Scene-Hole). We can see that the BIR from Listen2Scene-Hole significantly varies from the geometric-based approach.} %(DOES EACH COLOR REPRESENTS A SEPARATE OBJECT, WITH DIFFERENT ACOUSTIC MATERIAL CHARACTERISICS);
% CAN YOU SHOW THE IRs COMPUTED USING ORIGINAL AND SIMPLIFIED MESH. YOU WANT TO SHOW THAT THE IRs DO NOT CHANGE MUCH.
\label{hold_mesh}
% %\vspace{-0.3cm}
\end{figure}

We trained and evaluated our Listen2Scene network using the default 3D mesh with holes (Listen2Scene-Hole) and a closed mesh using our proposed approach (\S~\ref{mesh_preprocess}). We can see in Table~\ref{tab:plot} that the BIRs generated using Listen2Scene match the geometric-based sound propagation algorithm. Fig.~\ref{hold_mesh}, shows the left channel of the BIR from the geometric-based approach and the corresponding BIR from Listen2Scene-Hole. We can see that the BIRs from the geometric-based approach and Listen2Scene-Hole are significantly different. 

% This is because the sound won't be reflected to the listener due to significant holes present in the surface of the 3D environment mesh; a significant portion of early and late reflection is missed in the BIR generated using Listen2Scene-Hole.

\begin{figure}[h!] %t-top, b-bottom, h-exactly where I put %figure (These letters are called 'float')
\centering
%\vspace{-0.3cm}
\subfloat[Chamber music hall (Medium).]{\includegraphics[width=0.45\columnwidth]{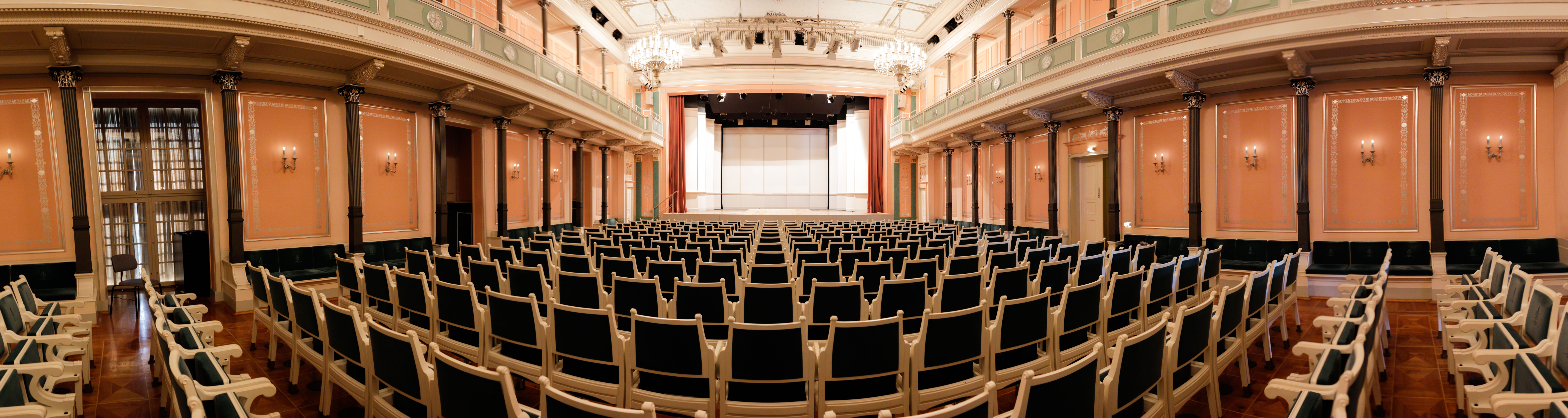}}
\quad
\subfloat[Auditorium (Large).]{\includegraphics[width=0.45\columnwidth]{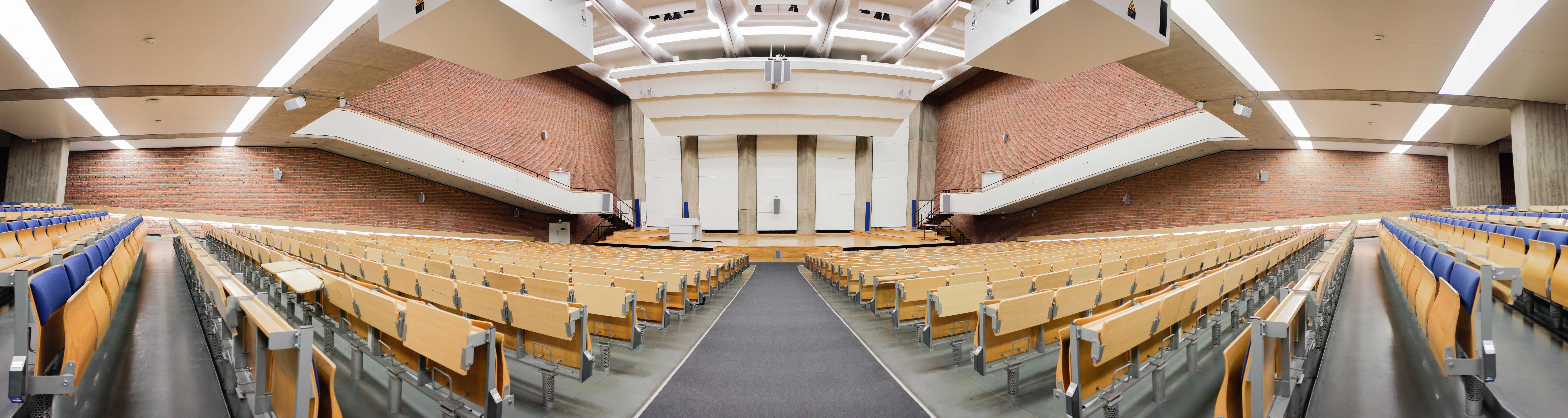}}
\quad
%\vspace{-0.3cm}
\subfloat[Left channel.]{\includegraphics[width=0.45\columnwidth]{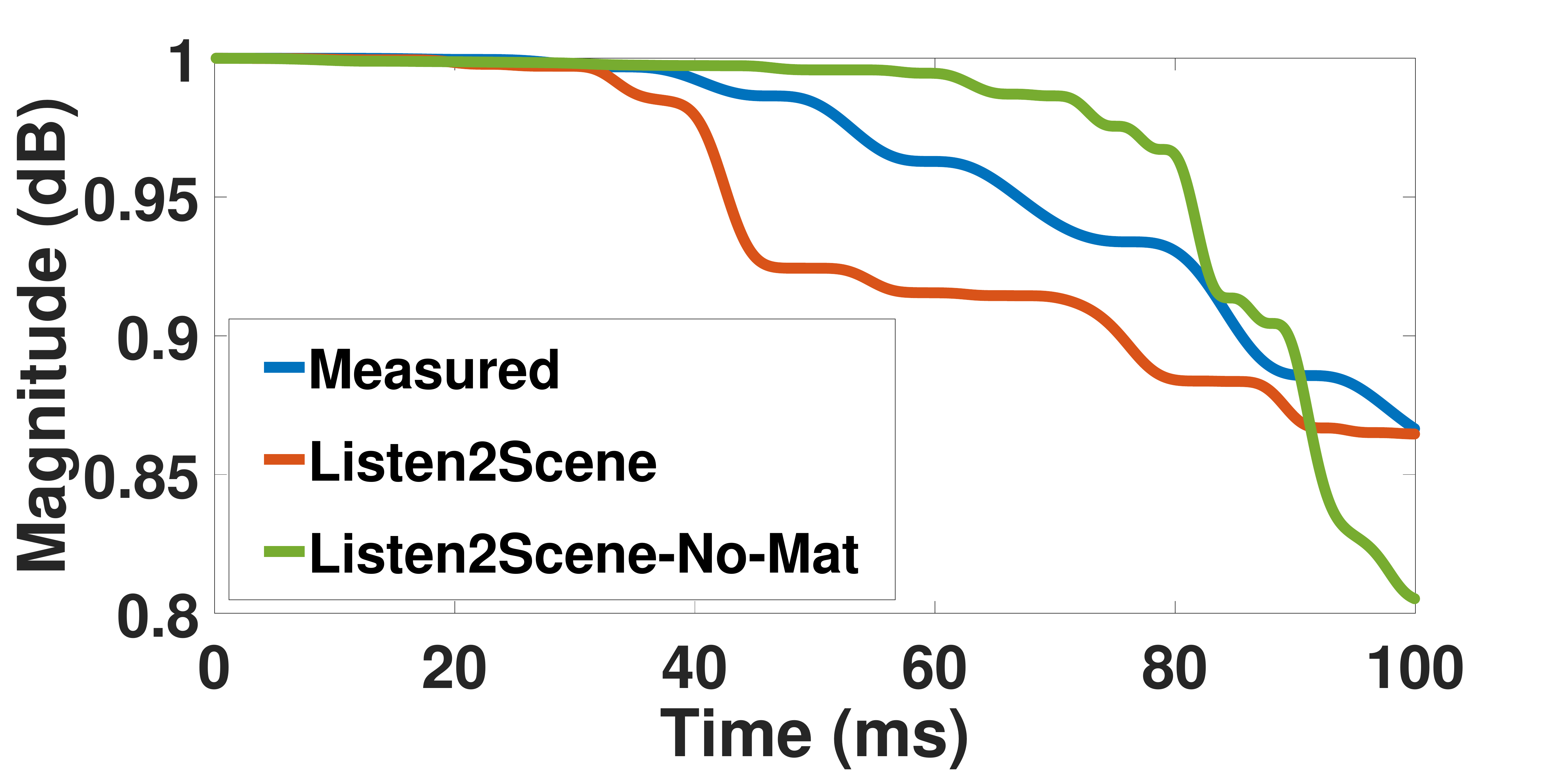}}
\quad
\subfloat[Right channel.]{\includegraphics[width=0.45\columnwidth]{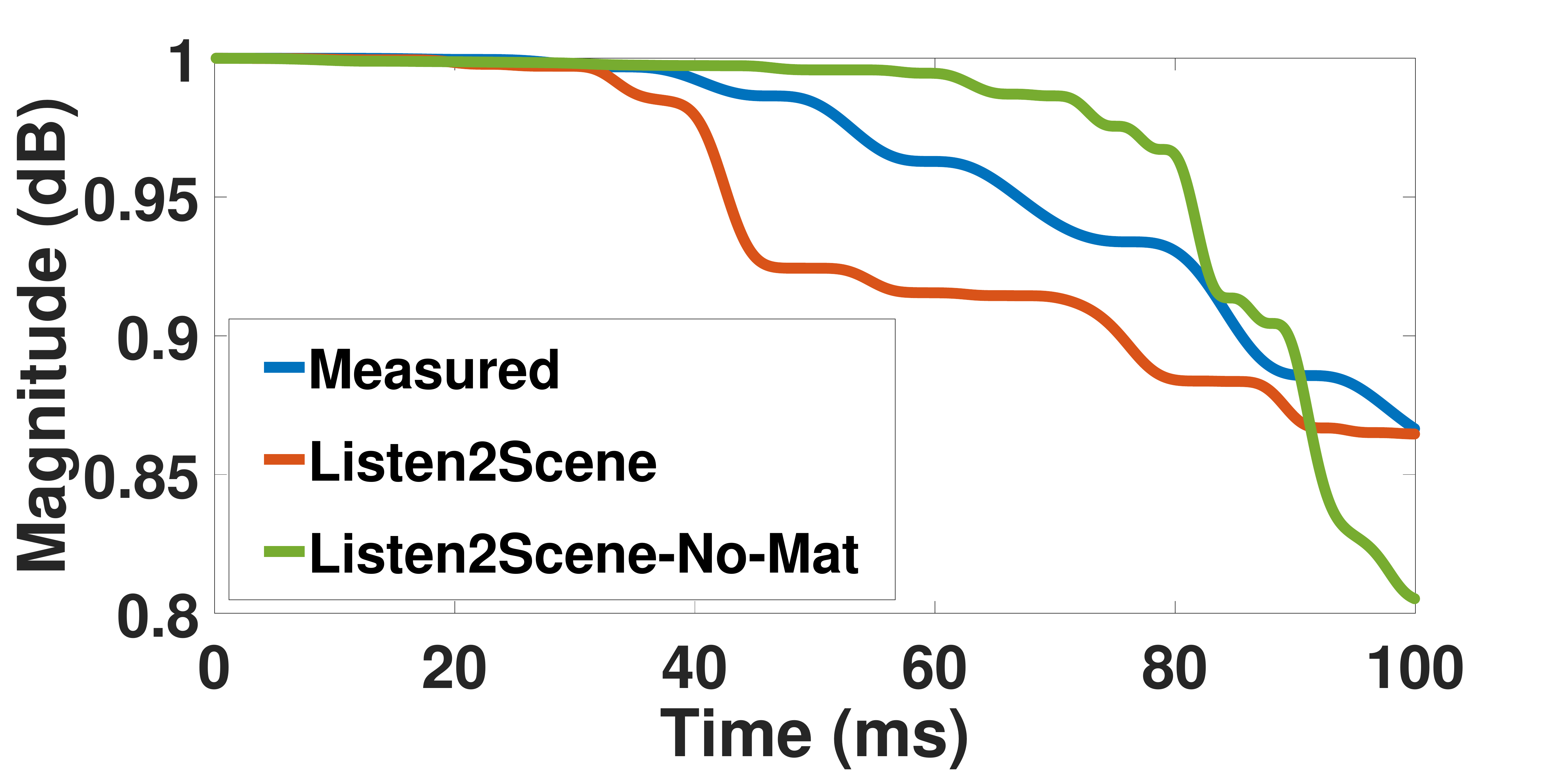}}
\quad
\subfloat[Left channel.]{\includegraphics[width=0.45\columnwidth]{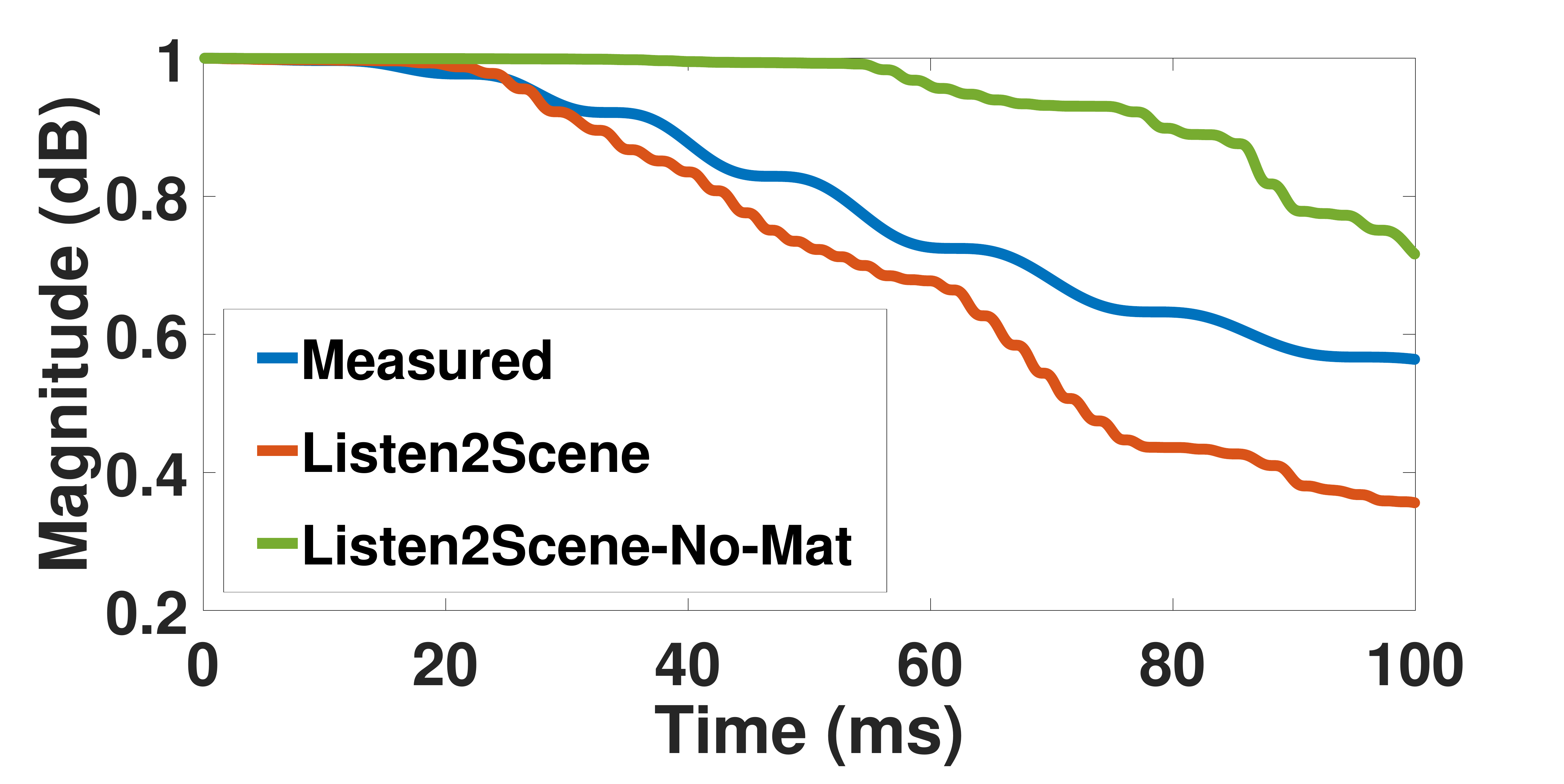}}
\quad
\subfloat[Right channel.]{\includegraphics[width=0.45\columnwidth]{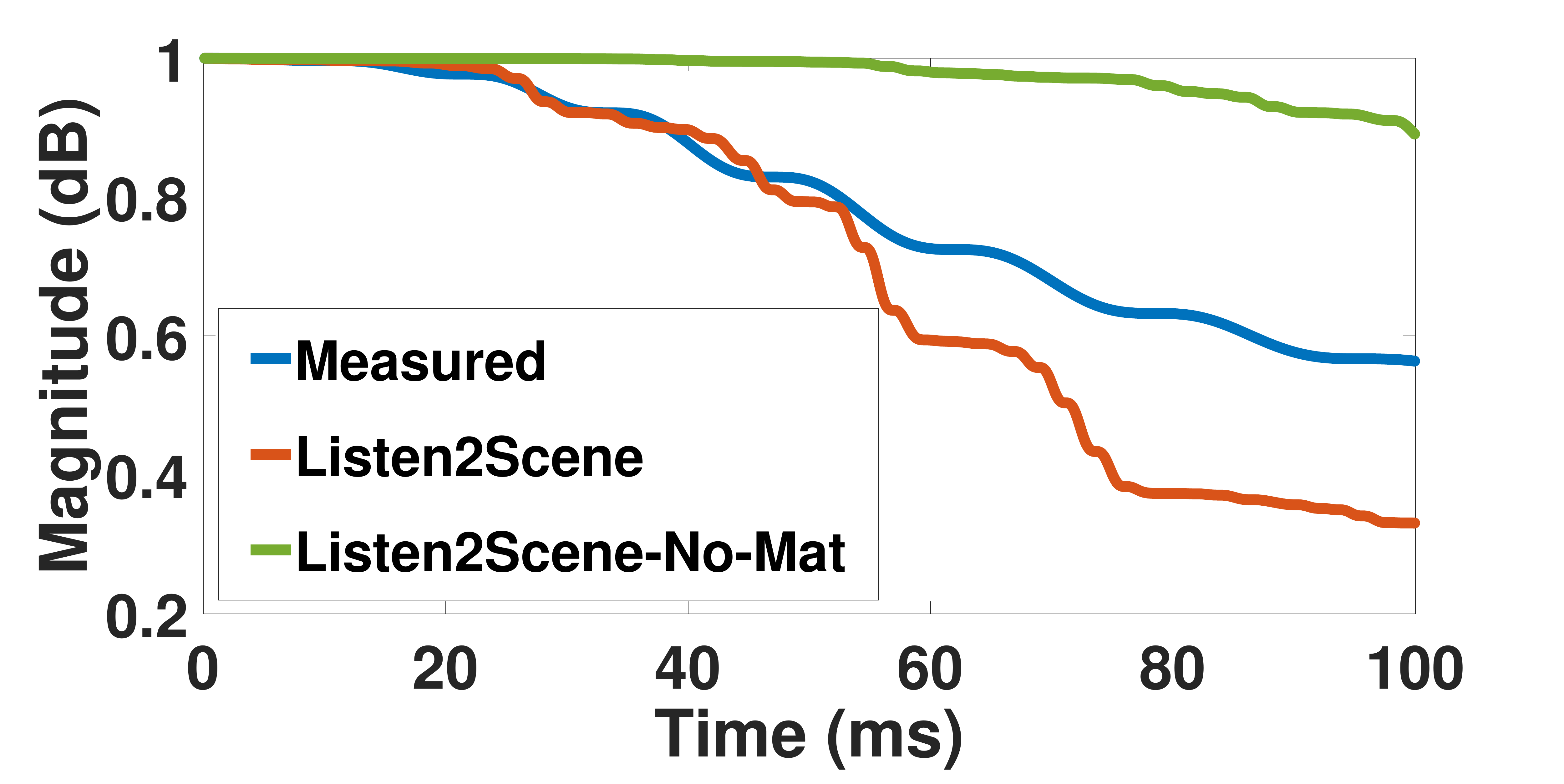}}
\caption{The normalized energy decay curves (EDC) of the captured BIRs and the BIRs generated using our approach with material (Listen2Scene) and without material (Listen2Scene-No-Mat) for the 3D scenes in BRAS ((a),(b)). We plot the EDC for the BIRs from the chamber music hall ((c),(d)) and auditorium ((e),(f)). We observe that the EDC of Listen2Scene is closer to the EDC of captured BIRs.}
% CAN YOU SHOW THE IRs COMPUTED USING ORIGINAL AND SIMPLIFIED MESH. YOU WANT TO SHOW THAT THE IRs DO NOT CHANGE MUCH.
\label{bras_becnhmark}
%\vspace{-0.3cm}
\end{figure}

\begin{table*}[h!]
\renewcommand{\arraystretch}{0.8}
\centering
\caption{The binaural impulse responses (BIR) synthesized for real-world 3D scenes. We compare the accuracy of our learning-based sound propagation method (Listen2Scene) with geometric sound propagation algorithms. These 3D reconstructed scenes were not used in the training data for Listen2Scene. Our Listen2Scene can synthesize BIRs corresponding to left and right channels by considering interaural level differences (ILD) and interaural time differences (ITD). We can see high-level structures of BIRs from our Lisen2Scene is similar to the geometric-based method. The mean absolute error of the normalized BIRs (MAE) is less than $0.5$ x $10^{-2}$.}
\label{tab:plot}
\resizebox{0.9\linewidth}{!}{
\begin{tabular}{L{0.19\textwidth}L{0.1775\textwidth}L{0.1775\textwidth}L{0.1775\textwidth}L{0.1775\textwidth}}
% {C{0.6in}L{1in}L{1in}L{1in}L{1in}L{1in}}
\toprule
                %   & Davis & 301   & 501   & 620   & 750   \\
&  \multicolumn{2}{c}{\textbf{Real-world environment 1}}      &       \multicolumn{2}{c}{\textbf{Real-world environment 2}}      \\
 % \textbf{3D Scene} &  \multicolumn{2}{l}{\tabfig{2.700}{benchmark/S1.png}}      &       \multicolumn{2}{l}{\tabfig{2.700}{benchmark/S2.png}}      \\
  \textbf{3D Scene} &  \multicolumn{2}{l}{\tabfig{2.700}{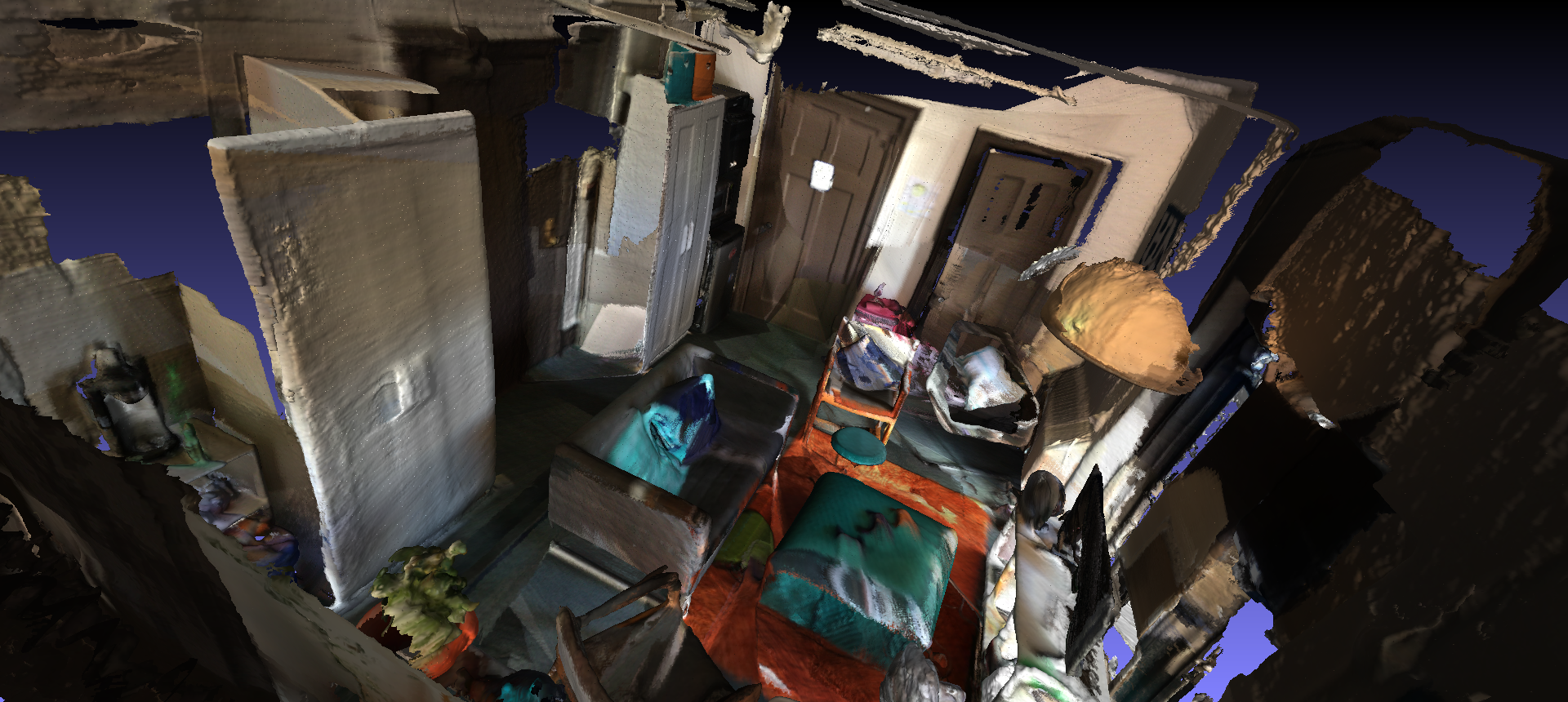}}      &       \multicolumn{2}{l}{\tabfig{2.700}{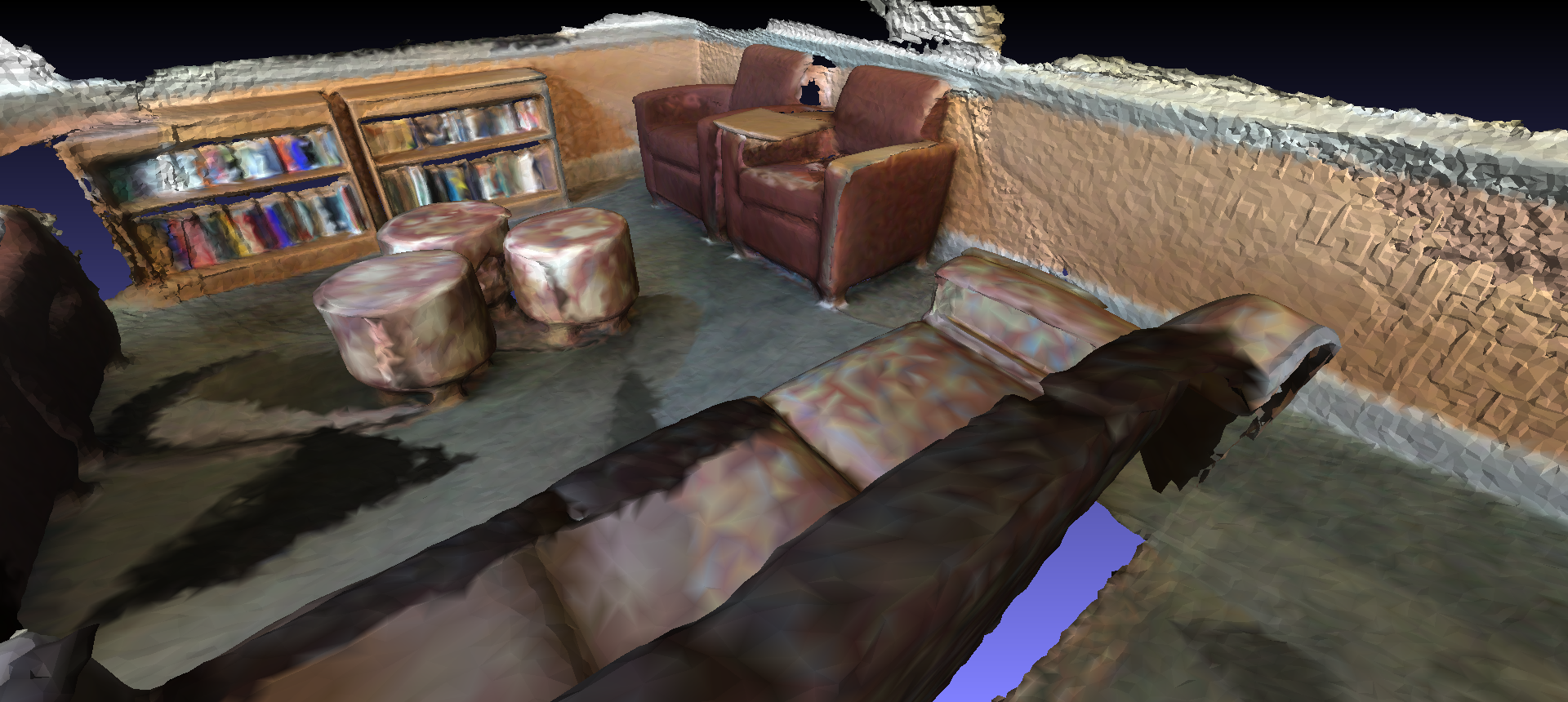}}      \\
% Size ($m^3$)           &  &  &  &  & (x$\times$y$\times$z) \\
 \textbf{Channel }             &  \multicolumn{1}{c}{\textbf{Left}}     & \multicolumn{1}{c}{\textbf{Right}}     & \multicolumn{1}{c}{\textbf{Left}}     & \multicolumn{1}{c}{\textbf{Right}}         \\
 \textbf{MAE (\boldmath{$10^{-2}$}) }             &  \multicolumn{1}{c}{\textbf{0.50}}     & \multicolumn{1}{c}{\textbf{0.49}}     & \multicolumn{1}{c}{\textbf{0.19}}     & \multicolumn{1}{c}{\textbf{0.21}}         \\
\textbf{Geometric-based BIR} & 
\scalebox{1.0}[1.00001]{\tabfig{1.1600}{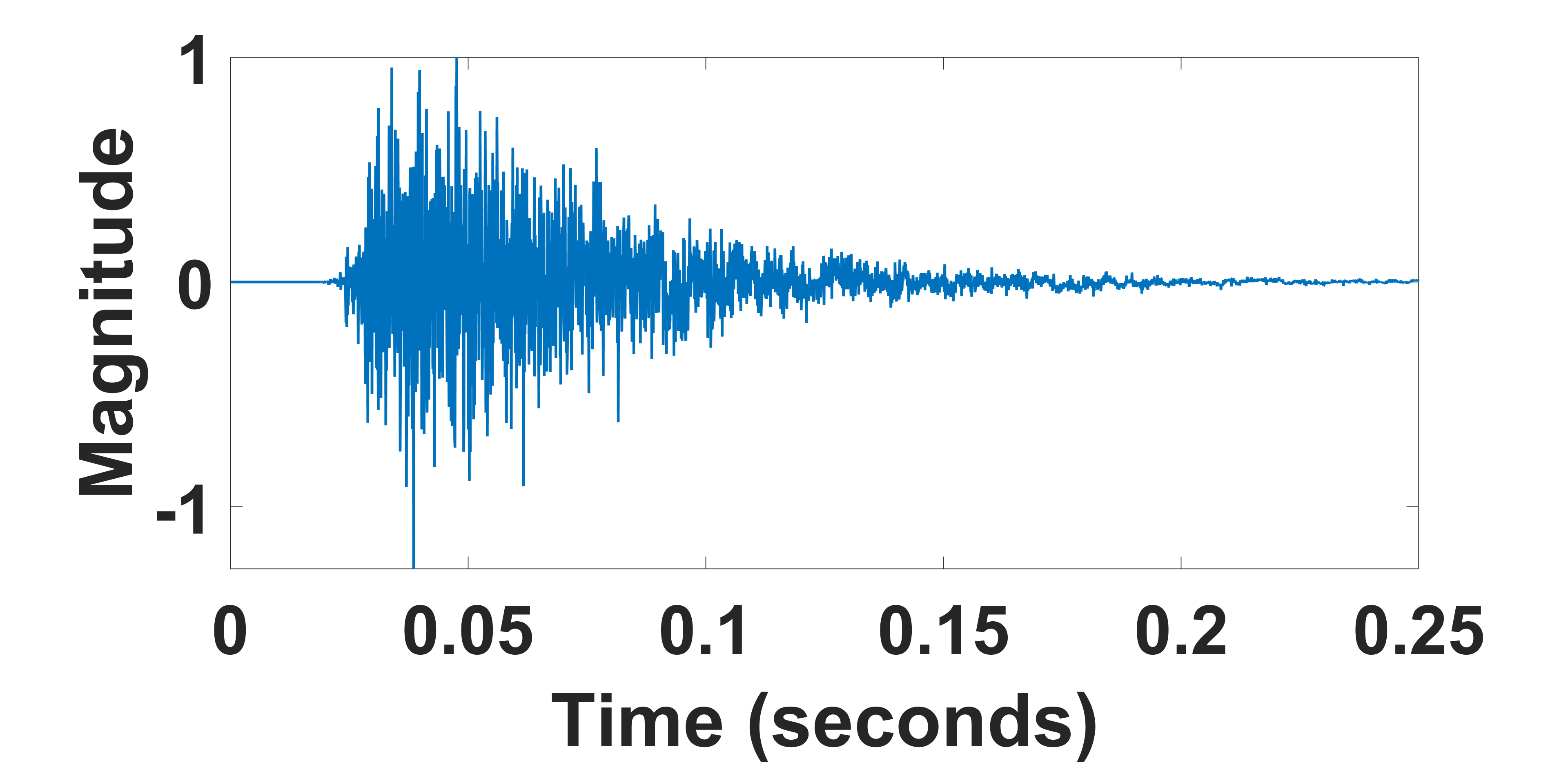}}  & \scalebox{1.0}[1.00001]{\tabfig{1.1600}{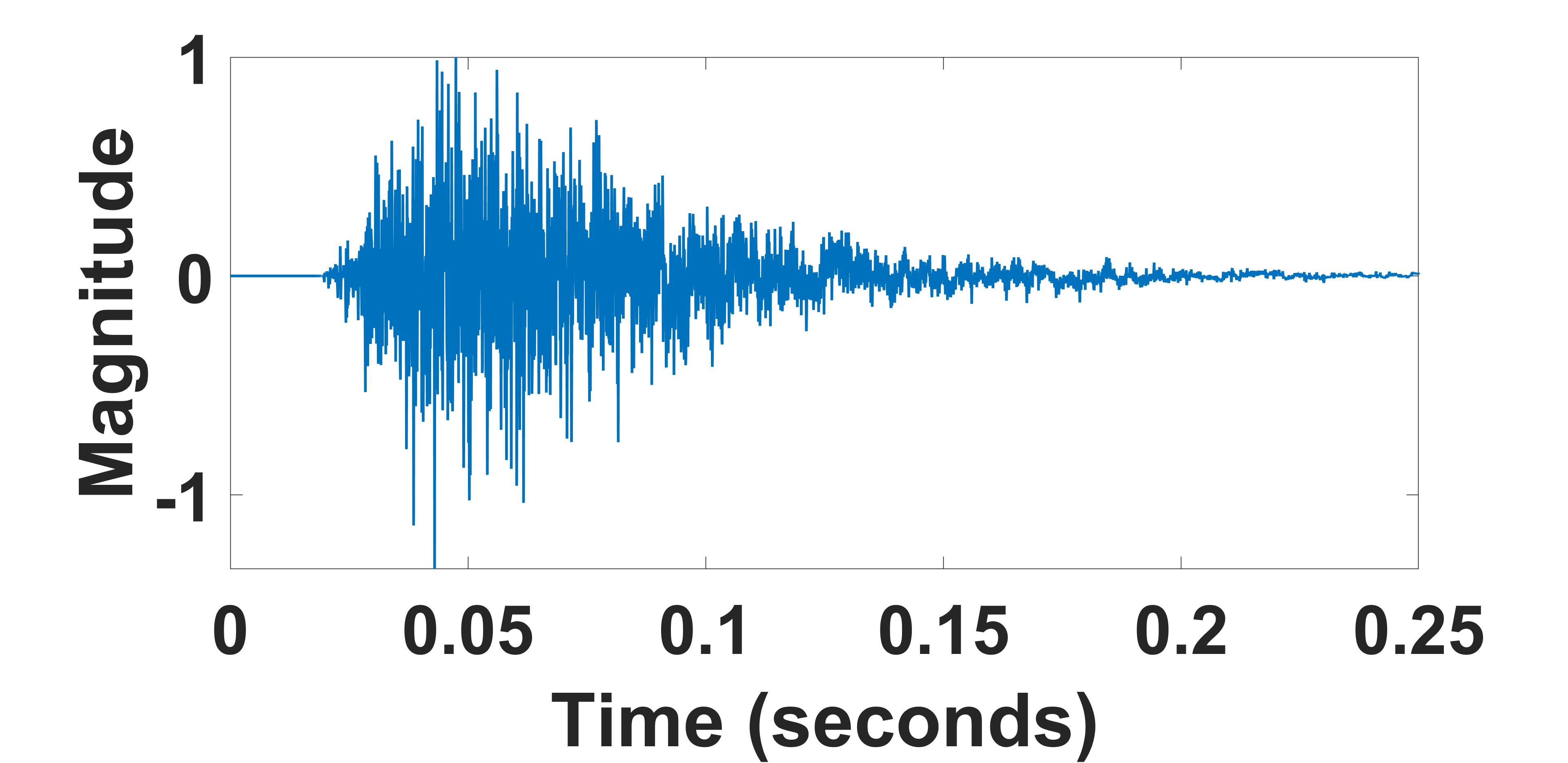}}   & \scalebox{1.0}[1.00001]{\tabfig{1.1600}{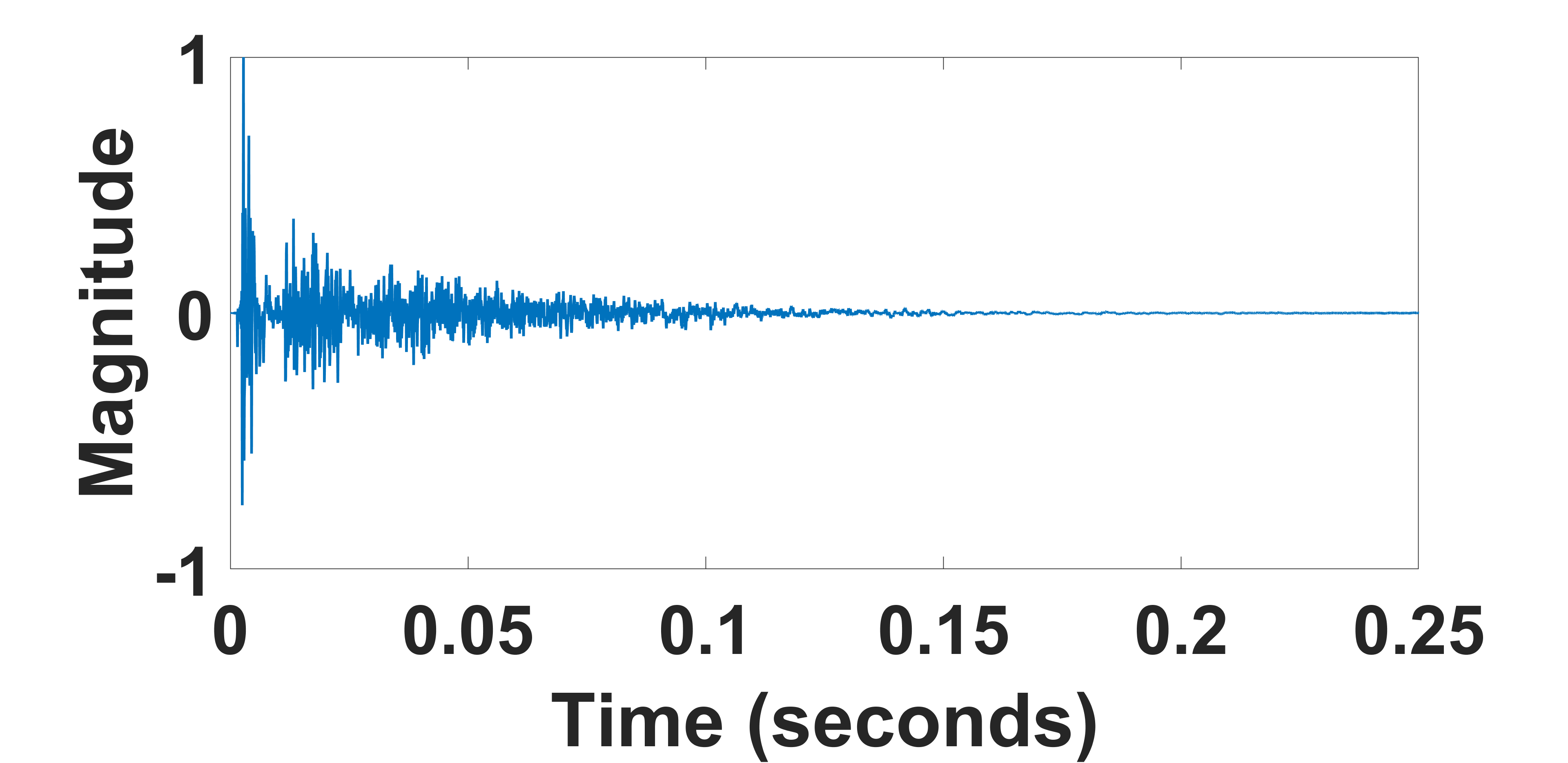}}   & \scalebox{1.0}[1.00001]{\tabfig{1.1600}{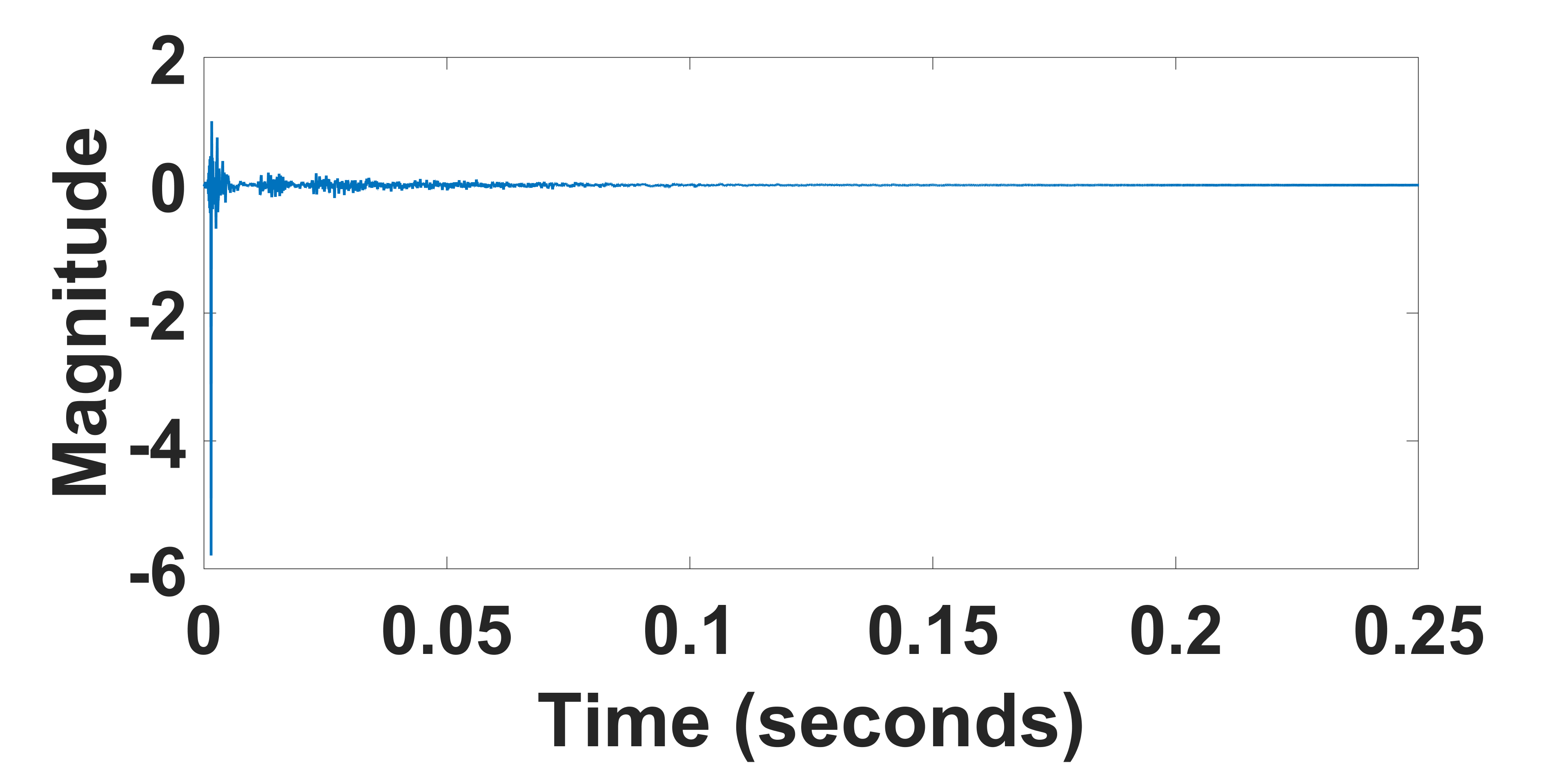}}    \\
\textbf{Our Listen2Scene} & 
\scalebox{1.0}[1.00001]{\tabfig{1.1600}{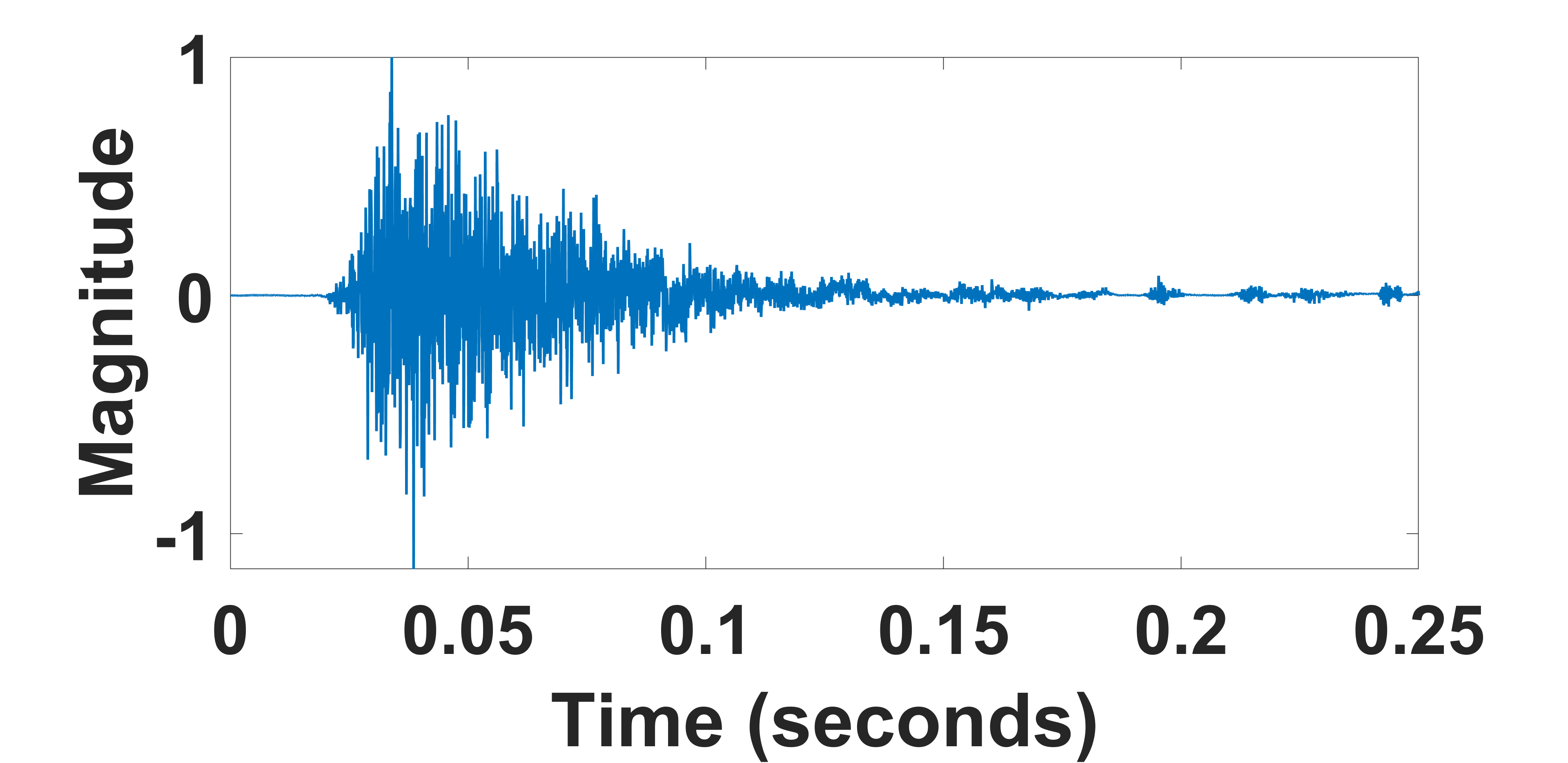}}  & \scalebox{1.0}[1.00001]{\tabfig{1.1600}{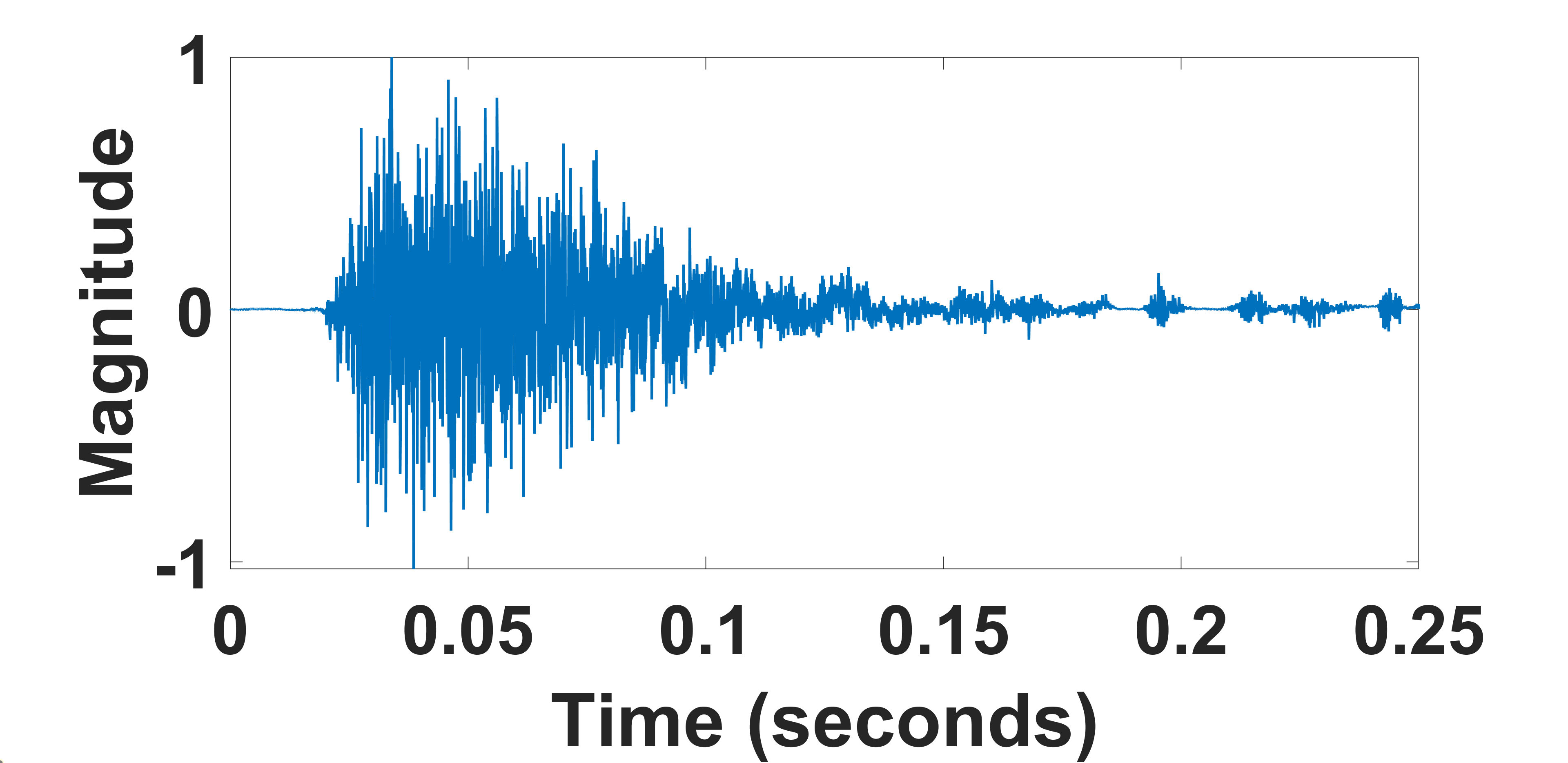}}   & \scalebox{1.0}[1.00001]{\tabfig{1.1600}{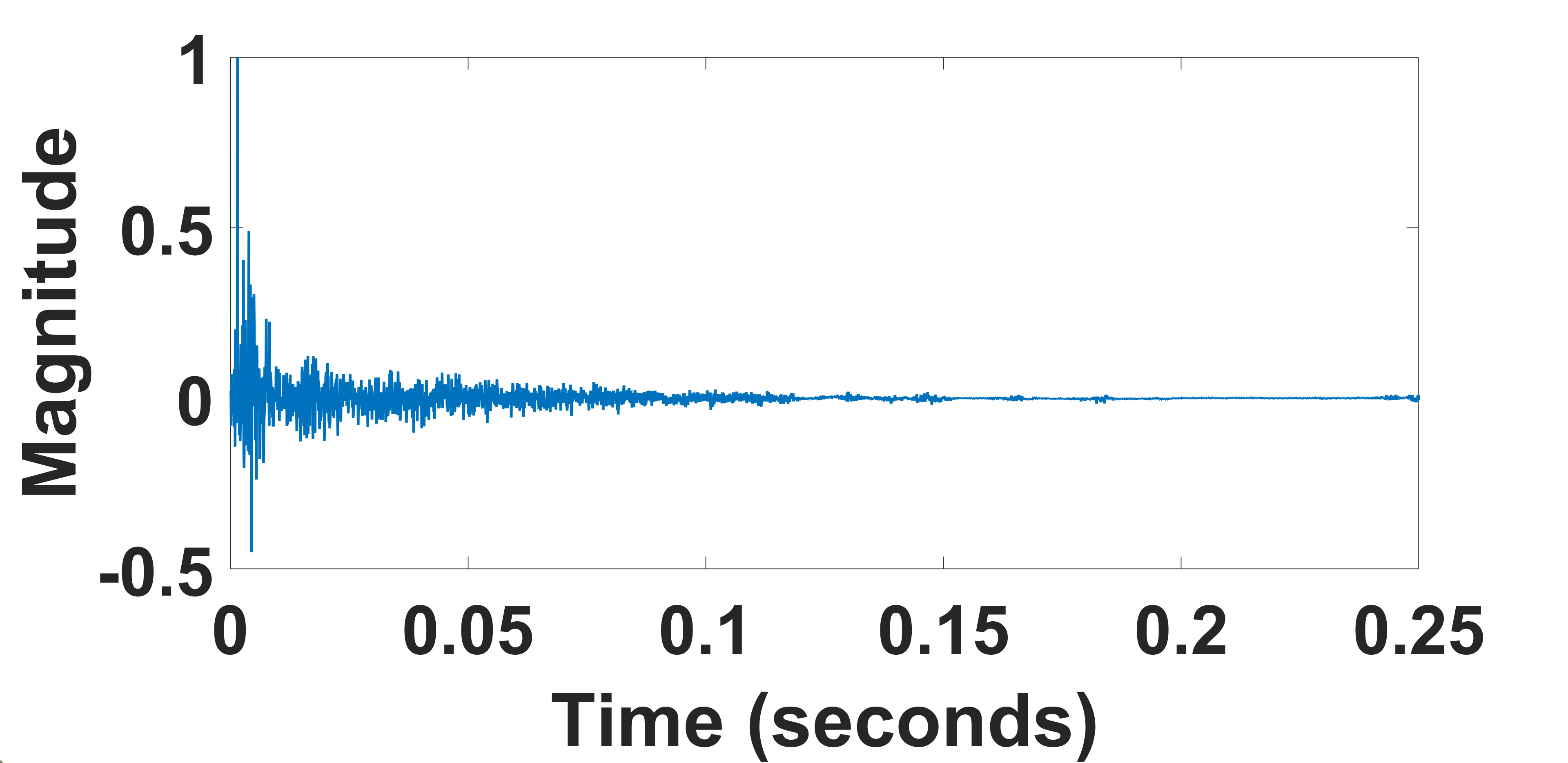}}   & \scalebox{1.0}[1.00001]{\tabfig{1.1600}{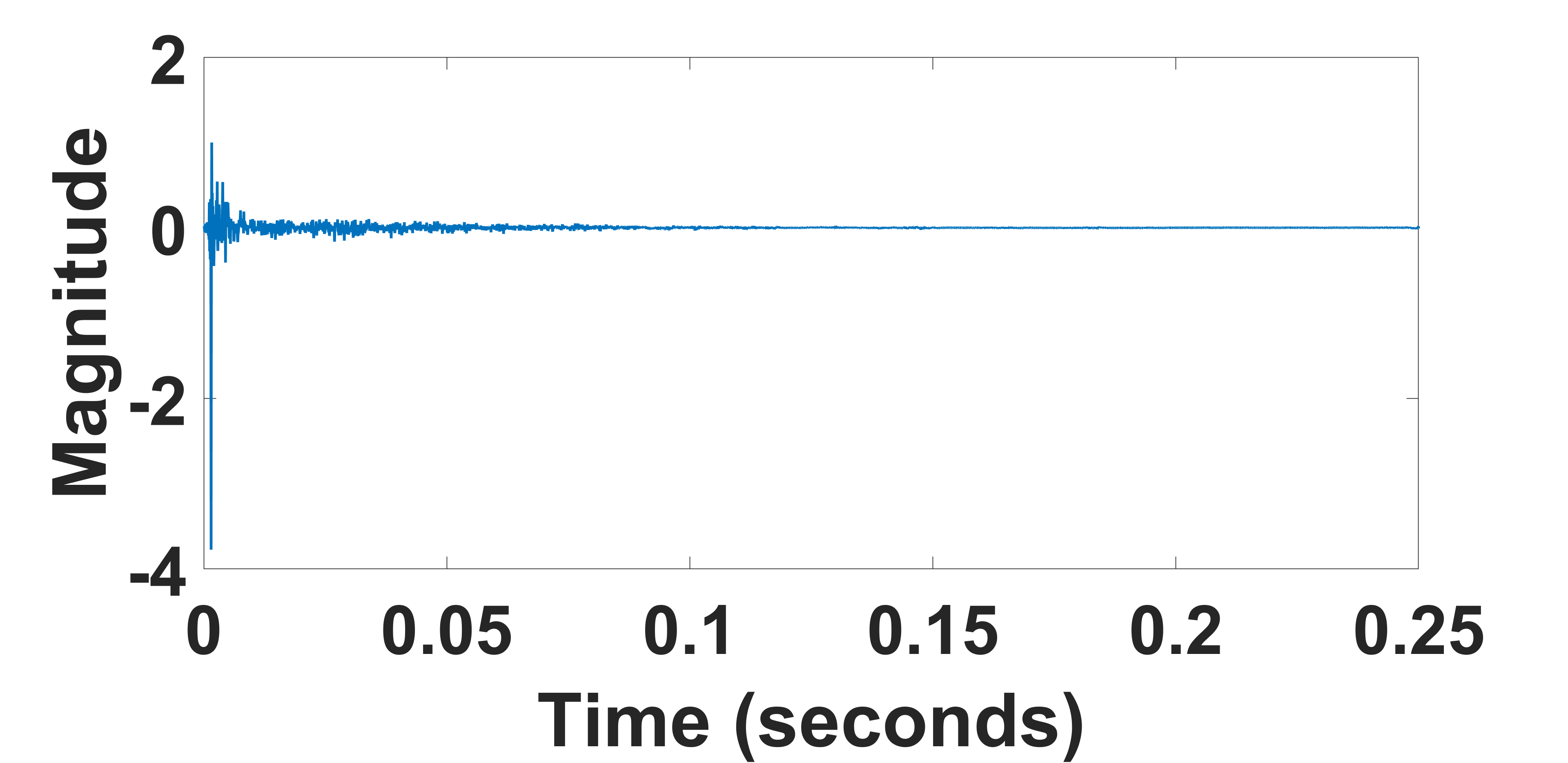}}     \\
\toprule
                %   & Davis & 301   & 501   & 620   & 750   \\
&  \multicolumn{2}{c}{\textbf{Real-world environment 1}}      &       \multicolumn{2}{c}{\textbf{Real-world environment 2}}      \\
 % \textbf{3D Scene} &  \multicolumn{2}{l}{\tabfig{2.700}{benchmark/S1.png}}      &       \multicolumn{2}{l}{\tabfig{2.700}{benchmark/S2.png}}      \\
  \textbf{3D Scene} &  \multicolumn{2}{l}{\tabfig{2.700}{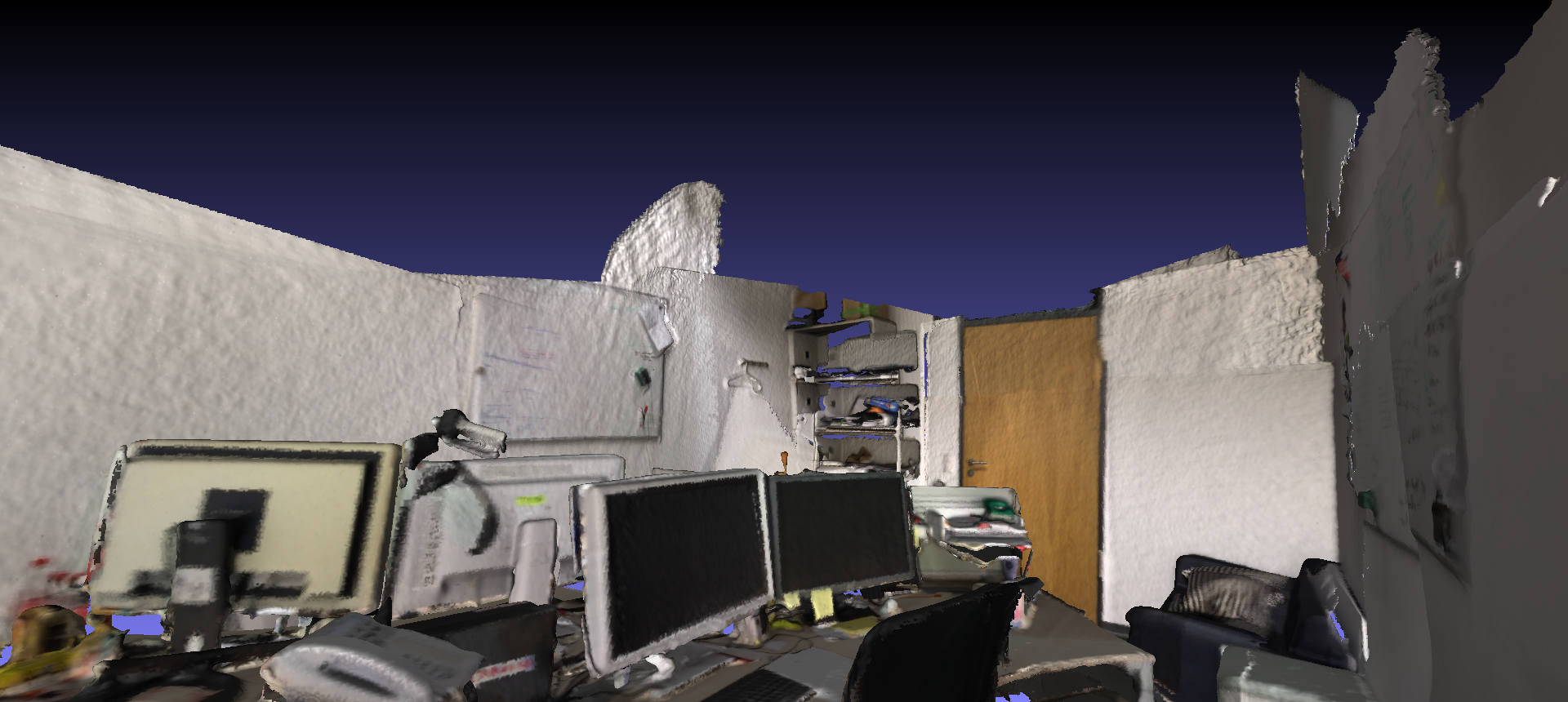}}      &       \multicolumn{2}{l}{\tabfig{2.700}{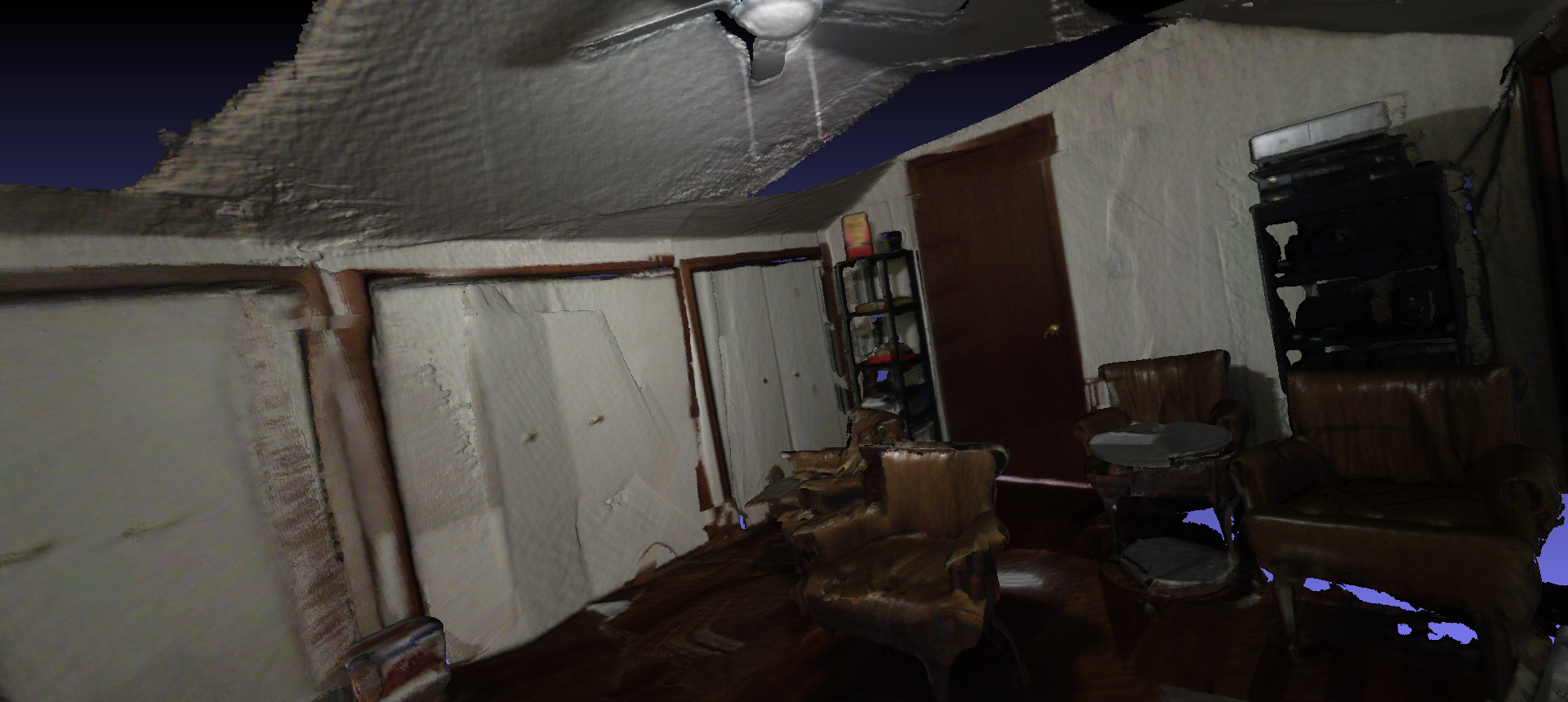}}      \\
% Size ($m^3$)           &  &  &  &  & (x$\times$y$\times$z) \\
 \textbf{Channel }             &  \multicolumn{1}{c}{\textbf{Left}}     & \multicolumn{1}{c}{\textbf{Right}}     & \multicolumn{1}{c}{\textbf{Left}}     & \multicolumn{1}{c}{\textbf{Right}}         \\
 \textbf{MAE (\boldmath{$10^{-2}$}) }             &  \multicolumn{1}{c}{\textbf{0.44}}     & \multicolumn{1}{c}{\textbf{0.40}}     & \multicolumn{1}{c}{\textbf{0.34}}     & \multicolumn{1}{c}{\textbf{0.37}}         \\
\textbf{Geometric-based BIR} & 
\scalebox{1.0}[1.00001]{\tabfig{1.1600}{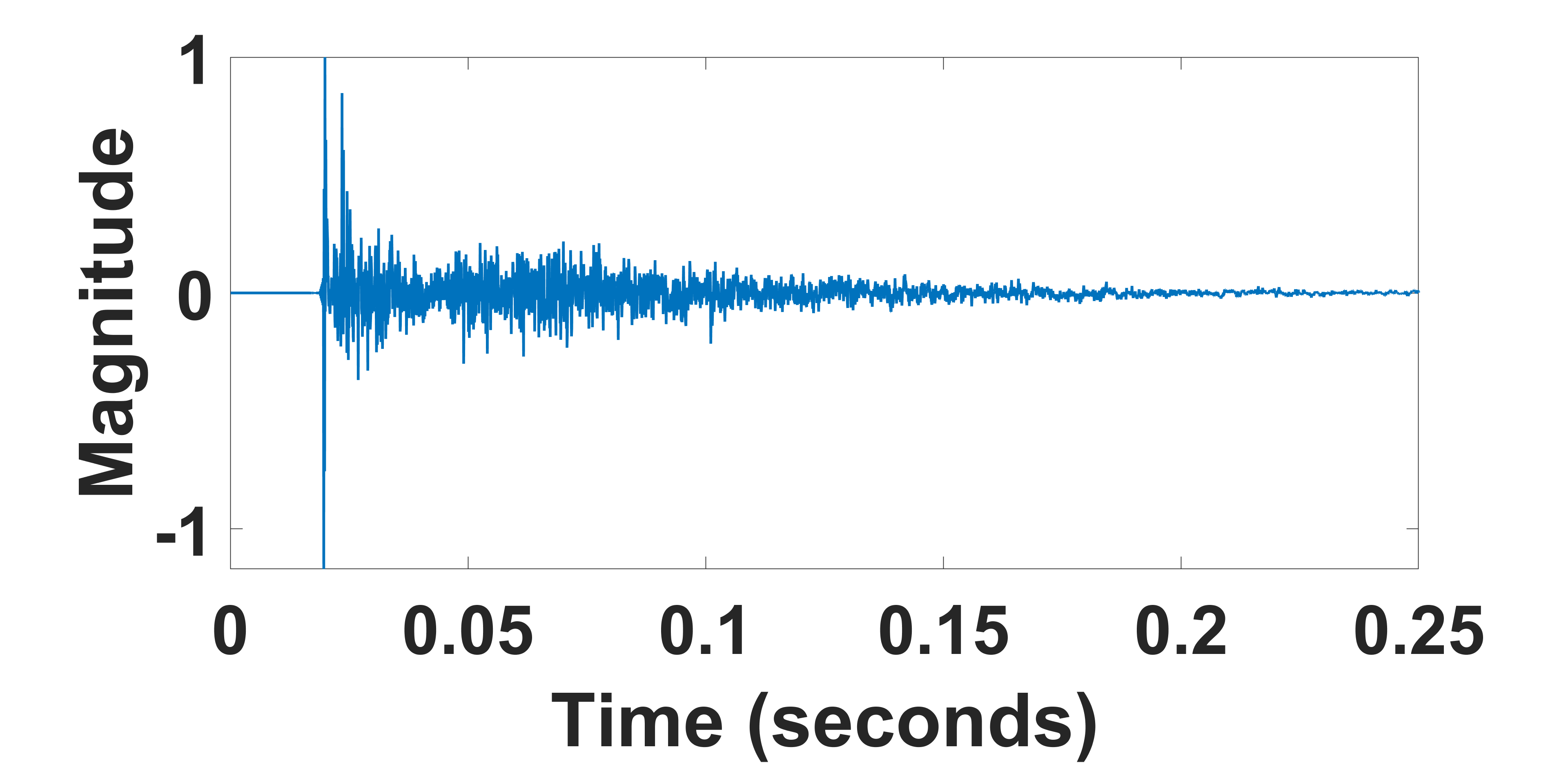}}  & \scalebox{1.0}[1.00001]{\tabfig{1.1600}{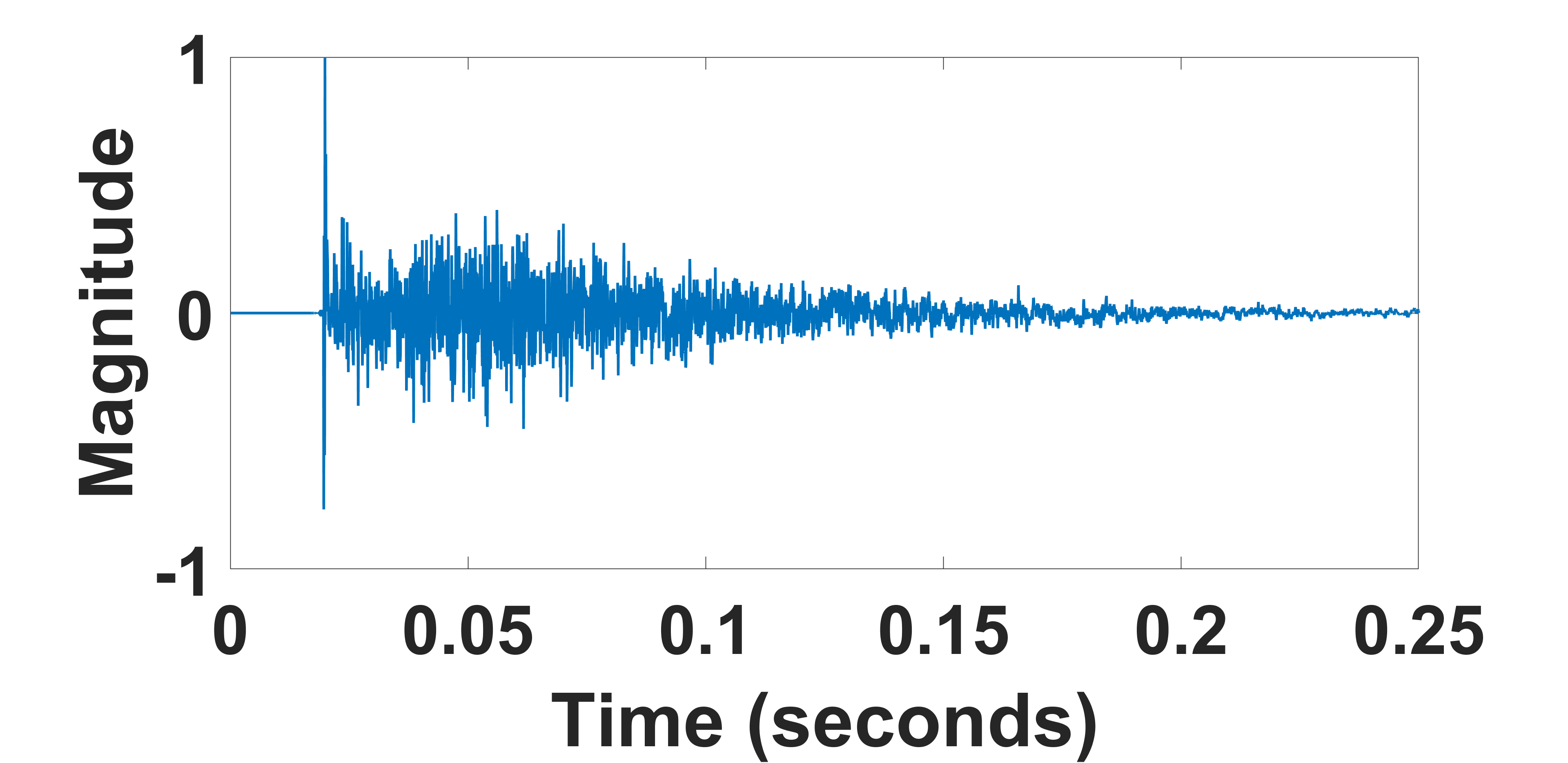}}   & \scalebox{1.0}[1.00001]{\tabfig{1.1600}{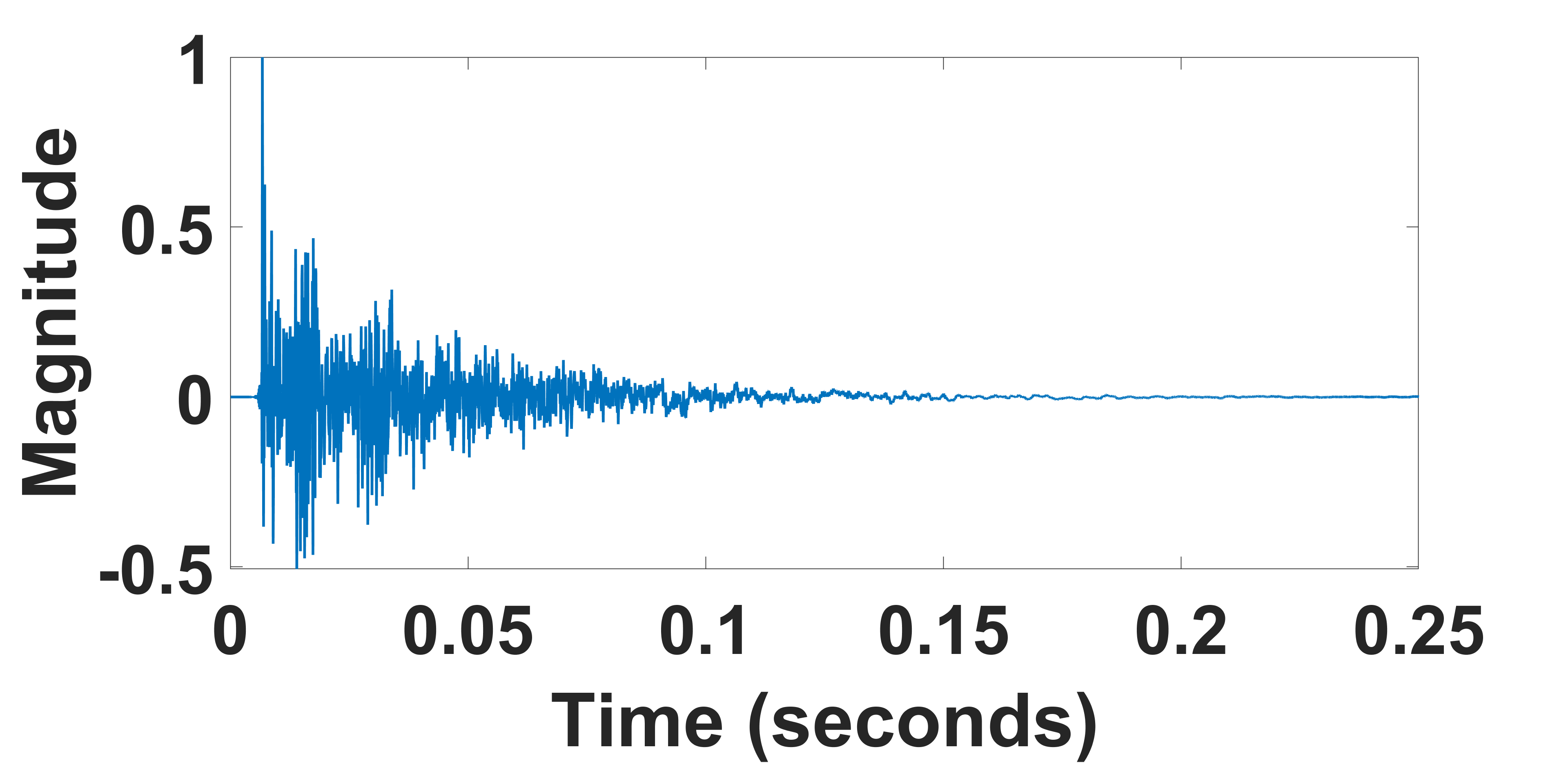}}   & \scalebox{1.0}[1.00001]{\tabfig{1.1600}{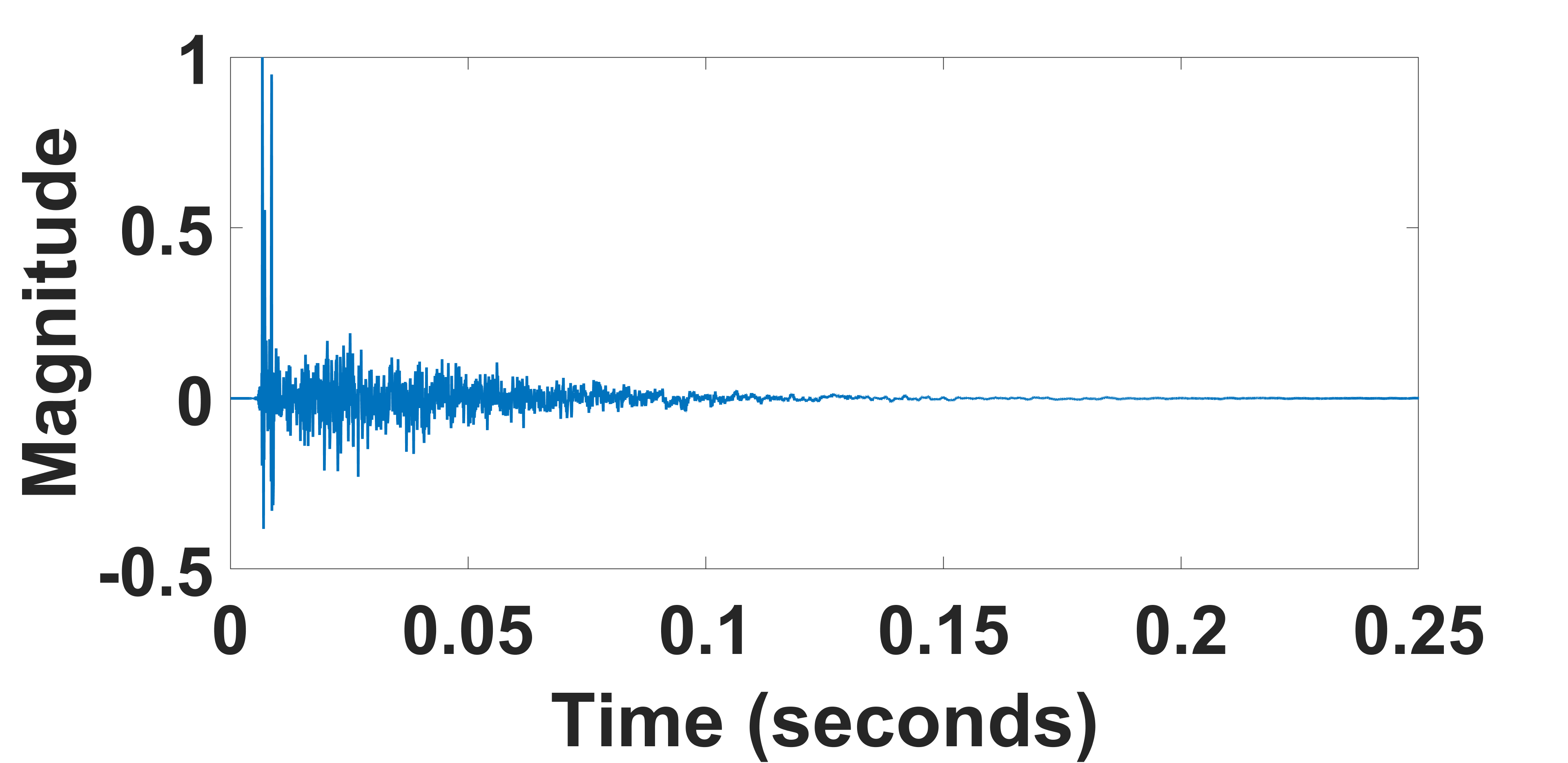}}    \\
\textbf{Our Listen2Scene} & 
\scalebox{1.0}[1.00001]{\tabfig{1.1600}{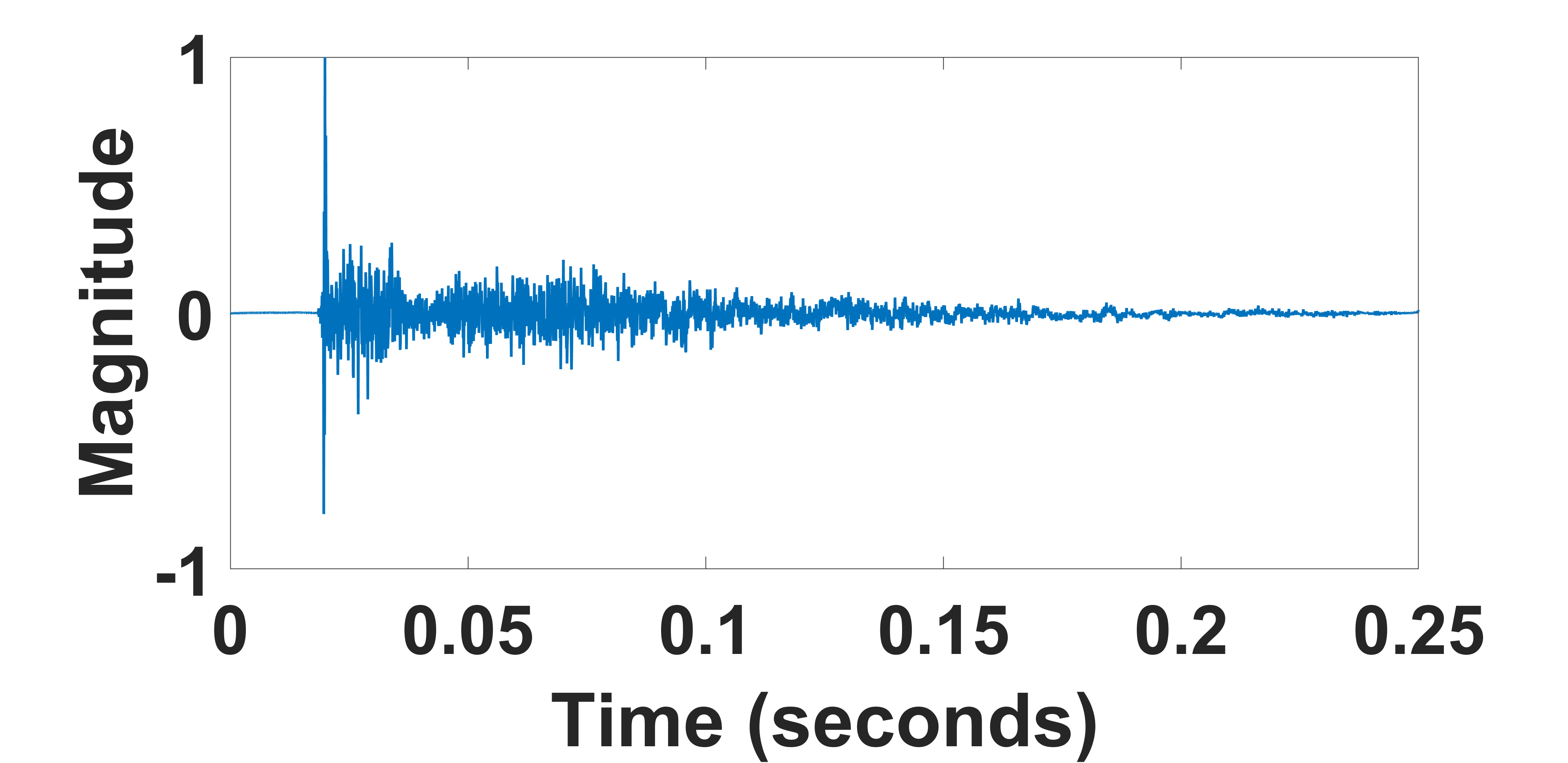}}  & \scalebox{1.0}[1.00001]{\tabfig{1.1600}{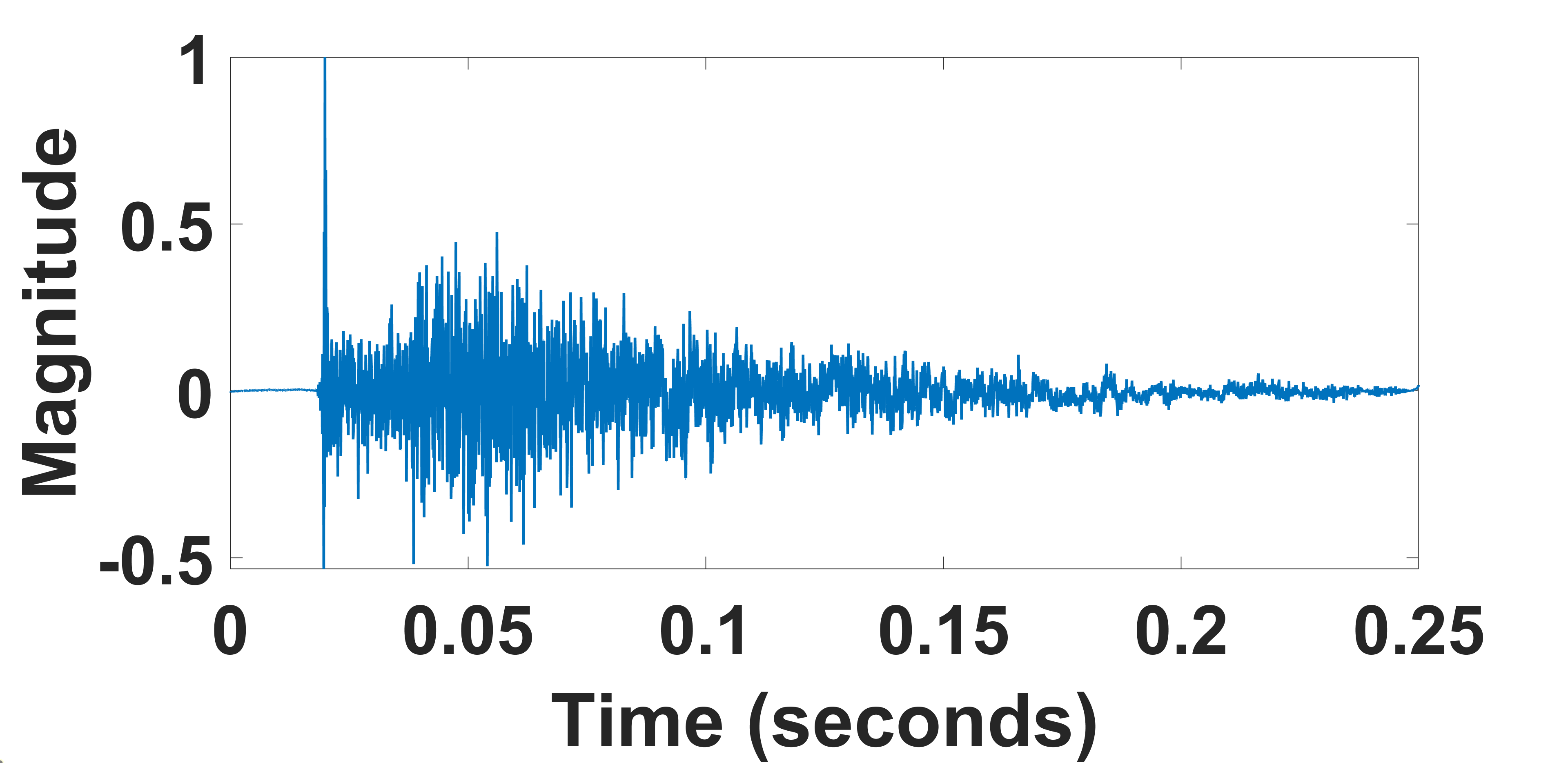}}   & \scalebox{1.0}[1.00001]{\tabfig{1.1600}{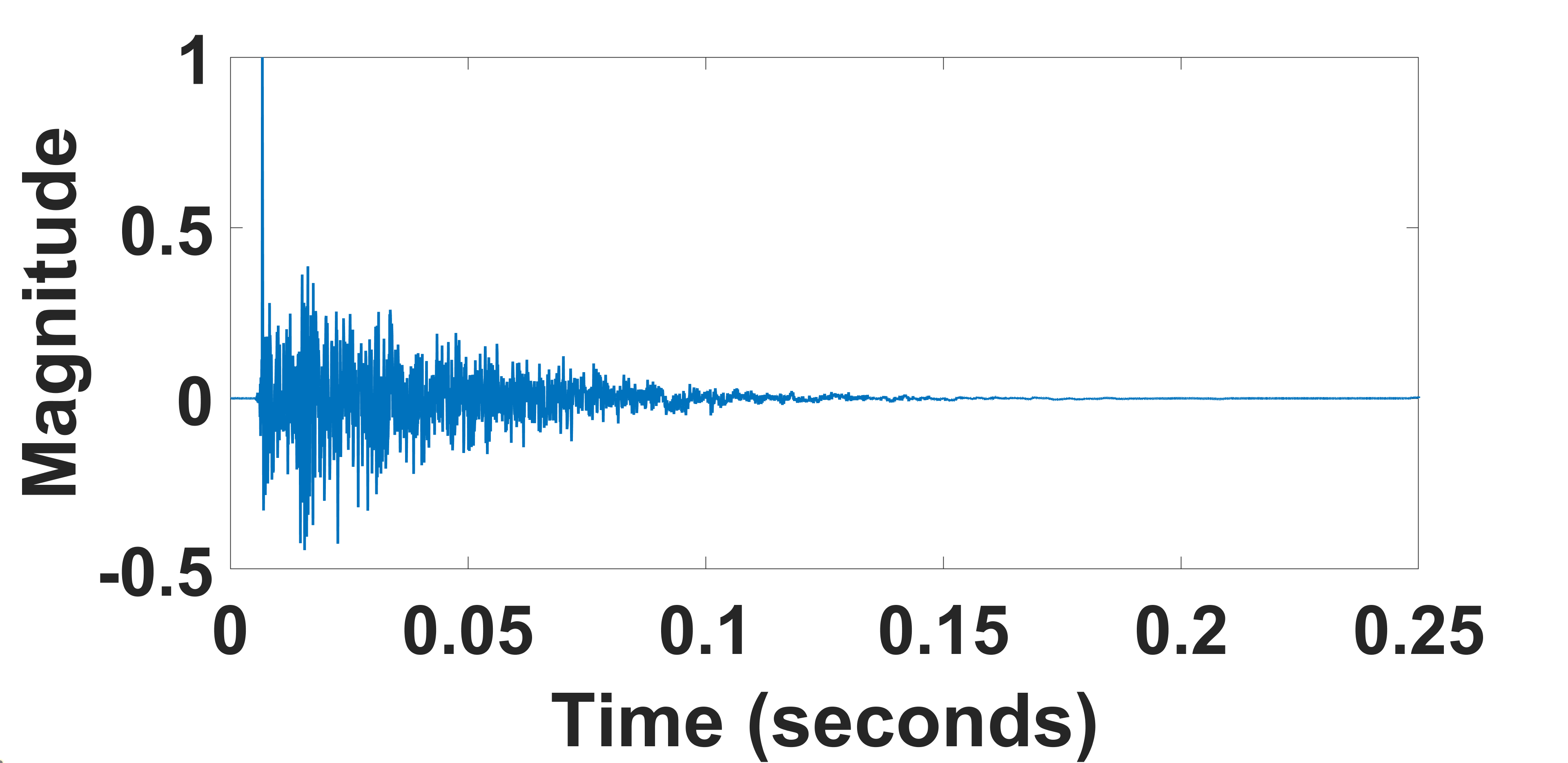}}   & \scalebox{1.0}[1.00001]{\tabfig{1.1600}{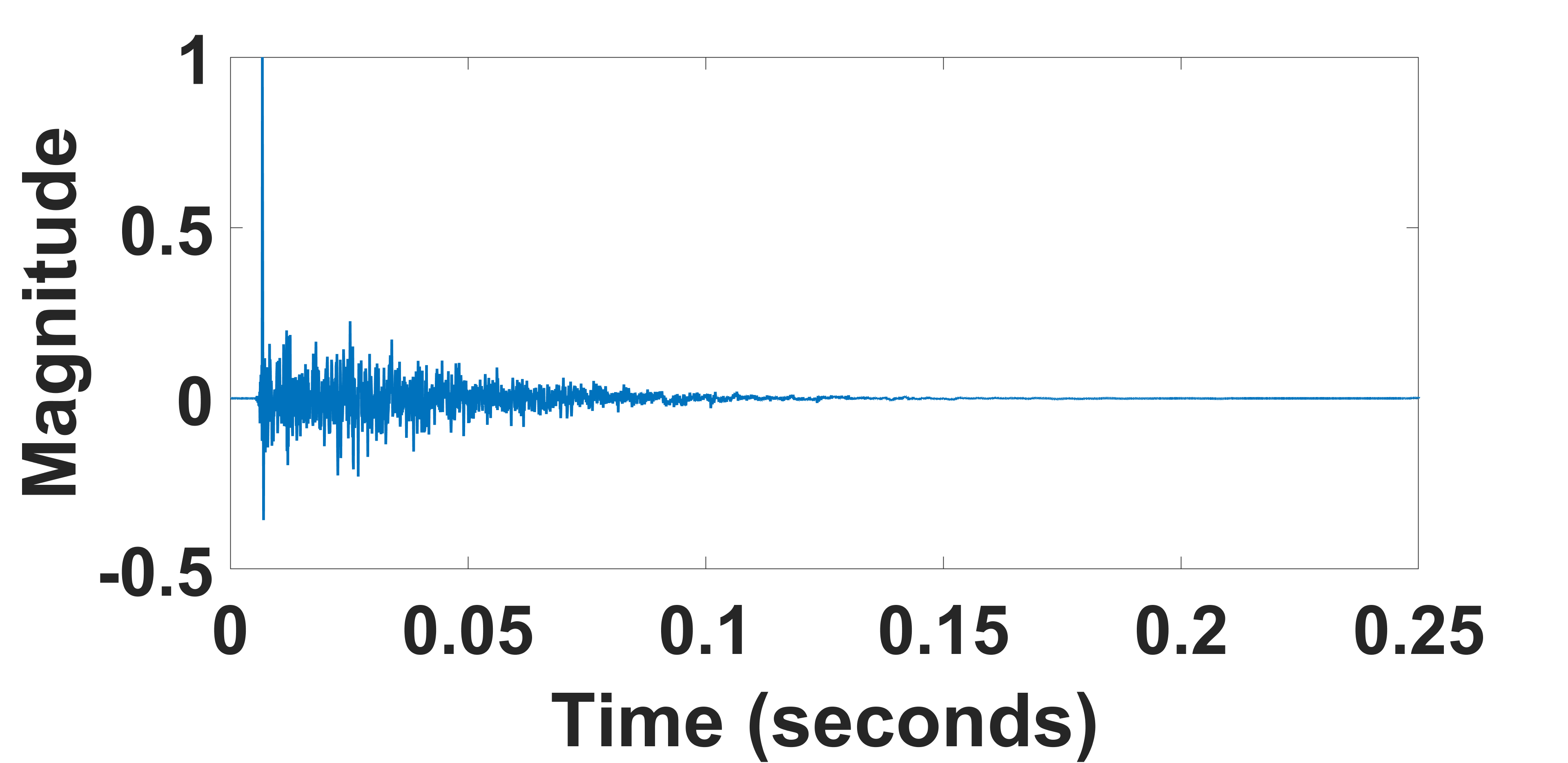}}     \\
\toprule
\toprule

\end{tabular}
}
\end{table*}
\begin{table*}[h!]
    \setlength{\tabcolsep}{1.6pt}
    \renewcommand{\arraystretch}{0.8}
	\caption{The responses from the acoustic experts and AMT participants on the plausibility of the sounds in each video created using 3D scenes in the ScanNet. We report the response from each age-category separately and the standard deviation (SD) of the combined results. We shifted our rating scale from -2 - 2 to 1 - 5 and calculated the SD (\S~\ref{exp_res}). We compare video auralized using our Listen2Scene approach with the videos auralized using clean speech, MESH2IR, Listen2Scene-No-Mat and geometric-method. We compare Listen2Scene-No-Mat using a single source in medium (M) and large (L) 3D scenes. We observe that 67\% of total participants prefer Listen2Scene when we play video generated using Listen2Scene and MESH2IR with a single source. The highest comparative percentage is~\textbf{bolded}.  }
    %\vspace{-0.3cm}
	\label{tab:userstudy}
	\centering
       \resizebox{0.93\linewidth}{!}{
	\begin{tabular}{@{}lccccccccccc@{}}	%c-no.of columns. |-verticle line between columns
		\toprule
       \textbf{Participants} &\textbf{}& \multicolumn{3}{c}{\textbf{Acoustic Experts (13 participants) [\%]}} & \multicolumn{3}{c}{\textbf{AMT (57 participants) [\%]}}& \multicolumn{4}{c}{\textbf{Combined (70 participants) [\%]}}\\
         \midrule
		\textbf{Baseline Method} & \textbf{No of} &\textbf{Baseline}  & \textbf{No}& \textbf{Listen2Scene}  &\textbf{Baseline}  & \textbf{No}& \textbf{Listen2Scene}  &\textbf{Baseline}  & \textbf{No}& \textbf{Listen2Scene} & \textbf{SD}\\
        \textbf{} & \textbf{Sources} & & \textbf{Preference}&  \textbf{}  & & \textbf{Preference}&  \textbf{}   & & \textbf{Preference}&  \textbf{}   & \textbf{}         \\
		\midrule
		Clean&                 1 &15.38 & 0.00  & \textbf{84.62}&29.82 & 1.76  & \textbf{68.42} &27.14 & 1.43  & \textbf{71.43}    & 1.44 \\
		        &              2 & 7.69 & 0.00  & \textbf{92.31}  & 19.30 & 3.51  & \textbf{77.19}  & 17.14 & 2.86  & \textbf{80}  & 1.39 \\
		\midrule
		Mesh2IR &              1& 23.08 & 0.00 & \textbf{76.92} & 31.58 & 3.51 & \textbf{64.91} & 30 & 2.86 & \textbf{67.14}  & 1.51 \\
	\cite{mesh2ir}&        2& 15.38 & 7.69 & \textbf{76.92} & 15.79 & 5.26 & \textbf{78.95} & 15.71 & 5.71 & \textbf{78.57}  & 1.26 \\ 
        \midrule
		    &              1 (M) & 30.77 & 23.08 & \textbf{46.15}  & 29.82 & 19.30 & \textbf{50.88}  & 30 & 20 & \textbf{50}  & 1.29 \\
		
		Listen2Scene-No-Mat& 1 (L) & 7.69 & 30.77 & \textbf{61.54}  & 26.31 & 7.02 & \textbf{66.66}  & 22.85 & 11.43 & \textbf{65.71}  & 1.15 \\
		                  &  2& 23.08 & 23.08 & \textbf{53.85} & 22.81 & 15.79 & \textbf{61.40} & 22.86 & 17.14 & \textbf{60}  & 1.30 \\ 
        \midrule
        Geometric-Method& 2  & 23.08 & 46.15 & \textbf{30.77}  & 38.60 & 12.28 & \textbf{49.12}  & 35.71 & 18.57 & \textbf{45.71}  & 1.43\\
		
% 		train-GAN+GAN & 1546   & train-clean-{100,360}\\
% 		dev-GAN+GAN  & 388  & dev-clean\\		
		\midrule

    \textbf{Age Category} &\textbf{}& \multicolumn{3}{c}{\textbf{18 - 24 (16 participants)  [\%] }} & \multicolumn{3}{c}{\textbf{25 - 34 (47 participants)  [\%] }}& \multicolumn{4}{c}{\textbf{35 or older (7 participants)  [\%]}}\\
   \midrule
		\textbf{Baseline Method} & \textbf{No of} &\textbf{Baseline}  & \textbf{No}& \textbf{Listen2Scene}  &\textbf{Baseline}  & \textbf{No}& \textbf{Listen2Scene}  &\textbf{Baseline}  & \textbf{No}& \textbf{Listen2Scene} & \textbf{}\\
        \textbf{} & \textbf{Sources} & & \textbf{Preference}&  \textbf{}  & & \textbf{Preference}&  \textbf{}   & & \textbf{Preference}&  \textbf{}     & \textbf{}       \\
		\midrule
		Clean&                 1 &31.25 & 0.00  & \textbf{68.75}      &19.15 & 2.13  & \textbf{78.72}                &\textbf{71.43} & 0.00  & 25.57   & \\
		        &              2 & 12.5 & 0.00  & \textbf{87.5}      & 14.89 & 2.13  & \textbf{82.98}               & 42.86 & 14.28  & \textbf{42.86}  & \\
		\midrule
		Mesh2IR &              1& 18.75 & 6.25 & \textbf{75.00}       & 34.04 & 0.00 & \textbf{65.96}                &28.57  & 14.28 & \textbf{57.14}  & \\
	\cite{mesh2ir}&        2& 12.5 & 6.25 & \textbf{81.25}           &14.89 & 4.26 & \textbf{80.85}                & 28.57 & 14.28 & \textbf{57.14}  & \\ 
        \midrule 
		    &              1 (M) & 18.75 & 0.00 & \textbf{81.25}     & 36.17 & 25.53 & \textbf{38.30}               & 14.28 & 28.57 & \textbf{57.14}  & \\
		
		Listen2Scene-No-Mat& 1 (L) & 12.5 & 12.5 & \textbf{75.00}    & 21.28 & 12.77 & \textbf{65.96}                 & \textbf{57.14} & 0.00 & 42.86  & \\
		                  &  2& 6.25 & 18.75 & \textbf{75.00}      & 27.66 & 17.02 & \textbf{55.32}                & 28.57 & 14.29 & \textbf{57.14}  & \\ 
        \midrule
        Geometric-Method & 1 (L) & 31.25 & 25.00 & \textbf{43.75}    & 27.66 & 23.40 & \textbf{48.94}                 & 42.86 & 14.23 & \textbf{42.86} & \\
% 		\midrule
% 		train-GAN+GAN & 1546   & train-clean-{100,360}\\
% 		dev-GAN+GAN  & 388  & dev-clean\\		
		\bottomrule
	\end{tabular}}
	\vspace{-0.3cm}
\end{table*}
\section{ACOUSTIC EVALUATION}
% In this section, we evaluate the performance of our learning method and compare it with prior methods.

\subsection{BRAS Benchmark}
We use the BRAS benchmark~\cite{bras} to evaluate the contribution of material properties to the accuracy of the BIR generated using our Listen2Scene method. The BRAS contains a complete scene description, including the captured BIRs (i.e. ground truth) and the 3D models with semantic annotations for a wide range of scenes.
We trained our approach without including the material properties (Listen2Scene-No-Mat) and including material properties (Listen2Scene). We evaluate our approach using recorded BIRs from the chamber music hall and auditorium (Fig.~\ref{bras_becnhmark}). We generated BIRs corresponding to the source and listener positions in the same 3D models and compared the accuracy.
%of the measured BIRs for 2 different 3D scenes using our models with and without including materials.
We plot the normalized early reflection energy decay curves (EDC) of the captured BIRs and the BIRs generated using our models (Fig.~\ref{bras_becnhmark}). The EDC describes the amount of energy remaining in the BIR with respect to time~\cite{EDC}. We observe that in 2 different scenarios, adding material improves the energy decay pattern of the BIRs. We calculated the mean absolute error (MAE) between the EDC of captured BIRs and generated BIRs. MAE decreases by 3.6\% for the medium room and 6.6\% for the large room. \\

{ 
\begin{table}[h!]
\caption{\label{tab:acoustic_param}We calculate the mean absolute reverberation time ($T_{60}$) error, direct-to-reverberant ratio (DRR) error and early-decay-time (EDT) error for monaural IRs generated using MESH2IR and BIRs generated using our approach with materials (Listen2Scene) and without material (Listen2Scene-No-Mat), Listen2Scene-Full, Listen2Scene-No-BIR, and Listen2Scene-ED. We compare them with BIRs computed using the geometric method (\S~\ref{dataset_create}). We compare the monaural IRs generated using MESH2IR with each channel in BIRs separately and compute the average. The best results of each metric are shown in \textbf{bold}.} % We observe a 48\% accuracy improvement of EDC error with our Listen2Scene when compared to MESH2IR.
%\vspace{-0.3cm}
\resizebox{0.9\columnwidth}{!}{
\begin{tabular}{lccc}
\hline
\multirow{2}{*}{\textbf{IR Dataset}} & \multicolumn{3}{c}{\textbf{Mean Absolute Error $\downarrow$}}                                                                                     \\
% \cline{2-4}
& \multicolumn{1}{l}{$\mathbf{T_{60}}$ \textbf{(s)}} & \multicolumn{1}{l}{\textbf{DRR (dB)}} & \multicolumn{1}{c}{\textbf{EDT (s)}} \\
\hline
MESH2IR~\cite{mesh2ir} & 0.16 & 5.06 & 0.25\\
% \hline
Listen2Scene-No-Mat& 0.10 & 3.15 & 0.14  \\
Listen2Scene-Full& 0.10 & 3.18 & 0.16  \\
Listen2Scene-Fix& 0.11 & 2.56 & 0.17  \\
Listen2Scene-No-BIR& \textbf{0.08} & 4.21 & 0.21  \\
Listen2Scene-ED& 0.10 & 3.49 & 0.16  \\
% \hline
\textbf{Listen2Scene} & \textbf{0.08} &\textbf{1.7}  & \textbf{0.13}  \\
\bottomrule
\end{tabular}
}
% \vspace{-0.3cm}
\end{table}
}

%we can see that our proposed approach outperforms the MESH2IR network. We also can see that adding material information reduces the mean absolute acoustic metrics errors and significantly improves the DRR error.

%CAN YOU CALCULATE THE PERCENTAGE IMPROVEMENT OVER PRIOR METHODS (MESH2IR) AND MENTION THAT IN SECTION 1, WHEN YOU TALK ABOUT ACCURACY IN MAIN CONTRIBUTIONS

\vspace{-0.6cm}
\subsection{Accuracy Analysis}
\label{acc_analysis}
We quantitatively evaluate the accuracy of our proposed approach using standard acoustic metrics such as reverberation time ($T_{60}$), direct-to-reverberant ratio (DRR), and early-decay-time (EDT). $T_{60}$ measures the time taken for the sound pressure to decay by 60 decibels (dB). The ratio of the sound pressure level of the direct sound to the sound arriving after surface reflections is DRR~\cite{drr_book}. The six times the time taken for the sound pressure to decay by 10 dB corresponds to EDT. We generate 2000 high-quality BIRs using many rays with the geometric method~\cite{geo_int1} for 166 real scenes not used to train our networks in the ScanNet dataset. We compare the accuracy of Listen2Scene with the BIRs computed using the geometric method on these scenes.

In our Listen2Scene network, we pass the average sound absorption and reflection coefficients at 500 Hz and 1000 Hz as input. In our Listen2Scene-Full variant, the average coefficient over the 8-octave bands between 62.5 Hz and 8000 Hz is given as input. Also, in our Listen2Scene, we simplify the mesh to 2.5\% of the original size. Our Listen2Scene-Fix variant simplifies all the meshes to have a constant number of faces (2000 faces). The motivation behind our approach is that we empirically observed that instead of having a fixed size if we simplify the meshes to 2.5\% of the original size, the contextual information is preserved better. We calculate the mean absolute acoustic metrics error of the BIRs generated using our approach with materials (Listen2Scene) and without materials (Listen2Scene-No-Mat), Listen2Scene-ED, Listen2Scene-No-BIR, Listen2Scene-Fix and Listen2Scene-Full. 
We report the average error from two channels in our generated BIRs (Table~\ref{tab:acoustic_param}). Many prior learning-based approaches are not capable of generating IRs for new scenes not used during training~\cite{luo} or generating BIRs for standard inputs taken by physics-based BIR simulators~\cite{fewshot}. MESH2IR~\cite{mesh2ir} can generate monaural IRs from 3D mesh models.  Therefore, we compare the acoustic metrics of MESH2IR separately with the left and right channels and report the average error. We highlight the accuracy improvements in Table~\ref{tab:acoustic_param}. We can see that our Listen2Scene outperforms MESH2IR and other variants of the Listen2Scene network.\\

\vspace{-0.6cm}

\subsection{Time-domain comparison}
We plot additional time-domain representation of BIRs generated using a geometric-based sound propagation approach~\cite{geo_int1} and our proposed Listen2Scene (Table ~\ref{tab:plot}) for two different 3D scenes. We can see that the amount of reverberation and the high-level structures of the BIRs generated using our approach match BIRs generated using the geometric-based method. Also, we can see that the ILD and ITD in our generated BIRs match the BIRs from a geometric method. The mean absolute error of the normalized BIRs generated using Listen2Scene is less than $0.5$ x $10^{-2}$.

\begin{table}[h!]
    \setlength{\tabcolsep}{0.8pt}
	\caption{The total participants' (acoustic experts and AMT participants) responses on which synthetic speech sample is closer to real-world speech created using captured IRs in the BRAS dataset. We created synthetic speech samples using Listen2Scene and Listen2Scene-No-Material for 2 different real-world environments. The highest comparative percentage is~\textbf{bolded}.  }
	\label{tab:bras_user}
	\centering
        \resizebox{0.95\columnwidth}{!}{
	\begin{tabular}{@{}lccr@{}}	%c-no.of columns. |-verticle line between columns
		\toprule
		\textbf{Environment} & \textbf{Listen2Scene-No-Material}  &  \textbf{Listen2Scene} \\
        
		\midrule
		Chamber music hall&        44.29\%  & \textbf{55.71\%}\\
	    Auditorium    &           21.43\%  & \textbf{78.57\%}\\
		\bottomrule

	\end{tabular}
 }
	\vspace{-0.5cm}
\end{table}

% \subsection{Run time}
% We generated 2500 BIRs for a given 3D scene to calculate the run time. Our network consists of a graph neural network (GNN) and a BIR generator network. For a given 3D scene, we perform mesh encoding using GNN only once and we generate BIRs by varying source and listener positions. On average, our network takes 0.21 seconds to encode the scene using GNN and 0.023 milliseconds to generate a BIR. Therefore on average, our network takes 0.1 milliseconds to generate 2500 BIRs for a given 3D scene.  

\subsection{Run time}
We generated 2500 BIRs for a given 3D scene to calculate the run time. Our network comprises a graph neural network (GNN) and a BIR generator network. For a given 3D scene, we perform mesh encoding using GNN only once, and we generate BIRs by varying source and listener positions. On average, our network takes 0.21 seconds to encode the scene using GNN and 0.023 milliseconds to generate a BIR. Therefore, on average, our network takes 0.1 milliseconds per BIR to generate 2500 BIRs for a given 3D scene. On average, interactive image-based geometric sound propagation algorithm~\cite{image-based} takes around 0.15 seconds to generate an impulse response~\cite{fastrir}. Therefore, our Listen2Scene is more than two orders of magnitude faster than image-based sound propagation methods~\cite{image-based}.

\section{PERCEPTUAL EVALUATION}
\begin{figure}[h!]
  \centering
  \includegraphics[width=0.7\linewidth]{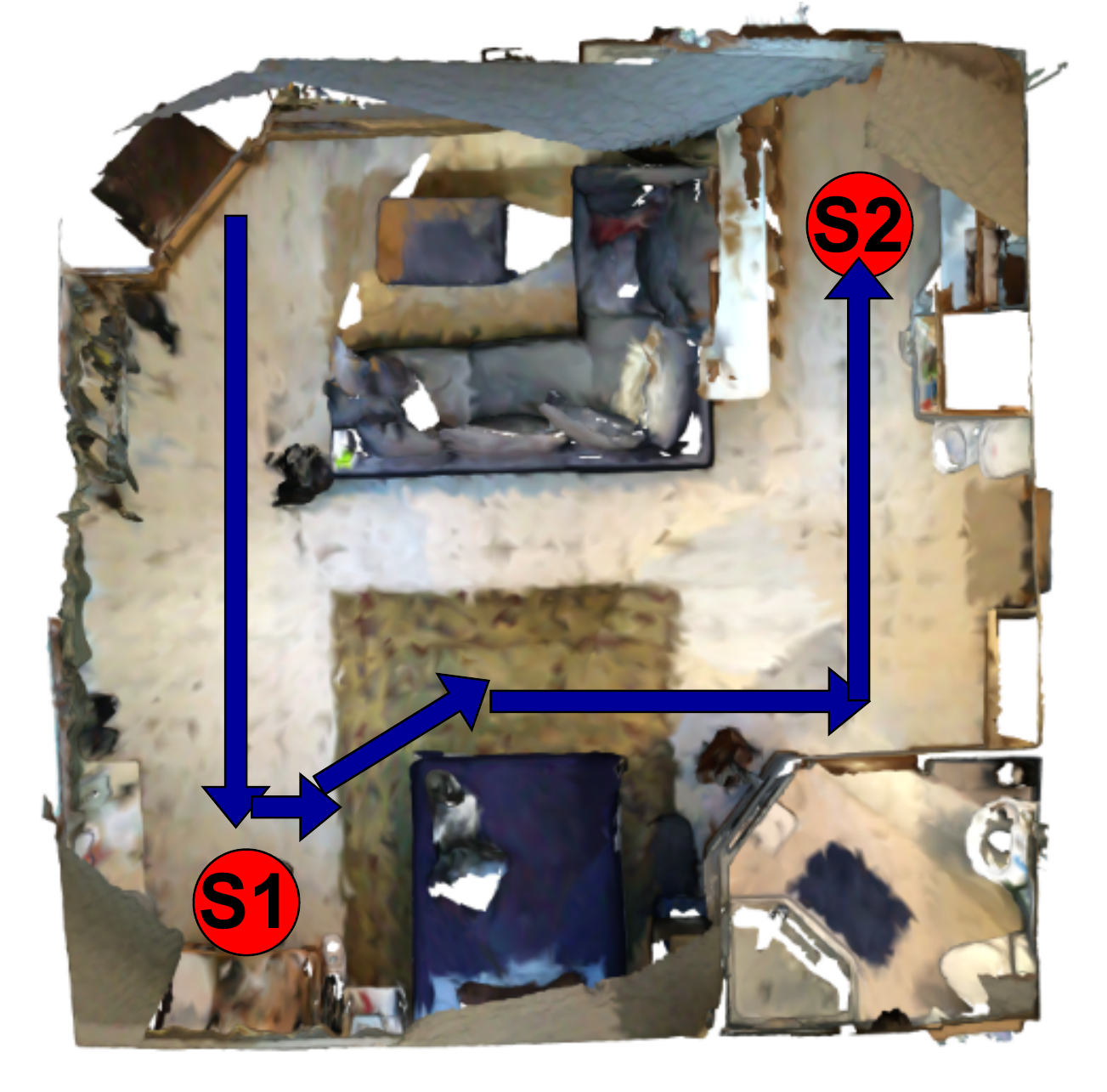} %mesh_net1.eps
  \caption{The path covered by the listener in a real-world 3D scene (studio apartment) with two sound sources. The listener path is shown in blue arrows. The red circle represent the two sound sources in the 3D scene. The source S1 is a speech signal from a speaker, and the source S2 is water pouring from the kitchen.} 
  \label{audio_path}
  \vspace{-0.4cm}
\end{figure}

We perceptually evaluate the audio rendered using Listen2Scene and compare them with prior learning-based and geometric-based sound propagation algorithms. Our study aims to verify whether the audio rendered using our Listen2Scene is plausible (with left and right channels). We auralized three scenes with a single sound source and two scenes with two sound sources from the ScanNet test dataset (more details in the video). Fig.~\ref{audio_scene} shows the snapshot of 5 scenes used to evaluate the quality of our proposed audio rendering method. We created a 40-second video of each scene by moving the listener around the scene. Fig.~\ref{audio_path} shows the listener path in a 3D scene with two sound sources. We evaluate our approach by adding sounds synthesized using different methods to the 3D scene walkthrough:  clean or dry sound (Clean),  sound propagation effects created using MESH2IR, Listen2Scene-No-Material, geometric-based method and Listen2Scene. We also compared the reverberant speech created using  Listen2Scene-No-Material and Listen2Scene with the captured / real-world IRs from two different scenes in the BRAS dataset (Fig.~\ref{bras_becnhmark}). 

\vspace{-0.4cm}
  % \vspace{-0.2cm}
\subsection{Participants}
We conducted our user study among the acoustic experts (13 participants) and the participants from Amazon Mechanical Turk (AMT) (57 participants), an online crowdsourcing platform that can be used to collect data from diverse participants. Since we have a limited number of acoustic experts to evaluate our approach, we also evaluated using AMT. We conducted our user study on $70$ participants ($47$ males and $23$ females), of which 16 participants were between 18 and 24 years of age, 48 participants were between 25 and 34 years of age, and 7 participants were above 35 years of age. We ensured the quality of our evaluation by pre-screening the participants. As our pre-screening questions, we asked the online participants whether they use headphones with a laptop/desktop and only allowed them to proceed with the survey if they answered yes.  The average completion time of our user study is 20 minutes for each user. The just-noticeable-difference (JND) relative reverberation time change is 5\% - 25\%~\cite{jnd}. On average on each rendered 40-second video, the reverberation time changes by 30\%. Therefore, under normal conditions, we expect the listeners to identify the relative changes in the audio correctly.      

% \vspace{-0.2cm}
\subsection{Benchmarks}
%We have included our sound-rendered videos used for our evaluation in the supplementary video.
% Among the participants, 14 were between 18-24 years of age, 44 were between 25-34, and 7 were older than 34. 
We performed the following five benchmark comparisons in perceptual evaluation. Our first four benchmarks compare 40-second-long audio-rendered 3D environment walkthrough videos from our Listen2Scene with baseline methods. In our last benchmark, we compare the real speech with speech rendered using our Listen2Scene and Listen2Scene-No-Material.\\
% \vspace{-0.2cm}
% \begin{itemize}
%to understand how plausible the sound rendered using our approach in 3D scenes.\\
%\item

\textbf{Clean vs. Listen2Scene:} We compared audio-rendered 3D scenes with and without acoustic effects from Listen2Scene. We created two different 3D scene walkthrough videos for our experiment with a single sound source and two sound sources. For a single sound source,  we evaluate whether our approach creates continuous and smooth acoustic effects when moving around the scene and whether the user can perceive the indirect acoustic effects. In the two sound sources walkthrough video, we evaluate whether the relative distance between the two sound sources in the rendered audio using our Listen2Scene matches the video. 
%If the binaural impulse responses do not change smoothly with changes in distance, the participants will prefer the auralized video with clean speech.
%In the 3D scene with 2 sound sources, we investigate whether the participants feel that the location of sound sources is synchronous in the video and audio.\\

%\item 
\textbf{MESH2IR vs. Listen2Scene:} We auralized a 3D walkthrough video each for a single source and two sources. We use the prior monaural audio rendering method MESH2IR and our proposed binaural audio rendering approach, Listen2Scene, for our comparison. We aim to investigate whether the participants feel that the acoustic effects in the left and right ears change smoothly and synchronously as the user walks into the real-world 3D scene. In addition to distance, we investigate whether our acoustic effects change smoothly with the direction of the source. We also evaluated whether our approach is plausible even when there is more than one source in the 3D scene.

%For example, when the source is close to the left ear, the sound level will be high in the left ear when compared to the right ear. 

%\item
\textbf{Listen2Scene-No-Material vs. Listen2Scene:} We auralized two real-world 3D scenes with a single source from a medium-sized and a large 3D scene, and another 3D scene with two sources. In this experiment, we evaluate whether the reverberation effects from Listen2Scene match closely with the environment when compared with Listen2Scene-No-Material. Our goal is to evaluate the perceptual benefits of adding material characteristics to our learning method. The amount of reverberation varies with the size of the 3D scene, therefore we compare the contribution of material to the plausibility of auralized medium and large 3D scenes. In real environments, the listener hears audio from multiple sound sources. Therefore, we evaluate the plausibility of our approach when more than one source is played in the 3D walkthrough video.

%\item
\textbf{Geometric-method vs. Listen2Scene:} We auralized one real-world 3D scene with two sources. In this experiment, we evaluate whether the participants feel the Listen2Scene or the geometric-based sound propagation~\cite{geo_int1} is more plausible for the corresponding 3D scene walkthrough video.

%\item 
\textbf{BRAS benchmark:} 
We played reverberant speech created using captured left channel IRs from the BRAS and left channel impulse responses synthesized using our Listen2Scene and Listen2Scene-No-Material in two different 3D scenes (Fig.~\ref{bras_becnhmark}). We use single-channel IRs to remove acoustic effects from ITD and ILD and make the participants focus on reverberation effects corresponding to the complexity and shape of the environment. We asked the participants to choose which speech sampled auralized using our BIRs is closer to the real speech from the BRAS.
% \end{itemize}

\subsection{Experiment and Results}
\label{exp_res}
In our experiment, we randomly choose the location of two videos (left or right) used for the comparison to eliminate bias from collected data and ask the participants to rate from -2 to +2 based on which video sounds more plausible, i.e. the way the sound varies in both ears when the listener moves towards and away from the sound source. The participants rate -2 if the left video sounds more plausible and vice versa. If the participants have no preference, they rate 0. We group the negative scores (-1 and -2) and positive scores (1 and 2) to choose the participants' preferences.

%Since we had access to a limited number of acoustic experts to conduct our user study, we used additional participants from AMT.
Table~\ref{tab:userstudy} summarises all the participants' responses.  We observe that 67\% - 79\% of the total participants find that the auralized scenes with 1-2 sources using Listen2Scene are more plausible than MESH2IR. Interestingly, 17\% - 27\% of total participants find that just adding clean sound to a 3D scene video is more plausible. When we further break down our results based on age, we observe that 42.86\% to 71.43\% of 35 or older participants prefer adding just clean sound to the video. We believe that this might be caused by an increase in volume from our approach when the listener moves too close to the speaker. All of our participants older than 35 are from AMT, therefore we were not able to get feedback from the participants after the studies. We also observed that when there is more than one source in the 3D scene, the relative sound variation of the sources based on their location is more plausible with Listen2Scene, as compared to using dry sound or  MESH2IR. In large 3D models, where the $T_{60}$ tends to be higher, 66\% of participants feel Listen2Scene is more plausible than Listen2Scene-No-Material. We also can see that 10\% more participants feel our learning-based approach is more plausible than the geometric-based method. The BIRs generated using the learning-based method smoothly change with the distance and listeners can feel a smooth transition in audio when they move to different positions in the 3D scene. From Table~\ref{tab:bras_user}, we can see that audio/speech rendered using our Listens2Scene approach is closer to the real-world speech. Overall, we notice that our approach creates plausible acoustic effects when there are one or more sound sources in the 3D scene.

\vspace{-0.1cm}
% \section{CONCLUSION, LIMITATIONS AND FUTURE WORK}
\section{CONCLUSION LIMITATIONS AND FUTURE WORK}
We present a material-aware learning-based sound propagation approach to render thousands of audio samples on the fly for a given real 3D scene. We propose a novel approach to handle material properties in our network. Moreover, we show that adding material information significantly improves the accuracy of BIR generation using our Listen2Scene approach and is comparable to geometric propagation methods or captured BIRs in terms of acoustic characteristics and perceptual evaluation.
%We perform perceptual evaluation and observe that acoustic effects generated using our approach are more plausible than the state-of-the-art learning-based IR generators. 
Overall, our algorithm offers two orders of magnitude performance improvement over interactive geometric sound propagation methods.

Our approach has some limitations.
The performance of our network depends on the training data. We can train our network with real captured BIRs, though it is challenging and expensive to capture a large number of such BIRs. Currently, we use BIRs generated using geometric algorithms for medium-sized 3D scenes in the ScanNet dataset for training, and the overall accuracy of Listen2Scene is also a function of the accuracy of the training data.
Our approach is limited to static real scenes.
%and its accuracy is governed by the underlying 3D mesh representation.
%to generate the most accurate BIRs for new 3D scenes not used for training. We show that adding average acoustic material coefficients significantly improves the performance of the network. 
Our material classification methods assume that accurate semantic labels for each object in the scene are known.
%Acoustic material coefficients are frequency dependent and specified for 8-octave bands between 62.5 Hz and 8000 Hz. 
It is possible to consider sub-band acoustic material coefficients to further improve the accuracy. However,  the complexity of the graph representation of the 3D scene drastically increases, and we are limited by the GPU memory in handling such complex graphs. Due to the limitation of the training dataset used for training, the performance of our network has been currently evaluated on small and medium-sized scenes. In future work, we like to train and evaluate our approach on very large scenes. Since the ScanNet dataset does not have the same 3D environment with different structural changes, we are not able to train and evaluate different structural detail resolutions. As part of future work, it would be useful to analyze our learning-based sound propagation approach on different structural detail resolutions.

%As part of our future work, we would like to evaluate the performance on other reconstructed models. 
% Furthermore, we would like to integrate our sound propagation and rendering algorithm with AR and VR systems.

% As part of our future work, we would like to evaluate the performance on other real reconstructed models. Furthermore, we would like to integrate our sound propagation and rendering algorithm with AR and VR systems.
% The main limitation of our network is that we assume a point sound source and we do not consider the orientation of the source. Also, our network cannot be explicitly controlled using the listener's head orientation. We are constrained to add these features because the geometric-based simulator we used to create the dataset cannot be explicitly controlled using head and source orientation. Controlling the BIR generator using the head and source orientations will improve accuracy in applications like 3D virtual tour. 

% \textcolor{blue}{}

%\bibliographystyle{abbrv}
\bibliographystyle{abbrv-doi}

\bibliography{template}
\end{document}